\documentclass[pre,aps]{revtex4}
\usepackage{epsfig}
\usepackage{amstext}
\begin{document}

\title{Inferring DNA sequences from mechanical unzipping data: 
the large-bandwidth case.}

\author{V. Baldazzi $^{1,2,3}$, S. Bradde $^{2,4}$, 
S. Cocco $^2$, E. Marinari $^4$, R. Monasson $^3$}
\affiliation{$^1$ Dipartimento di Fisica, Universit\`a di Roma
{\em Tor Vergata}, Roma, Italy\\
$^2$ CNRS-Laboratoire de Physique Statistique de l'ENS, 24 rue Lhomond,
75005 Paris, France\\
$^3$ CNRS-Laboratoire de Physique Th\'eorique de l'ENS, 24 rue Lhomond,
75005 Paris, France\\
$^4$ Dipartimento di Fisica and INFN, Universit\`a di Roma
{\em La Sapienza}, P.le Aldo Moro 2,  00185 Roma, Italy}

\begin{abstract}
The complementary strands of DNA molecules can be separated when
stretched apart by a force; the unzipping signal is correlated to the base content of the sequence but is affected by
thermal and instrumental noise. We consider here the ideal case where
opening events are known to a very good time resolution (very large
bandwidth), and study how the sequence can be reconstructed from the 
unzipping data. Our approach relies on the use of statistical
Bayesian inference and of Viterbi decoding algorithm.
Performances are studied numerically on Monte Carlo generated data,
and analytically.  We show how multiple unzippings of the same
molecule may be exploited to
improve the quality of the prediction, and calculate analytically
the number of required unzippings as a function of the bandwidth, the
sequence content, the elasticity parameters of the unzipped strands.
\end{abstract}
\maketitle

\section{Introduction }

As DNA molecules are the support for the genetic information,
the knowledge of their sequence content is very important both
from the biological and medical points of view. Over the last decade 
the sequencing of various genomes, in particular the 
human one, was done at the price of intense efforts.
A traditional strategy for reading a DNA molecule is based on the 
so-called Sanger method \cite{mb,mb2}. The DNA molecule is divided 
into fragments (with $N\sim 100-1000$ base pairs); each fragment 
is amplified through PCR. The copies of each fragment are denaturated, 
and double-stranded DNA subfragments are synthesized under the action
of DNA polymerases. The key point is that each of the four nucleotides 
$A,T,C,G$  is present in solution 
under its normal form at high concentration and under a modified form, 
tagged with a base-specific fluorescent label and inadequate for 
further polymerization, at low concentration. At the end of the
polymerization step many copies of each fragment are obtained.
The copies of a fragment have a common extremity and have various
lengths $L$, with a base-specific fluorescent base $B$ at the end.
The entire population of copies is sorted by length using gel
electrophoresis and the sequence of the fragment is reconstructed 
from the list of terminal bases $B(L)$, $1\le L\le N$. 
The method correctly predicts 99.9\% of the bases of a fragment, but 
additional errors may arise during the reconstruction of the whole 
sequence from its fragments.

Despite the success of conventional sequencing the quest for alternative 
(faster or cheaper) methods is  an active field of research. 
Recently  various single molecule experiments were  carried out, allowing 
a direct investigation of DNA mechanics and protein-DNA interaction
~\cite{Bus03,Coc21,Smi92,Clu96,Smi96,Ess97,Boc98,Boc02,Boc04,Felix06,Lip01,
Dan03,Har03,Van03,Per03,Wui00,Mai00,Lev03,Lan03,Sau03,Mat04}.
These experiments provide dynamical information usually hidden in large scale 
bulk experiments, such as intermediate metastable states or fluctuations 
at the scale of the individual molecule.
Remarkably, these dynamical effects are largely sequence--dependent
in various experimental situations e.g.
the opening of the double helix under a
mechanical stress \cite{Ess97,Boc98,Boc02,Boc04,Felix06,Lip01,Dan03,Har03},
the digestion of a DNA molecule by an exonuclease \cite{Van03,Per03},
DNA polymerization \cite{Wui00,Mai00,Lev03}, translocation 
through nanopores \cite{Sau03,Mat04}.  Understanding how much 
information about the sequence is contained in the measured signals
is important.

\begin{figure}
\begin{center}
\psfig{figure=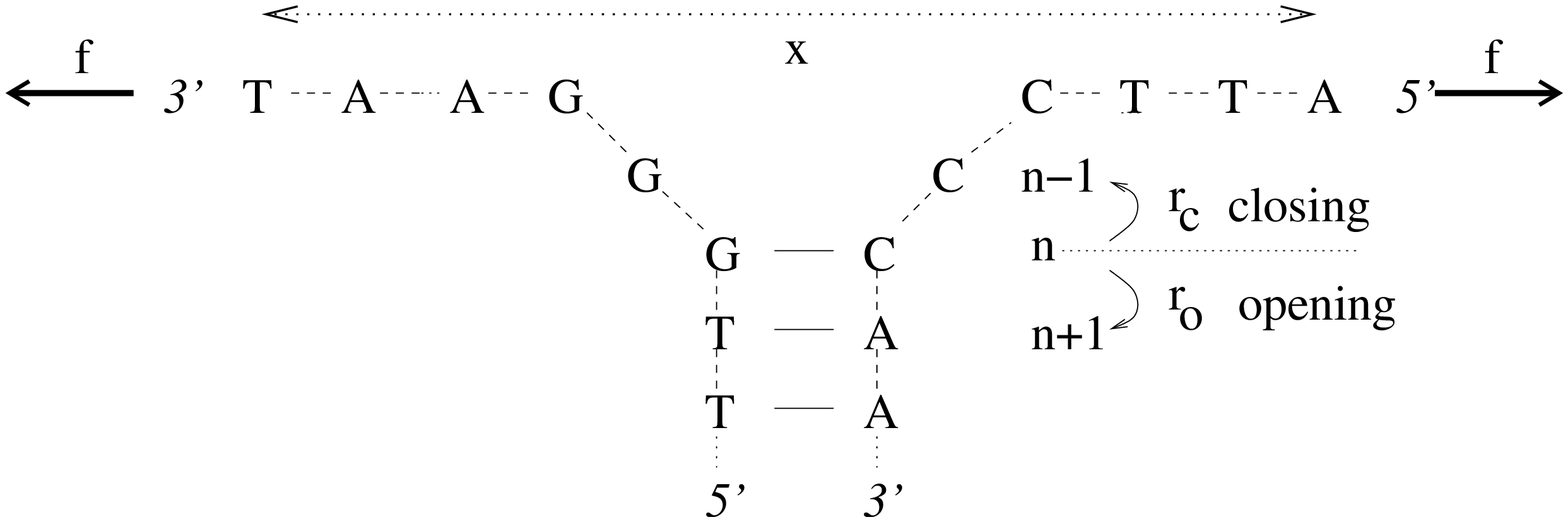,height=4cm,angle=0}
\end{center}
\caption[]{Sketch of a fixed-force unzipping experiment: the adjacent 5' 
and 3' extremities of a DNA molecule are submitted to a constant force 
$f$. The distance between the extremities, $x$, is measured as
a function of time. $x$ is proportional to the number $n$ of
open base pairs (bp) up to some fluctuations due to the floppiness of the
unzipped strands. The number $n$ of open bp increases or decreases by one
with rates $r_o$ and $r_c$ respectively, see dynamical model in
Section \ref{direct}. }
\label{fig1}
\end{figure}

Hereafter,we focus on mechanical unzipping experiments,
first introduced by Bockelmann and Heslot in 1997 \cite{Ess97}.
The complementary strands are pulled apart  at a constant velocity
while the force necessary to the opening is measured.
The average opening force for the $\lambda$ phage is of  about 15 pN, with
fluctuations around this value that depend on the particular sequence
content. In a more recent experiment, 
Bockelmann, Heslot and collaborators have shown that the force 
signal is correlated to the average sequence on the scale of ten base pairs 
but could be affected by the mutation of 
one base pair adequately located along the sequence \cite{Boc02}.

Liphart et al. \cite{Lip01} and Danilowicz et al. \cite{Dan03}  have 
performed an analogous experiment, using a constant force setup, on a 
short RNA and a long DNA respectively. As sketched in Fig \ref{fig1}, the  distance between the two 
strands extremities is measured as  a function of the time while the
molecule is submitted to a constant force. The dynamics 
is  characterized by rapid zipping or unzipping jumps followed by long 
pauses where the unzipped length  remains constant. Several repetitions  
have shown that positions and duration of these plateaus are largely 
reproducible, thus providing a 'fingerprint' of the sequence.
The theoretical description of the DNA mechanical unzipping, at 
constant velocity and constant force, has been extensively developed 
 \cite{Boc98,Felix06,Coc3,Bun06,Lub,Hwa,Coc4, Ger04,Felix,mar,siggia} .
Models have been able to reproduce the force (for constant velocity 
experiments) or position (for constant force experiments) signals
given the DNA sequence. It is a natural question to ask  whether one 
could, inversely, get information about the sequence from experimental 
data \cite{batta}. 

This question was addressed by us in a recent letter \cite{corto}. It 
was found that the error in the prediction e.g. the probability that a base
is erroneously predicted decreases exponentially with the amount of 
available data. The decay rate was shown to depend on the sequence 
content, the applied force, the time and space resolution, ...  
The goal of the present paper is to provide a complete presentation of 
the numerical and analytical work supporting the results of \cite{corto}
in the idealized case of perfect time and space resolutions. 
Though this case is not realistic from an experimental point of view, it
can be studied in great detail. We show that the most important result,
the exponential decay of the probability of misprediction with the
amount of collected data, holds in more realistic situation where 
the bandwidth and the fluctuations in the extension of the DNA strands
are taken into account. Our analysis focuses 
on the fixed force device data which is somewhat simpler from a 
theoretical point of view. 

In Section II we  first introduce the dynamical model that, given
a sequence, determines the unzipping signal. The inverse problem
is then introduced and treated within the Bayesian inference framework.
Section III reports the numerical results for the quality of
prediction from numerical data obtained from the Monte Carlo simulation 
of the unzipping of a $\lambda$-phage DNA. The analytical study of 
inference performances is presented in section IV. While the above
study assumed the existence of infinite temporal and spatial
resolution over the fork location the effects of realistic limitations
are studied in Section~V. A summary and discussion of the results is
presented in Section~VI.

\section{Bayesian inference framework}

The direct problem of fixed-force DNA unzipping is to determine, given the 
sequence of the molecule, the distribution of the stochastic measured
signal, that is, the extension between the two strands extremities as
a function of time. The direct problem 
is considered in Section \ref{direct}, and results are used in
Section \ref{inverse} to address the inverse problem, that is, the
prediction of the sequence given a measured extension signal.

Throughout this section we consider that the experimental signal gives access to
the number of open bases itself rather than the distance
between the extremities of the unzipped strands. This is merely an approximation
since, due to the fluctuations in the extension of strands, the number of open bases 
is not in one-to-one correspondence with the distance between the strands. Corrections to
this simplifying  assumption will be discussed in Section \ref{flucsec}. 

\subsection{From sequence to signal: the direct problem}
\label{direct}

\begin{figure}
\begin{center}
\psfig{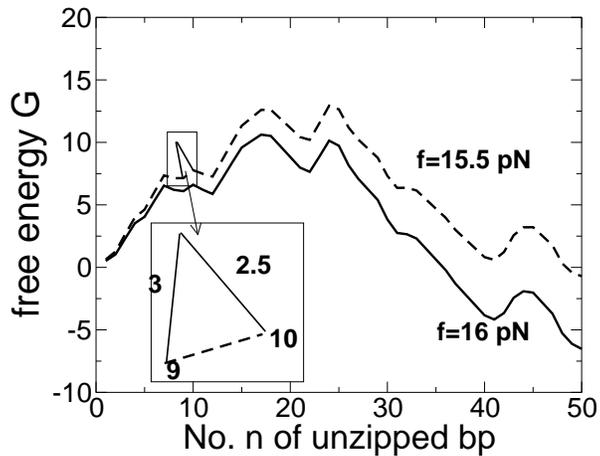} 
\end{center}
\caption[]{Free energy $G$ (units of k$_B$T) 
to open the first $n$ base pairs, for the first
50 bases of the DNA $\lambda$--phage at forces 15.9 (dashed curve) and 16.4 pN
(full curve). For $f=15.9$ pN the two minima at bp 1 and bp 50 are separated 
by a barrier of 12 k$_B$T. Inset: additional barrier representing the 
dynamical rates (\ref{ratemd}) to go from base 10 to 9 (barrier equal to
$g_s$=2.5  k$_B$T), and from base 9 to 10 (barrier equal to
$g_0(b_9,b_{10})$=3  k$_B$T), see text. }
\label{gf50}
\end{figure}

\begin{figure}
\begin{center}
\psfig{figure=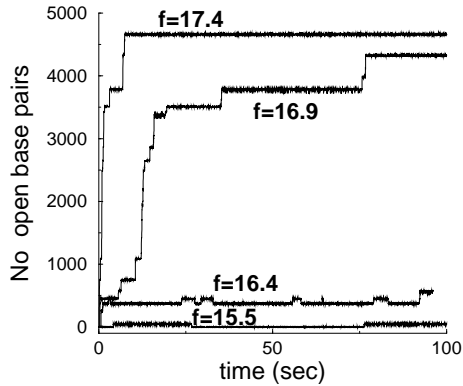,height=6cm,angle=-90}
\end{center}
\caption{Number of open base pairs as a function of the time for
  various forces (shown on Figure). 
Data show one numerical unzipping (for each force) obtained from a Monte Carlo
simulation of the random walk motion of the fork with rates (\ref{ratemd}).}
\label{dinf15.5}
\end{figure}

\begin{table}
\begin{center}
\begin{tabular}{|c|c|c|c|c|}
\hline $g_0$ &A&T&C&G \\ \hline A &1.78 &1.55 &2.52 & 2.22\\ 
                   \hline T &1.06 &1.78 &2.28 &2.54 \\
                   \hline C& 2.54 &2.22 &3.14& 3.85  \\
                    \hline G & 2.28& 2.52 & 3.90&3.14\\ \hline
\end{tabular}
\end{center}
\caption{Binding free energies  $g_0(b_i,b_{i+1})$ (units of k$_B$T) obtained from the
 MFOLD server  \cite{Zuk,San} for 
DNA at room temperature, pH=7.5, and ionic concentration of
 0.15 M. The base values $b_i, b_{i+1}$ are given by the line and
column respectively.} 
\label{tableg0}
\end{table}

In a previous work we have developed a theoretical description of the 
dynamics of DNA and simple RNA molecules under a constant 
unzipping force \cite{Coc4}. Despite its simplicity this model is
capable of reproducing the unzipping data for a given sequence~\cite{Lip01,Dan03} 
 and the rezipping dynamics of a partially unzipped DNA \cite{Boc04}.

Let $b_i=A,T,C$, or $G$ denote the $i^{th}$ base along the $5'\to3'$
strand (the other strand is complementary), and
${B}=\{b_1,b_2,\ldots , b_N\}$.  The free energy
excess when the first $n$ bp of the molecule are open with respect to
the closed configuration ($n=0$) is
\begin{equation} \label{p}
G(n,f;{B})= \sum_{i=1} ^n g_0(b_i,b_{i+1})- n\, g_{s} (f) \ .
\end{equation}
and involves two contributions.
The first free energy, called $g_0(b_i,b_{i+1})$ is the binding energy of 
base pair (bp) number $i$; it depends on $b_i$ (pairing interactions)
and on the neighboring bp $b_{i+1}$ due to stacking interactions.
$g_0$ is obtained  from the MFOLD server \cite{Zuk,San}, and listed
in Table~\ref{tableg0}. 
The second contribution, called $g_{s}(f)$ is the work to stretch
the two opened single strands when one more bp is opened. The elasticity 
of DNA strands is described by a 
modified freely  jointed chain with a Kuhn length $\ell _0=15${\AA} and an 
effective  nucleotide length $\ell= 5.6$ {\AA}
\cite{Smi96}. The corresponding free energy for forces up to 20 pN is
\begin{equation}
{g}_{s}(f)= 2\, f\, \ell\,\ln \big[\sinh(z)/z\big]/z 
\quad \hbox{\rm with}\quad z\equiv f\, \ell _0  / (k_B T) \ .
\end{equation}

As an illustration the free energy $G(n,f;\Lambda )$ of the first 
50 bases of the $\lambda$ phage sequence, $\Lambda = (\lambda _1,
\lambda _2, \ldots ,\lambda_N)$, is plotted in
Fig~\ref{gf50} for forces $f=15.9$ and $16.4$ pN. At these forces the two
global minima are located in $n=1$ (closed state) and $n=50$
(partially open state).
Experiments on a small RNA molecule, called P5ab, \cite{Lip01} have shown that,
at the critical force $f_c$  such that the closed state  has the same
free energy than the open one: $G(0,f_c; B)=G(N,f_c;B),$
 the barrier between these two minima is not too high, the
molecule then switches between these two states.
For long molecule e.g. $\lambda$--DNA the  barrier
 between the closed and open states mya become very large e.g.
$\sim 3000$ k$_B$T for the $\lambda$--DNA at the critical force 
$f_c=15.5$ pN \cite{Coc4}. The time it takes to cross this barrier is
 huge and full opening of the molecule never happens during
 experiments (unless the force is chosen to be much larger than its
 critical, infinite time value). The experimental opening signal is 
 characterized by pauses at local
minima of the free energy $G(n,f;\Lambda )$ and rapid jumps between
them \cite{Dan03}.
  This dynamical behavior is reproduced (Fig~\ref{dinf15.5}) when one
considers that the fork separating the closed from the open regions
along the molecule undergoes a random walk motion in the free energy
landscape $G(n,f;\Lambda )$ \cite{Coc4}. The fork, located at position $n$,
can move forward ($n\to n+1$) or backward ($n\to n-1$) with rates
(probability per unit of time) equal to, respectively,
\begin{equation}
r_o(b_n,b_{n+1})= r\; \exp \big[ g_0(b_n,b_{n+1})\big] \ , \quad 
r_c= r\; \exp  \big[{g_{s} (f)}\big] 
\label{ratemd}
\end{equation}
see Fig \ref{fig1}. The value of the attempt frequency $r$ is of the order
of $10^{6}$ Hz \cite{Coc4,Felix,Boc04}. Notice that the free-energies are measured in units of k$_B$T. 

The expression (\ref{ratemd}) for the rates is derived from the
following assumptions. First the rates should satisfy detailed balance.
Secondly we impose that the opening rate $r_o$ depends on the binding 
free energy, and not on the force, and vice-versa for the closing 
rate $r_c$. This choice is motivated by the
fact that the range for base pairs interaction is very small: the
hydrogen and stacking bonds are broken when the bases are kept apart
at a fraction of \AA~, while the force work is appreciable on the
distance of the opened bases ($\approx 1$ nm).  On the contrary, to
close the base pairs, one has first to work against the applied force,
therefore the closing rate $r_c$ depends on the force but not on
the sequence. This physical origin of the rates is reported in the
the inset of Fig~\ref{gf50}. Notice that, as room temperature is much smaller than the thermal denturation temperature, 
we safely discard the existence of denatured bubble in the zipped DNA portion.

\subsection{From signal to sequence: the inverse problem.}
\label{inverse}

\begin{figure}
\begin{center}
\psfig{figure=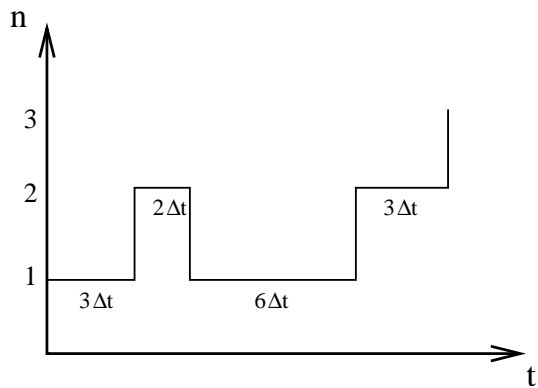,height=5cm,angle=0} 
\caption[]{Fork position $n$ as a function of time 
$t=i\times \Delta t$ with $i$ 
integer--valued; the sojourn times on each base are given. We call
$t_i$ the total time spent on base $i$, and $u_i,d_i$ the numbers of
$i\to i+1,i\to i-1$ transitions respectively. 
Assuming the fork does not come back to $n=1$ or 2 at
later times, we have: $t_1/\Delta t=9$, $u_1=2$, $d_1=0$, and
$t_2/\Delta t=5$, $u_2=1$, $d_2=1$.}
\label{schemadatiseq}
\end{center}
\end{figure}

We consider here the ideal case where the experimental 
setup is not affected by any instrumental noise: 
data are acquired with a infinite temporal resolution, and, in addition,
the unzipped strands do not fluctuate in length.
The latter assumption will be lifted in Section \ref{flucsec}, while
the case of a large but not infinite bandwidth will be
studied in Section \ref{mstepsec}. 

In the absence of DNA strands fluctuations the distance between
extremities is exactly proportional to the number $n$ of unzipped bases.
The measured signal is thus the time trace $T = (i_0,i_1,i_2, 
\ldots,i_M)$ where $i_m$ is the position of the fork at time 
$m\times \Delta t$, and $t_{exp}=M\, \Delta t$ is the duration of the
experiment. The infinite bandwidth assumption amounts to postulate 
that the delay $\Delta t$ between two measures is 
smaller than the sojourn time on a base. Therefore successive positions
$i_m,i_{m+1}$ differ by $\pm 1$ at most.   
A typical result of this idealized experimental situation is sketched in  
Fig.~\ref{schemadatiseq}. The signal is stochastic due to the thermal
motion of the fork in the landscape of Fig \ref{gf50}: 
two repetitions of the experiment do not yield the same time-traces.
The probability of a time-trace $T$, given the sequence $B$, reads
\begin{equation}
{\cal P}({T}|B) = \prod _{m=1} ^{M-1} \left\{
\begin{array} {c c c}
\Delta t\; r_o(b_{i_m},b_{i_{m+1}}) & \hbox{\rm if} & i_{m+1}=i_m+1 \\
\Delta t\; r_c & \hbox{\rm if} & i_{m+1}=i_m-1 \\
1 - \Delta t\; ( r_o(b_{i_m},b_{i_{m+1}})+r_c) & \hbox{\rm if} & i_{m+1}=i_m
\end{array} \right. \ .
\end{equation}
This probability can be conveniently rewritten through the introduction
of the numbers $u_i$ and $d_i$ of, respectively, up 
($i_m=i\to i_{m+1}=i+1$) and down ($i_m=i\to i_{m+1}=i-1$) transitions
from base $i$, as well as
the total time $t_i$ spent on base $i$ (number of sojourn events
$i_m=i \to i_{m+1}=i$, multiplied by $\Delta t$) in the time-trace
${\cal T}$,
\begin{equation}
{\cal P}({T}|B) = \prod _{i} \big[ \Delta t\; r_o(b_i,b_{i+1})
\big]^{u_i} \; \big[ \Delta t\; r_c \big]^{d_i}
\; \big[ 1 -  \Delta t\; ( r_o(b_i,b_{i+1})+r_c) \big] ^{t_i/\Delta t} 
= C({T}) \times  \prod _{i}  M\,(b_i,b_{i+1}; u_i,t_i)
\label{pdetb}
\end{equation}
where 
\begin{equation} \label{defmatrixmbayes}
M (b_i,b_{i+1} ;t_i,u_i)= \exp \big[ g_0(b_i,b_{i+1}) \; u_i -  r \,  
 e^{g_0(b_i,b_{i+1})}\; t_i \big]
\end{equation}
and $C({T})=\Delta t^{u+d}\, r_c^d \exp(-r_c\,t_{exp})$, $u=\sum
_i u_i$,  $d=\sum_{i} d_i$, and 
we have used the fact that $\Delta t$ is small with respect to
the average sojourn time on a base, $(r_o+r_c)^{-1}$. Up to
the multiplicative factor $C({T})$ (which does not depend on the sequence 
$B$), the probability ${\cal P}({T}|B)$ 
is equal to the product of terms $M$ 
expressing the interactions between adjacent bases (\ref{defmatrixmbayes}).

The probability that the DNA sequence is $B$ given the observed 
time-trace ${T}$ is, in the Bayesian inference framework \cite{mckay}, 
\begin{equation} \label{bayes}
{\cal P}( B | {T})= \frac{{\cal P} ({T}|B) \,{\cal P}_0(B)}
{{\cal P}({T})}
\end{equation}
The value $B^*(T)$ of the sequence maximizing this probability, for
a given time-trace $T$, is our prediction for the sequence. In the 
absence of any knowledge over the sequence $B$ the {\em a priori} 
distribution over the sequences, ${\cal P}_0$, is uniform and equal to 
$4^{-N}$.  A straightforward albeit important consequence of 
(\ref{bayes}) is that $B^*(T)$  can be found from the maximization of 
${\cal P}(T|B)$ (\ref{pdetb}). We will briefly see in Section 
\ref{shannonsec} an alternative way of predicting sequences from 
the probability (\ref{bayes}). 

In practice $B^*(T)$ can be exactly found in a time growing linearly 
with $N$  only with the Viterbi algorithm
\cite{viterbi,mckay}. The principle of the algorithm is equivalent to a zero
temperature  transfer matrix technique. We start from the 
first base and choose the optimal value of this base for each possible value
of the second one; in this way we assign a probability $P_2$ to each value 
$b_2$ of the second base through
$P_{2}(b_{2})=\max_{b_1 } M(b_1,b_{2};t_1,u_1)$. 
Then we optimize on the second base,  and obtain
$P_{3}(b_{3})= \max_{b_2} \, M(b_2,b_{3}; t_2,u_2)  \; P_2(b_2)$, 
and so on,
\begin{equation} \label{recur}
P_{i+1}(b_{i+1})= \max_{b_i} \,M(b_i,b_{i+1};t_i,u_i)\;P_i(b_i)
\end{equation}
until we reach the last base $N$ of the sequence. At each step, the 
maximum of (\ref{recur})  is reached for some base $b_i^{max}(b_{i+1})$
that depends on the choice of the next base $b_{i+1}$.
Once the value $b_N^{*}$ that optimize  $P_N(b_N)$ has been calculated, 
one obtains the whole optimal sequence using the recursive relation  
$b_{i-1}=b^{max}_{i-1}(b_i^*)$ until the first base of the chain.

A direct application of the procedure may produce substantial numerical
errors due to the product of a large number of terms. It turns out
convenient to introduce the logarithms of the probabilities, 
$\pi_i (b_i) =-\ln P_i(b_i)$, and solve the recurrence relation
\begin{equation}
\pi_{i+1}(b_{i+1})=\min_{b_i} \big[ \pi_i(b_i) - 
g_0(b_i,b_{i+1}) \; u_i +  r \,  e^{g_0(b_i,b_{i+1})}\; t_i
\big] \ ,
\label{recurpi}
\end{equation}
obtained from (\ref{recur}).

If more than one unzippings are performed on the same molecule,
several time-traces $T_1,T_2, ...,T_R$ are available. As all
unzippings are independent of each other we have
\begin{equation}\label{bayesrun}
{\cal P} ( T_1,T_2, ...,T_R | B) = \prod _{\rho=1} ^R 
{\cal P} ( T_\rho| B) 
\end{equation}
where the distribution of a single time-trace is given by (\ref{pdetb}).
It is immediate to check that equations (\ref{recur}) and (\ref{recurpi}) are
still valid provided $u_i$ and $t_i$ are, respectively, the total 
number of transitions $i\to i+1$ and the total time spent on base
$i$. Total means that these numbers have to be computed from the all $R$
time-traces taken together.

\subsection{Estimators of performances}

As in the previous Section, we consider a time-trace $T$, and call 
$B^*(T)$ the 
sequence with maximal probability given those data. The true sequence
is denoted by $B^L$; in most applications $B^L=\Lambda$, the phage sequence
but we will consider other e.g. repeated sequences. We focus on the 
indicators
\begin{equation}
v _i (T)  = \left\{ \begin{array} {c l}
1 & \hbox{\rm if base $i$ is correctly predicted
{\em i.e.} } b^*_i(T)= b^L_i \\ 0 & \hbox{\rm
otherwise}
\end{array} \right.
\end{equation}
As the time-trace $T$ is stochastic, so are the $v_i(T)$s. Our numerical and
theoretical analysis
aim at calculating some statistical properties of these indicators.
For instance the probability that base $i$ is not correctly predicted
is given by
\begin{equation} \label{defomi}
\epsilon _i = 1- \langle v_i (T)\rangle \ ,
\end{equation}
where the average value $\langle . \rangle$ is taken over the probability
${\cal P}(T|B^L)$ of time-traces given the true sequence $B^L$.
The two-points connected correlation function,
\begin{equation} \label{defcorrel}
\chi _{i,j} =  \langle v_i (T)\, v_{j}
(T)\rangle - \langle v_i (T)\rangle\,  \langle v_{j}(T)\rangle  \ ,
\end{equation}
tells us how much a correct prediction on base $i$ influences the
quality of prediction on base $j$. 
From this local quantities we define the global error and correlation
functions through, respectively,
\begin{equation} \label{defom}
\epsilon = \frac{1}{N} \sum_{i=1}^N \epsilon _i\quad , \quad
\chi _d = \frac{1}{N-d} \sum_{i=1}^{N-d} \chi _{i,i+d}\ .
\end{equation}
Note that the zero-distance correlation
function is simply $\chi _{0} = \epsilon (1- \epsilon)$ in the limit of
large sequences.

\section{Numerical Analysis}

\subsection{Maximum probability prediction}
\label{mlp}

To test this inference method we have generated ideal opening data
from the sequence $\Lambda$ of the $\lambda$--phage with a Monte
Carlo procedure.  Once a time-trace $T$ has been produced a
second program ignoring the phage sequence and based on the Viterbi
algorithm allows us to make a prediction on the sequence, $B^*(T)$.  

\subsubsection{Generation of numerical time-traces}

The unzipping signal $T$ is obtained through a Monte
Carlo (MC) simulation with opening and closing rate defined by the model
(\ref{ratemd}). To save time, at each MC step, the fork moves by one base 
pair, either forward or backwards, without remaining on the same base.
Prior to the move the sojourn time $t$ on the base where the fork is, say, 
$i$, is randomly chosen according to an exponential distribution with
characteristic time $\tau = 1/(r_o(i)+ r_c)$. Then, the fork moves
backward ($i\to i-1$) with probability $q=r_c \,\tau $, and forward ($i\to 
i+1$) with probability $1-q$.

\begin{figure}
\begin{center}
\psfig{figure=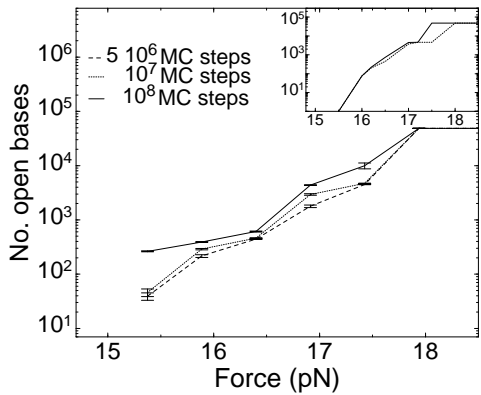,height=6cm,angle=-90}
\caption[]{Number of open bases as a function of applied force, and for
$5\times 
10^6,10^7,10^8$ Monte Carlo steps. Data are averaged over 100 samples. 
The durations of the unzippings are, respectively, of 7, 15, and 140  
seconds. The DNA $\lambda$-phage includes 48,502 bp.
In inset we report the theoretical estimate of the number of open base 
pairs, for $10^7$ and 1$0^8$ MC steps, of Section \ref{avuisec}.}
\label{basi1run}
\end{center}
\end{figure}

The total number of open base pairs increases with the duration of the
opening experiments {\em i.e.} with the number of MC steps as shown in Fig
\ref{basi1run}. The higher the force the more tilted the free energy
landscape, and the larger is the number of open bases. With $10^7$ MC
steps we typically open 290 bp at 15.9 pN, 450 bp at 16.4 pN, and 4700
bp at 17.4pN; each numerical unzipping lasts for $\sim 15$ sec. 

The temporal resolution is introduced by filtering the output dynamics 
with a time step $\Delta t$. Fork positions $n_i$ are registered at
times $t_i=i\times\Delta t$. Each time-trace is then
preprocessed to obtain the numbers $u_i$ of $i\to i+1$ transitions and
the set of times $t_i$ spent on each base $i$. 
The set of data $\{u_i,t_i\}$ is then passed to the Viterbi procedure. 

\subsubsection{Results for global estimators}

\begin{figure}
\begin{center}
\psfig{figure=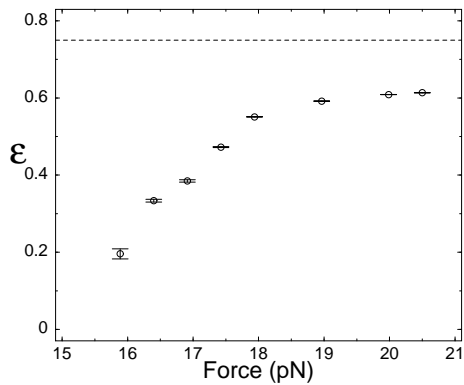,height=6cm,angle=-90}
\caption[]{Fraction $\epsilon$ (\ref{defom}) of mispredicted bases 
as a function of the force for the $\lambda$-phage sequence. Data
are averaged over 100 samples and shown with standard deviations. 
The dotted line $\epsilon =0.75$ shows the failure rate for a random 
choice of one base among the four base values. }
\label{omega1run}
\end{center}
\end{figure}

We show in Fig~\ref{omega1run} the average fraction of mispredicted
bases, $\epsilon$ (\ref{defom}), as a function of the force. For each
time-trace we calculate the fraction of the opened bases that were
incorrectly predicted, and then average over MC time-traces (samples).
$\epsilon$ increases with the force because the number of predicted
(open) base pairs (Fig~\ref{basi1run}) increases, and the time the
opening fork spends on each base decreases.  At a force of 16 pN 80\%
of the predicted bases are correct.  As the force increases $\epsilon$
approaches 0.75, which corresponds to a random guess among four possible
bases.

\begin{figure}
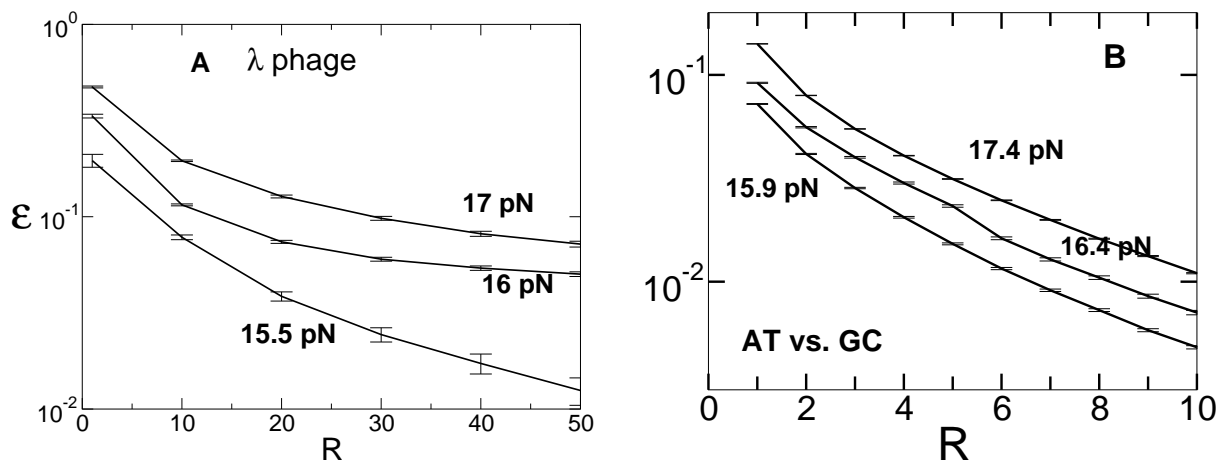

\begin{center}
\psfig{figure=fig7a.eps,height=6cm,angle=0}
\hskip .5cm
\psfig{figure=fig7b.eps,height=6cm,angle=0}
\caption[]{{\bf A}. Error $\epsilon$ as a function of the number of
unzippings for the phage. {\bf B}. Same as A but without
distinguishing $A$ from $T$ and $G$ from $C$, see text. }
\label{sigmarun}
\end{center}
\end{figure}

The quality of prediction is, not surprisingly, greatly improved by
the repetition of the numerical unzipping on the same molecule.
Let $R$ denote the number of time-traces (of the same duration)
available. We show in Fig.~\ref{sigmarun} how the error $\epsilon$
decreases with $R$. Notice that the error is calculated over the
bp that have been opened at least once in all $R$ unzippings.
When opening and closing several times the molecule, the
opening fork makes multiple passages through the same portion of the
sequence; in this way more information on the waiting and transition
times on each base are collected, and processed altogether by the
Viterbi algorithm. Figure \ref{sigmarun} indicates that the error
decreases exponentially with $R$, an observation that will
find theoretical support in Section \ref{theory}.

\subsubsection{Results for local estimators}

\begin{figure}
\begin{center}
A. \psfig{figure=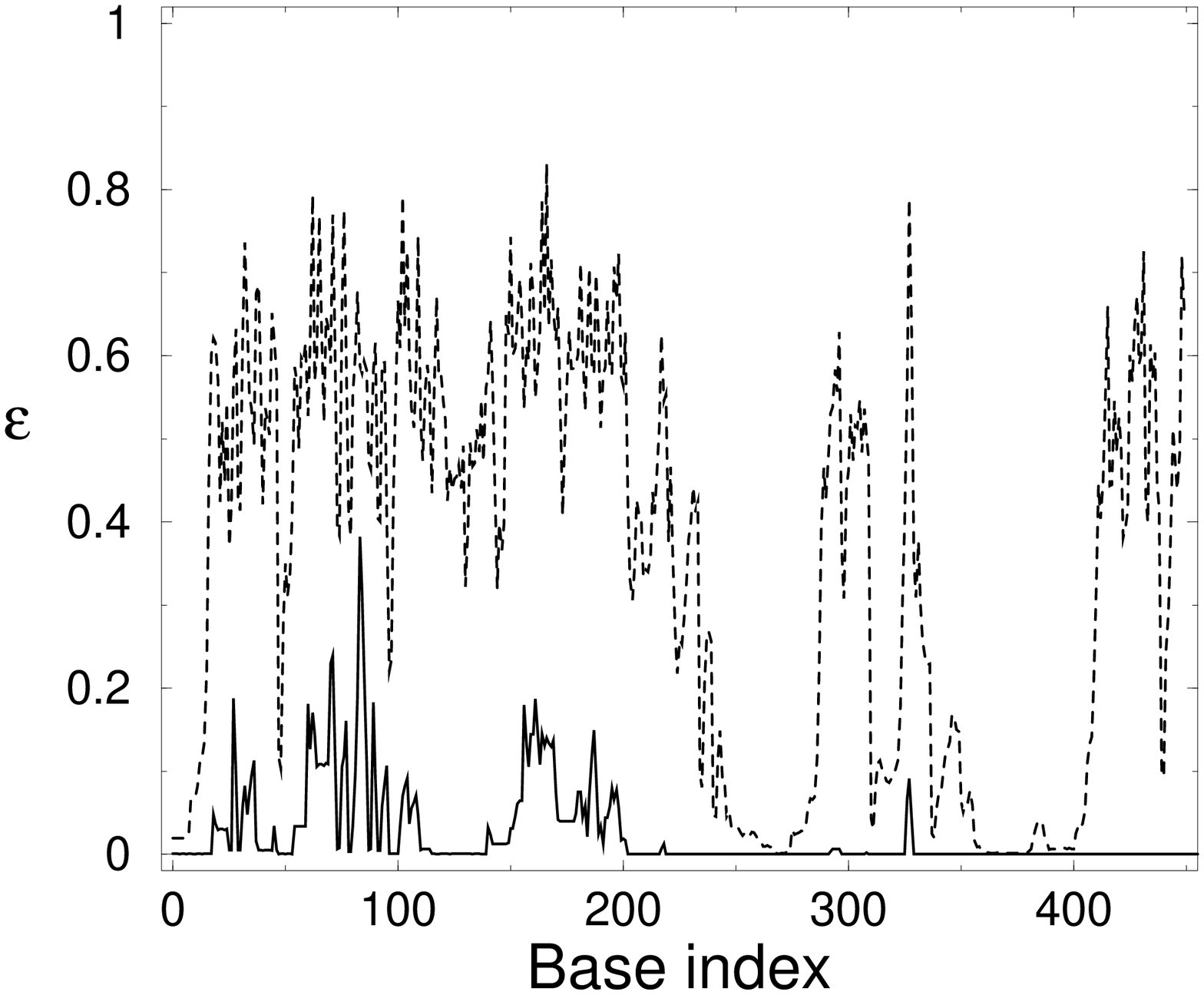,height=6cm, angle=-90}
\hskip .5cm
B. \psfig{figure=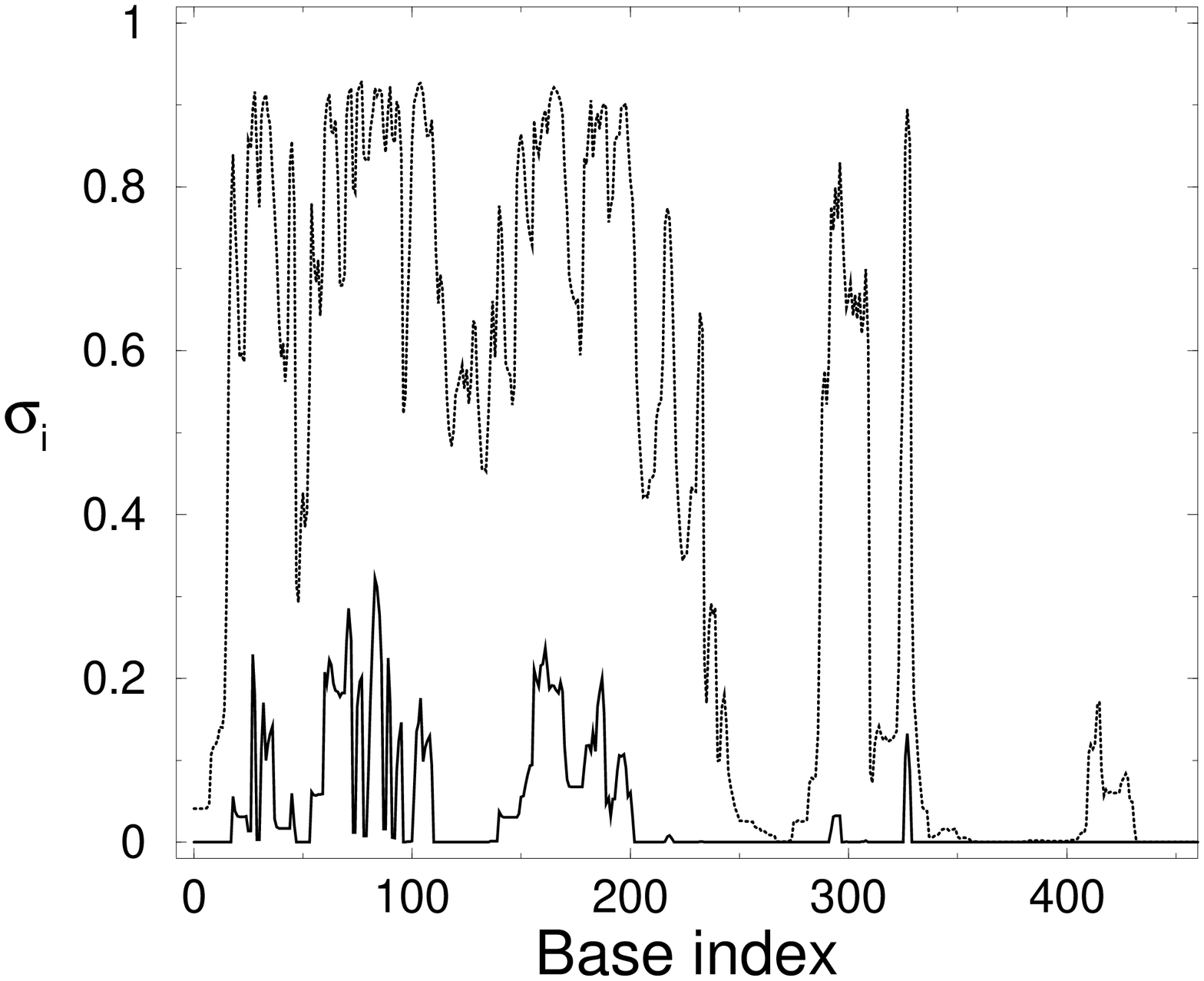,height=6cm,angle=-90}
\caption[]{Probability $\epsilon _i$ ({\bf A}) that bp $i$ is not correctly
predicted and Shannon entropy  $\sigma _i$ ({\bf B}) for the 
first 450 bp of the DNA $\lambda$--phage. Inference is made from
$R=1$ unzipping (dashed line) and $R=40$ unzippings (full
line). The force is $f=16.4$ pN, and data are averaged over 1000 MC samples.}
\label{sigmabase}
\label{ombaserun}
\end{center}
\end{figure}

\begin{figure}
\begin{center}
\psfig{figure=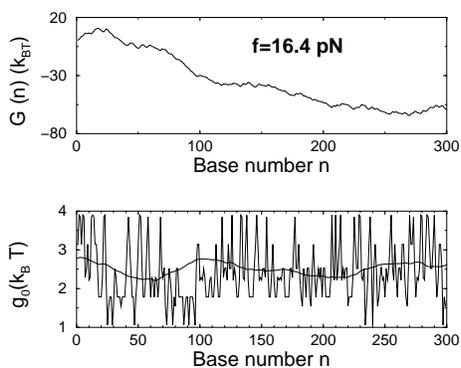,height=6cm,angle=-90}
\caption[]{Top: free energy landscape for unzipping at force $f=16.4$ pN.
Local minima correspond to the portion of the sequence that are best
predicted. Bottom: pairing free energy as a function of the base pair index,
without and with window-average (Gaussian weight over 20 base pairs).}
\label{ombaserun2}
\end{center}
\end{figure}

Figure \ref{ombaserun}A (dashed curve) show the errors $\epsilon_i$
for the first 450 bases of the $\lambda$--phage at $f=16.4$ pN.
Comparison with the free energy landscape $G(n,f;\Lambda)$
(\ref{p}) at the same force shows that the best predicted bases
correspond to valleys  (Fig~\ref{ombaserun2} top),
in which the fork spends a lot of time, while prediction for
bp located on the top of barriers are much poorer. 
In addition Fig.~\ref{ombaserun}A shows that the errors $\epsilon_i$
sharply decrease when the prediction is made from $R=40$ unzippings.

\begin{figure}
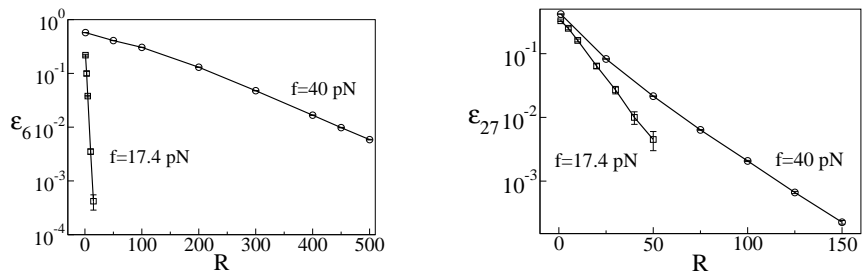

\begin{center}
\psfig{figure=fig10a.eps,height=3.5cm, angle=0}
\hskip 1cm
\psfig{figure=fig10b.eps,height=3.5cm, angle=0}
\caption[]{
Error rate  $\epsilon_i$ (semilog scale)
as a function of the number of repeated unzippings for 
base pairs $i=6$ (left) and $i=27$ (right) arbitrarily selected, for
forces $f=17.4$ and $40$ pN. 
Numerical data are averaged over 25000 to $10^7$ samples, see error bars.
}
\label{omrun}
\end{center}
\end{figure}

We have investigated in detail the decay of the error $\epsilon _i$ with $R$
for two arbitrarily selected bases $i=6$ and $i=27$. 
Figure~\ref{ombaserun2}(top) shows that  bp 6 is located in a valley 
of the free energy landscape at force $f=16.4$ pN
while base pair 27 is located on a barrier at the same force.
Figure \ref{omrun} shows that
the error decays exponentially with $R$, $\epsilon _i \simeq 
\exp(-R/R_c(f,i))$. The value of the decay constant $R_c(f,i)$ strongly 
depends on the force and the bp index. At large force, $f=40$~pN, 
bp 27 is more easily predicted than bp 6. Fitting of the numerical data
yields $R_c(f= 40,i=6)=113\pm 2$ and  $R_c(f= 40,i=27)=25\pm 1$. 
Correspondingly about 400 and 75 unzippings, collected and
analyzed together, are needed to make the error smaller than $1\%$. 
At moderate force, $f=17.4$~pN, predictions for bp 6 require
less unzippings than for bp 27. We obtain $R_c(f=17.4,i=6)=2.2\pm 0.1$, 
meaning that about 6 unzippings are sufficient to reduce the failure rate 
 to $1\%$, while  $R_c(f= 17.4,i=27)=13\pm 1$ and  about 40 unzippings
are needed to reduce the error to the same amount. 
 
\begin{figure}
\begin{center}
\psfig{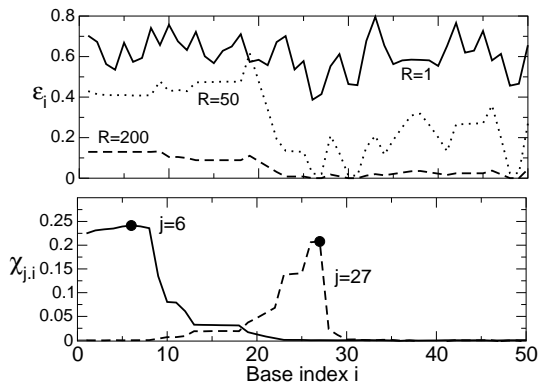}
\vskip .3cm
\caption{Top: error $\epsilon _i$ for the first 50 bases of the
  $\lambda$-DNA for $R=1,50,200$ unzippings. 
Bottom: connected correlation $\chi_{j,i}$ for bases $j=6$ and $j=27$
  (black dots) 
for $R=50$ unzippings. $\chi_{27,i}$ is multiplied by 10 to be more
visible; data correspond to $f=40$ pN (large force). }
\label{epsilonif40}
\end{center}
\end{figure}

The quality of predictions exhibit strong correlations from base to
base. We show in Fig~\ref{epsilonif40}(top)
the error $\epsilon _i$ for the first 50 bases of $\lambda$-DNA at
high force $f\geq 40$ pN. We
observe that groups of neighboring bases are locked-in in that their
errors decay at the same rate when increasing 
the number $R$ of unzippings. See for instance  in
Fig~\ref{epsilonif40} the blocks containing
base 6, extending from bases 1 to 9, and base 27, including bases
26 and 27 only. All the bases $i$ in a
block have the same decay constant $R_c(f,i)$. The lock-in phenomenon
is visible 
from the connected correlation function $\chi_{j,i}$ (\ref{defcorrel}),
shown for bases $j=6$ and $j=27$ in  Fig~\ref{epsilonif40}(bottom).
$\chi_{i,j}$ is essentially a step-wise function, with
highest valuea for the bases $i$ in the same block as $j$, and
smaller values for neighboring blocks.
The values of the decay constants at finite force as well as the 
blocks of locked-in bases will be found back analytically  by the theory.

\subsubsection{Entropy of predictions on a base}

The error $\epsilon$ is defined from the exact
knowledge of the true sequence. In practice one would like to be able
to assess the quality of prediction $b^*_i$ over base $i$ 
without referring to the unknown true sequence. 
To do so we calculate the four optimal sequences for each of the four 
possible choices of $b_i=A,T,G,C$ using the above Viterbi algorithm,
starting from base $i$ and going backward until the first base $b_1$ is
reached and optimized over; we call $P_1(b_1^*|b_i)$ the probability 
(\ref{recur}) corresponding to this left part of the sequence. 
Then we repeat the process starting from
base $i$ and going forward until the last base of the molecule is
reached and optimized over, and we obtain the probability 
$P_N(b_N^*|b_i)$  corresponding to the right part of the sequence. 
Hence we obtain the most likely sequence
constrained to have base $i$ equal to $b_i$, together with its weight
$W(b_i)=P(b_0^*|b_i)\times P(b_N^*|b_i)$. After a proper normalization
we define the probability 
\begin{equation}
\mu (b_i) = \frac{W(b_i)}{W(A)+W(C)+W(T)+W(G)} 
\end{equation}
for each of the four base values at location $i$. The base with the
highest value of $\mu$ is the one predicted by the usual Viterbi procedure.
The Shannon entropy, once averaged over MC data,  
\begin{equation} \label{defsigmai}
\sigma_i=- \langle \sum_{b_i}\, \mu(b_i)\,\log _4 \mu(b_i) \rangle
\end{equation}
is small when one of the four possible bases has much higher probability
than the other ones, and high (close to 1) when bases are equiprobable.
Figure \ref{sigmabase}B shows that
the behavior of $\sigma_i$ follows the one of $\epsilon_i$ along the
sequence (fig \ref{ombaserun}A). In other words, if a base has a much
higher probability $\mu$ than the other three bases it is very likely
to be the correct one. The
Shannon entropy is a good estimator of the quality of the prediction.

\subsection{Average Bayesian prediction}
\label{shannonsec}

Instead of the maximum likelihood probability $\mu(b_i)$ we can
compute the probability $\mu ^A _i (b)$ that base $i$ is of type $b=A,T,C,G$ through
the expression (\ref{bayes}),
\begin{equation}
\mu^A _i (b_i)= \sum_{B' | b'_i = b} {\cal P}(B'|T)
\end{equation}
where we have summed over all sequences constrained to have the value
$b$ for base $i$.  This corresponds to an average Bayesian
prediction in contrast with the maximum probability prescription of
Section \ref{mlp}. We construct our predicted
sequence $B^A$, assigning to each base $i$ the argument $b$ which
maximizes probability $\mu^A _i$.

As in Section \ref{mlp} we have studied the quality of the prediction
for different values of the applied force and of the number of
unzippings.
The fraction of mispredicted bases in $B^A$ as a function of the force and of the number of
unzippings shows a similar behaviour (not shown) to its maximum probability case counterpart
(Fig. \ref{omega1run} and \ref{sigmarun}); a
theoretical discussion of this equivalence in the case of homogeneous
sequences will be given in Section \ref{finitetemp}.
In order to better understand this similarity for the $\lambda$--phage
we have considered three ten bp long portions of its sequence, $B_i
^{(10)} = (b_i,b_{i+1},b_{i+2},b_{i+3},b_{i+4},b_{i+5},b_{i+6},
b_{i+7},b_{i+8},b_{i+9})$, located at $i=200$, $i=140$, and $i=90$.
The choice of the locations corresponds to low ($\sigma \simeq 0$), medium
($\sigma \simeq 0.5$) and high ($\sigma \simeq 1$) entropy regions
(Fig \ref{sigmabase}B). 
We  obtain complete sequences of length $N$ by setting the bases
outside the 10 bp window to the values they have in $B^*$.
For each of the three locations we have
calculated the probability (\ref{bayes}) of the $4^{10}\simeq 10^6$ 
sequences $B$ with the recursive formula (\ref{recur}),
divided by the largest probability {\em i.e.} the one of the sequence
$B^*$. These ratios $r(B)\le 1$ are called relative
probabilities. Even in a high entropy
region most of the sequences have a very small relative probability
$r(B) \ll 1$, meaning that the average sequence $B^A$ is actually
very close to the most likely one, $B^*$. It is interesting to notice 
that smaller and smaller relative probabilities $r$ do not necessarily
correspond to higher and higher `mutations' from $B^*$.
The average Hamming distance (number of bases $b_i$ not equal to
their values $b_i^*$ in $B^*$) of sequences with relative probabilities 
in $[r;r+dr]$ is not a monotonic function of $r$. Less and less likely sequences are not
obtained from the ground sequence through the mutation of a larger
and larger number of bases. Due to stacking interactions, in fact, 
bases are not independent and  it can be energetically favorable to 
flip a group of bases instead of a single one.

\section{Analytical study of inference performances}
\label{theory}

In this section, we present the theoretical studies carried out to
better understand how the quality of the prediction depends on
parameters e.g. force, sequence content, number of repetitions of the
unzipping on the same molecule, ... We start with the high force case
where closing basically never occurs. The analytical study of this
situation is performed first in the absence of stacking interactions
between bases, then in the presence of stacking interactions. We show
that the overall quality of the prediction crucially depends on the
number of repetitions of the unzipping. Later on we turn to the case
of finite force where closing and opening both take place, and show
how the finite force study can be exactly reduced to the high force
one with a stochastic number of unzippings whose distribution is
calculated.

Throughout Section~\ref{pairing} and Section~\ref{stacksec1} only two
types of bases, called weak ($W$) and strong ($S$) have been
considered instead of the four types $A,T,G,C$.  The real case of four
type of bases is taken back into account from
Section~\ref{stacksec2}. Considering two instead of four base types
allows us to make calculation shorter; we however stress that there
is, in principle, no obstacle to the extension of our calculation to
the four bases case.  It is also justified {\em a posteriori} by our
finding. The error in predicting the true value of a base $b$, say,
$b=A$, is the sum of the probabilities of predicting the other three
bases, here $b=G$, $b=T$, and $b=C$. We show that, when a large amount 
of data is
collected, one of these three probabilities, say, $b=G$, is much
larger than the other two probabilities, turning the four base type problem
into an effective two base types problem.

\subsection{High force theory: no stacking interactions}
\label{pairing}

A quick calculation shows that, for forces equal or larger than 40 pN,
the fork separating open and closed regions never goes backward in the
course of unzipping.  Indeed, $g _s(f=40$~pN$)\simeq - 8.6$, and thus
even for strong bases with pairing free energy $g_0 \simeq - 3.6$, the
ratio of closing over opening rates equals $\exp(g_s(f)-g_0)\simeq
e^{-5}$, and is less than one percent. Bases essentially never close,
and the matrix $M\,(b_i,b_{i+1}; u_i,t_i, d_i\,)$ (\ref{defmatrixmbayes}) simplifies
since $d_i=0$, and $u_i=1$ for all open base pairs. We hereafter
calculate the quality of prediction in this case.

Let us simplify further the problem and assume that base pair
interactions are essentially due to the presence of hydrogen bonds,
and not to stacking effects. In other words, we replace
$g_0(b_i,b_{i+1})$ with $g_0(b_i)$ where $b_i$ can take two values:
$W$ (weak) or $S$ (strong).  The free energies are $g_0(S)<g_0(W)<0$,
and $\Delta = g_0(W) -g_0(S) >0$ denotes their difference.

Consider an unzipping experiment (one run of our Montecarlo program)
which opens $N$ base pairs: $d_i=0$ for all $i$, $u_i=1$ for $i< N$
and $u_i=0$ for $i\ge N$. The times $t_i$ spent on the bases
$i=1,\ldots,N$ are uncorrelated and exponentially distributed:
\begin{equation} \label{pnostack1}
P(t_i|b_i^L)= r\, e^{g_0(b_i^L)} \, \exp \left(-r\, e^{g_0(b_i^L)}\,
t_i\right)
\end{equation}
The distributions corresponding to $W$ and $S$ bases are plotted in
Fig.~\ref{ptws}. We define the mean sojourn time on
base $i$,
\begin{equation} \label{taui}
\langle t _i \rangle = \frac{1}{ r} \, \exp( -g_0(b_i^L)) \ .
\end{equation}
and the normalized time
\begin{equation} \label{deftaui}
\tau _i = \frac{t_i}{\langle t _i\rangle} \ .
\end{equation}
Obviously neither $\langle t _i \rangle$ nor $\tau_i$ are accessible
from  the measure which gives access to $t_i$ only.
From (\ref{pnostack1}), the distribution of the normalized time is
exponential with average value unity,
\begin{equation} \label{dist1}
P_1( \tau_i ) = \exp\left(- \tau_i \right) \ .
\end{equation}

\subsubsection{Maximum a posteriori prediction}
\label{nostackingtheory}

\begin{figure}
\begin{center}
\psfig{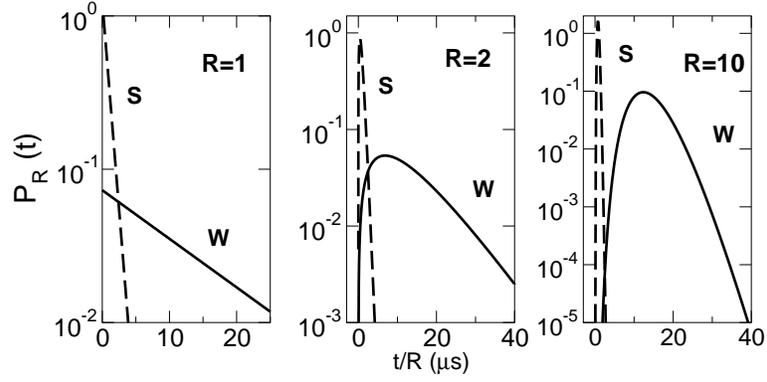}
\caption{Probability distribution $P_R$ of the sojourn time $t$ spent
on a weak ($g_0(W)=-1.06 $, $\langle t\rangle_W= 0.8 \mu$s, dashed
line) and strong ($g_0(S)=-3.9$, $\langle t\rangle_S=13.7\mu $s, full
line) bases. Time is rescaled by $1/R$ (see horizontal axis).  The
number of unzippings is $R=1$ (left), $R=2$ (middle), and $R=10$ (right).  The
probability $\epsilon$ (\ref{defomi}) that a $W$ (resp. $S$) base is
not correctly predicted is the area under the dashed (resp. full)
curve right (resp. left) to the crossing point.  As $R$ increases time
distributions are more and more concentrated, and the error gets
smaller and smaller.}
\label{ptws}
\end{center}
\end{figure}

Given a random value for $\tau_i$ drawn from distribution
(\ref{dist1}), the most likely value for the base, $b^*_i$, is
obtained from Bayes formula (\ref{bayes}) by maximizing
\begin{equation} \label{pnostack2}
P(b_i |\tau_i) \propto r\, e^{g_0(b_i)} \, \exp \left(-r\,
e^{g_0(b_i)} \langle t_i\rangle \, \tau_i \right) \propto
\exp\left(g_0(b_i)- e^{g_0(b_i) - g_0(b_i^L)} \tau_i \right)
\end{equation}
An immediate calculation leads to the conclusion that a weak base
(respectively a strong base) will be correctly predicted if $\tau _i <
\tau ^W$ (resp.  $\tau _i > \tau ^S$) where
\begin{equation}\label{fr}
 \tau ^{W} = \frac{\Delta}{1-e^{-\Delta}} \quad \hbox{\rm and}\quad
\tau ^{S} = \frac{\Delta}{e^{\Delta}-1}
\end{equation}
Therefore, the probability that a base is wrongly predicted depends
on whether the base is weak or strong, and reads
\begin{eqnarray} \label{pnostack3}
\epsilon _1 ^W &=& \int_{\tau ^W} ^\infty d\tau \, P_1(\tau) =
\exp\left(-\frac{\Delta}{1-e^{-\Delta}}\right) \nonumber \\ \epsilon
_1 ^S &=& \int_0^{\tau ^S} d\tau \, P_1(\tau) = 1-
\exp\left(-\frac{\Delta}{e^{\Delta}-1}\right) \ .
\end{eqnarray}
Plots of $\epsilon _1^W$ and $\epsilon _1^S$ as functions of the free
energy difference $\Delta$ shows that the latter probability is
smaller than the former.  At high force, maximum likelihood prediction
works better on weak bases than on strong bases. The two limiting
cases are:
\begin{itemize}
\item $\Delta \to 0$: we find $\epsilon _1^W=\frac{1}{e}=0.368$, while
$\epsilon _1^S=1-\frac{1}{e}=0.632$.  This result is, at first sight,
surprising since both bases should become equivalent when the free
energy difference tends to zero. It is a consequence of the maximal
likelihood principle: the reduced time $\tau$ has a higher probability
to be smaller than its average value ($\tau ^W=\tau ^S=1$ when $\Delta
\to 0$), and therefore weak bases are predicted with higher
probabilities than strong bases independently of the true base
$b_i^L$. We shall see in Section \ref{finitetemp} that this artifact disappears
when prediction are carried out from the average Bayesian framework
of Section \ref{shannonsec}.
\item $\Delta \to \infty$: when the difference in free energies
between both bases gets very large, both are asymptotically perfectly
predicted. The convergence to 100\% correct prediction is faster for
weak than for strong bases: $\epsilon _1^W \simeq e^{-\Delta},
\epsilon _1^S\simeq \Delta \, e^{-\Delta}$.
\end{itemize}
The above analysis can straightforwardly be extended to the case of
predictions made from repeated experiments. Let us call $R$ the number
of unzippings, and $\tau _i^{(1)}, \tau _i^{(2)}, \ldots,
\tau _i^{(R)}$ the (normalized) times spent on base $i$. Using formula (\ref{bayesrun}), we have to maximize
\begin{eqnarray}
P_R\big(b_i |\{ \tau _i^{(1)}, \tau _i^{(2)}, \ldots, \tau _i^{(R)} \}
\big) &\propto& \left[ r\, e^{g_0(b_i)} \right]^ R\, \exp \left[-r\,
e^{g_0(b_i)} \langle t_i\rangle \, \big( \tau _i^{(1)} + \tau _i^{(2)}
+ \ldots, + \tau _i^{(R)} \big) \right] \nonumber \\ &\propto& \exp
\left[ R \, g_0(b_i) -r\, e^{g_0(b_i) - g_0(b_i^L)} \, \tau _i \right]
\end{eqnarray}
where
\begin{equation} \label{totaltau}
\tau _i = \tau _i^{(1)} + \tau _i^{(2)} + \ldots + \tau _i^{(R)}
\end{equation}
is the total time spent on base $i$. The maximization over $b_i$ is
very similar to the one carried out from eqn (\ref{pnostack2}).  We
find that formula (\ref{pnostack3}) for the probabilities of correct
prediction holds for $R$ unzippings provided the single time distribution
$P_1$ is replaced with the distribution $P_R$ of the total time
$\tau_i$ (see Appendix \ref{appa}),
\begin{equation} \label{dist2}
P_R( \tau _i) = \frac{\tau _i ^{R-1}}{(R-1)!} \, \exp( - \tau _i ) \ ,
\end{equation} 
and the times $\tau^W,\tau^S$ (\ref{fr}) are multiplied by $R$. The
distribution of (not normalized) sojourn times after $R$ unzippings
are shown in Fig.~\ref{ptws} for $W$ and $S$ sequences.  An important
remark is that the distributions become more and more concentrated as
$R$ grows; in other words the times become less and less stochastic
and are faithful signatures of the thermodynamic nature of the
attached base. The probabilities that weak and strong bases are not
correctly predicted after $R$ unzippings are given by
\begin{eqnarray} \label{pnostack4}
\epsilon _R ^W &=& \int_{R\,\tau ^W} ^\infty d\tau \, P_R(\tau) =
\gamma \left(R , \frac{R\, \Delta}{1-e^{-\Delta}}\right) \nonumber \\
\epsilon _R ^S &=& \int_0 ^{R\, \tau ^S} d\tau \, P_R(\tau) = 1-\gamma
\left(R, \frac{R\, \Delta}{e^{\Delta}-1}\right) \ .
\end{eqnarray}
where
\begin{equation} \label{defgammaincomplete}
\gamma(a,x)=\int_x^\infty dt \,\frac{t^{a-1}\, e^{-t} }{(a-1)!}
\end{equation}
is the normalized incomplete Gamma function. 

\begin{figure}
\begin{center}
\psfig{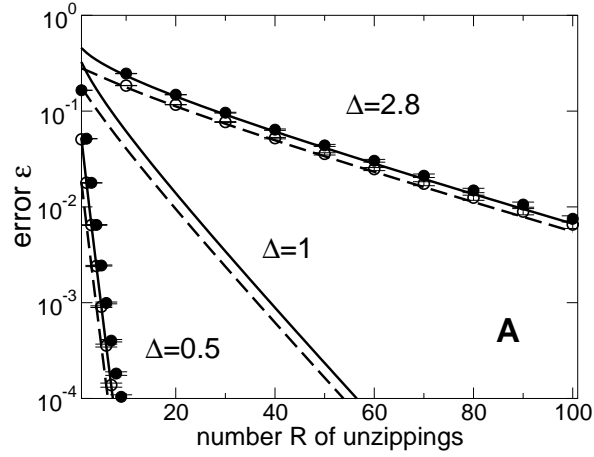} 
\caption{Errors on sequences of, respectively, strong (full line) and
weak (dashed line) bases as a function of the number $R$ of unzippings
in the infinite force limit and without stacking
interaction. The difference of pairing free-energies $\Delta$ is,
from bottom to top, 0.5, 1, and 2.8. We show the results of numerical
simulations for $\epsilon _R ^W,\epsilon _R^S$ with the error bars for
$\Delta=0.5$, 2.8 (full dots: $S$ sequence, empty dots: $W$ sequence).}
\label{omteo}
\end{center}
\end{figure}

To better understand how the quality of predictions improves with the
number of unzippings, we have analytically calculated the asymptotic
expansion of $\epsilon$ in Appendix \ref{app2c}.  From expression
(\ref{pnostack4}), we have when $R\gg 1$,
\begin{equation}\label{nostackomega}
\epsilon _R \simeq \frac{e ^{- R \, (\tau -1 -\ln \tau )}}{\sqrt
{2\pi R } \; (\tau -1)}
\end{equation}
with $\tau = \tau ^W$ or $\tau^S$ (\ref{fr}) depending on the type of base.
As a consequence, achieving good recognition requires a number of 
unzippings (much) larger than
\begin{equation}\label{fr2}
R_c = \frac{1}{ \tau - 1 - \ln \tau} \ .
\end{equation} 
This crossover number depends on the free energy difference $\Delta$,
but not on the type of base: $R_c (\tau ^W)=R_c(\tau ^S)$.  Fig
\ref{omteo} shows that $R_c$ is all the more large than $\Delta$ is
small.  Definitions (\ref{fr2}) for $R_c$ and (\ref{fr}) for
$\tau^W,\tau^S$ yield
\begin{equation} \label{rcnostack}
R_c \simeq \frac{8}{\Delta^2} \quad , \qquad \Delta \to 0 \ .
\end{equation} 
This expression is a good quantitative approximation for $R_c$ up to
$\Delta \simeq 3$.  We have checked the validity of these theoretical
results through numerical experiments using the Viterbi
procedure of Section \ref{stacksec}, where the free energy matrix
$g_0$ was modified to avoid stacking interaction.  Figure~\ref{omteo}
shows the perfect agreement between numerical and theoretical results.

That the effort (number of unzippings) necessary to ensure an
excellent prediction essentially depends on the difference of pairing 
free energies between the two types of bases one wishes to distinguish 
justifies {\em a posteriori} the simplification of taking into account only
two types of bases.  The cases of interest are:
\begin{itemize}
\item Weak bases represent $A$ or $T$, and strong bases $G$ or $C$:
the free energy difference is estimated to be $\Delta \simeq 2.8$
(obtained from $g_0(T,A)=-1.06,g_0(G,T)=-3.9$).  The probability of
wrong prediction for strong bases, $\epsilon^{S}_R$, is plotted in Fig
\ref{omteo}, as a function of the number $R$ of unzippings. $R=5$
unzippings are enough to achieve excellent base recognition.
\item Weak bases are $A$, strong bases are $T$: the free energy
difference is $\Delta\simeq 0.5$ (obtained from $g_0(T,A)=-1.06,
g_0(A,T)=-1.55$).  Figure \ref{omteo} shows it takes about 100 unzippings to
reach 99\% confidence in the prediction. Thus, the number of unzippings
considerably increases if we want to precisely resolve all base pairs.
\end{itemize}
Sequence prediction can be then done in a hierarchical manner. A
small number of unzippings $R \simeq 5$ is sufficient to distinguish
between A,T and G,C bases, in agreement with numerical simulation data
shown in Fig~\ref{sigmarun}A\&B, while more unzippings $R \simeq
100$ are necessary to clearly separate $A$ from $T$, and $G$ from $C$
bases. In this regard, our prediction procedure always amounts to
distinguish between two types of bases.

\subsubsection{Average Bayesian prediction}
\label{finitetemp}

Average Bayesian prediction consists in estimating the
the probability of the correct base $P(b_i^L|t_i)$ (thermal average) and 
averaging over $t_i$ (quenched average) rather than 
looking for the most likely base $b_i$ given the time $t_i$ spent on
base $i$ (\ref{shannonsec}). This procedure gives, in the general case of $R$ 
unzippings, 
\begin{equation} \label{omegat=1}
\epsilon _R ^A = \int_0^{\infty} d\tau \; \frac{P_R (\tau)}
{1+\exp(-R\, \Delta +\tau\,(e^{\Delta }-1))} \ .
\end{equation}
We stress that the above expression gives the value of $\epsilon _R
^A$ for both $W$ and $S$ bases. The quality of prediction does not
depend on base $b_i^L$, in contradistinction with the maximal
likelihood case, see eqn (\ref{pnostack4}).  This independence is a
direct consequence of Bayes inference formula. By definition indeed,
\begin{equation}
\epsilon ^{W,A}=\int_0^{\infty} d\tau \; P (\tau | W) \; P \big(
S|\tau\big) = \int_0^{\infty} d\tau \; P (\tau | W) \; \frac{ P (\tau
| S)}{P (\tau | W)+P (\tau | S)} \ .
\end{equation}
This expression is left unchanged when we exchange $S$ and
$W$. Therefore
\begin{equation}
\epsilon ^{S,A}  = \epsilon ^{W,A} 
\end{equation}
Notice that this proof is quite general: it not only holds for any
number $R$ of unzippings, but also for any microscopic model yielding an
explicit expression for $P (\tau | b^L)$. In particular, it remains
true at finite force.  As the number $R$ of unzippings increases, the
prediction approaches perfection, see Appendix~\ref{app2c},
\begin{equation} \label{nostackomegabis}
\epsilon _R ^A \simeq \frac{\pi \sigma}{\sin ( \pi \sigma)} \, \frac
{e ^{- R \, (\tau -1 -\ln \tau )}} {\sqrt {2\pi R } \; (1-\tau) }
\end{equation}  
with
\begin{equation} \label{deftau84}
\tau = \frac{\Delta}{e^\Delta -1} \quad \hbox{\rm and}\quad \sigma=
\frac{1}{\Delta} - \frac{1}{e^\Delta -1}\ .
\end{equation}
This asymptotic scaling is, to the exponential order, identical to the
one obtained in the maximum likelihood case (\ref{nostackomega}).
Therefore average and maximum likelihood predictions are
asymptotically equivalent.

\subsubsection{Relationship with Shannon entropy}

The above findings explains the similarity between the error 
(\ref{defomi}) and the Shannon entropy (\ref{defsigmai}) observed in
Fig.~\ref{ombaserun}A\&B. Let us 
call $\epsilon$ and $1-\epsilon$ the
probabilities that the prediction on a base is correct and erroneous
respectively. The Shannon entropy reads
\begin{equation}
\sigma = -  \epsilon\; \ln \epsilon - (1-\epsilon) \ \ln (1-\epsilon) 
\simeq -  \epsilon\; \ln \epsilon \simeq \hbox{\rm cst}\times
\sqrt R \ e^{-R/R_c}
\end{equation}
when the number of unzippings is large with respect to $R_c$. This explains
why the error and the Shannon entropy on a base roughly behave in the same way,
and essentially vanish when the number of unzippings is far above its
critical value $R_c$. This result is left unchanged in the case of
four, and not two base types.

\subsection{High force theory: stacking interactions} \label{stacksec}

Let us now study how the presence of stacking interactions modify the
above findings. With two kinds of bases, the pairing free energy
matrix is a $2\times2$ matrix $g_0(b,b')$.  Strong bases ($S$) are
chosen to be 'average' bases from a repeated GCGCGC... sequence while
weak bases ($W$) represent a repeated ATATAT...  sequence. The values
of the interactions are the average values of the pairing free energy
in each of the four quadrants of the original $4\times4$ matrix:
$g_0(W,W)=-1.42$, $g_0(S,W)=g_0(W,S)=-2.39$, and $g_0(S,S)=-3.50$. We
define the free energy differences
\begin{equation}\label{fediff}
\Delta ^W = |g_0 (W,W)-g_0(W,S)| \ , \quad \Delta^S=
|g_0(W,S)-g_0(S,S)| \ .
\end{equation} 
whose values are $\Delta ^W = 0.98, \Delta ^S = 1.11$.  The
calculation of the probability of correct base prediction is more
difficult than in the absence of stacking but can be carried out using
techniques issued from the statistical mechanics of one dimensional
disordered systems \cite{Dyson,Luck}.

We start from the recursive eqn (\ref{recur}) for the probability
$P_i(b_i)$ that the $i^{th}$ base of the sequence is equal to $b_i$.
As in the no--stacking case, we
introduce the normalized time $\tau_i$ through eqn (\ref{deftaui})
where the average sojourn time on base $i$ now reads
\begin{equation} \label{tauibis}
\langle t _i \rangle = \frac{1}{ r} \, \exp( -g_0(b_i^L,b_{i+1}^L))
\end{equation}
Defining $\pi_i(b_i)=-[\ln P_i(b_i)]/R$ and introducing the local fields,
\begin{equation}
h _i =  \pi _i(S) - \pi_i (W) 
\end{equation}
we rewrite eqns (\ref{recur},\ref{recurpi})  under the form
\begin{equation}
h _{i+1} = F_i \big(h _i, \tau_i\big)
\end{equation}
where function $F_i$ depends on base $b_i^L$ through the average
sojourn time (\ref{tauibis}),
\begin{eqnarray} \label{deffuncf}
F_i(h, \tau ) &=& \max \big[ h + g_0 (W,W)- g_0(S,W) - r \,
\frac{\langle t_i\rangle}{R} \, \big( e^{g_0 (W,W)}-e^{g_0(S,W)}\big)\,
\tau, 0\big] \nonumber \\&+& \min \big[ - h ,g_0 (W,S) - g_0(S,S) - r
\, \frac{\langle t_i\rangle}{R} \,\big(
e^{g_0(W,S)}-e^{g_0(S,S)}\big)\, \tau\big]
\end{eqnarray}
As $\tau _i $ is a stochastic variable with distribution $P_R$
(\ref{dist2}) (for $R$ repetitions of the experiment), $h _i$ is
itself a stochastic variable. Its probability distribution, $Q_i$,
obeys the recursion
\begin{equation} \label{recurq}
Q_{i+1} (h _{i+1}) = \int _0^{\infty} d\tau _i\,P_R(\tau _i) \int
_{-\infty} ^\infty dh _i \, Q _i( h_i)\; \delta \big( h _{i+1} -
F_i(h_i, \tau _i) \big) \ .
\end{equation}

\subsubsection{Repeated sequences}
\label{stacksec1}

The stationary solution $Q=Q_i$ of eqn (\ref{recurq}) is calculated in
Appendix \ref{appstack} for the three repeated sequences
$B^L=WWWW...$, $SSSS...$, and $SWSW...$ referred to as $WW$, $SS$, and
$SW$ sequences respectively. These sequences differ from each other
through their sojourn times $\langle t\rangle$ (\ref{tauibis}).  When
the condition $\Delta^W \le \Delta^S$ is fulfilled as is the case for
the example considered above, the stationary field distribution is
better written in terms of its cumulative function
\begin{equation} \label{defqhat}
\hat Q (h) \equiv \int _{h} ^{\infty} dh' \, Q(h') \ ,
\end{equation}
with the result
\begin{equation} \label{expressq}
\hat Q(h) = \left\{ \begin{array}{c c c} A(h) &\hbox{\rm if} &h <
-\Delta^S \\ \frac{A(h) - A(-h)\, B(h)} {1- B(-h) \, B(h)} & \hbox{\rm
if} & -\Delta^S < h < \Delta^S \\ 0 & \hbox{\rm if} & h > \Delta^S
\end{array} \right.
\end{equation}
where
\begin{equation} \label{defab}
A (h) = 1-\gamma \left( R,\frac{R( \Delta ^S - h)}{x
(1-e^{-\Delta^S})} \right) \ , \quad B (h) = \gamma \left( R, \max
\left( \frac {R(\Delta^W-h)} {x (e^{\Delta^W}-1)} ,0 \right) \right) -
\gamma \left( R,\frac { R( \Delta ^S - h)}{x (1-e^{-\Delta^S})}
\right) \ ,
\end{equation}
and $\gamma$ is the incomplete Gamma function
(\ref{defgammaincomplete}).  The parameter $x$ is defined as the ratio
of the average sojourn time $\langle t\rangle$ over its value for the
$SW$ sequence,
\begin{equation} \label{tauiter}
x = \frac{\langle t\rangle}{\langle t\rangle ^{SW}} \ .
\end{equation}
Knowledge of the field distribution allows us to calculate the average
fraction $\epsilon$ of mispredicted bases (\ref{defom}) and the
nearest-neighbor ($d=1$) disconnected correlation function 
\begin{equation} \label{ccdis}
\chi_1 ^{dis} = \chi _1 + (1-\epsilon)^2 
\end{equation}
where the connected correlation function is defined in eqn
(\ref{defcorrel}). The calculations are reported in Appendix
\ref{app2b}.  Results are
\begin{itemize}
\item {\em WW sequence}: we have $x=e^{-\Delta ^W}$, and
\begin{equation} \label{omegaww}
\epsilon^{WW} _R = 1- \int _{-\Delta ^S} ^{\Delta ^S} dh \, \hat Q
(-h) \, Q(h) \ , \quad (\chi ^{dis}_1) _ R ^{WW}= \int _0^\infty d\tau\,
P_R(\tau) \; \hat Q \left( -\Delta ^W + {\tau}{R} \big(1 -
e^{-\Delta^W} \big) \right) ^2 \ .
\end{equation}
\item {\em SS sequence}: we have $x = e^{\Delta ^S}$, and
\begin{equation} \label{omegass}
\epsilon _R ^{SS}= \int _{-\Delta ^S} ^{\Delta ^S} dh \, \hat Q (-h)
\, Q(h) \ , \quad (\chi ^{dis}_1)_R ^{SS}= \int _0^\infty d\tau\, P_R(\tau)
\bigg[ 1- \hat Q \left( -\Delta ^S + \frac{\tau}{R} \big(e^{\Delta^S}
-1\big) \right) \bigg]^2 \ .
\end{equation}
\item {\em SW sequence}: we have $x=1$; the probabilities that bases
$S$ and $W$ are not correctly predicted are, respectively,
\begin{equation} \label{omegasw}
\epsilon _R ^{SW,S}= \int _{-\Delta ^S} ^{\Delta ^S} dh \, \hat Q (-h)
\, Q(h) \quad , \qquad \epsilon _R ^{SW,W}= 1 - \epsilon _R ^{SW,W} \
,
\end{equation}
while the correlation function reads
\begin{eqnarray} \label{chisw}
(\chi^{dis}_1) _R ^{SW}=\int _0^\infty d\tau\, P_R(\tau) &\bigg[& \hat Q
  \left( -\Delta ^S + \frac{\tau}{R} \big(1-e^{-\Delta^S}\big) \right)
  \nonumber \\ &-& \frac{1}{2} \, \hat Q \left( -\Delta ^S + \frac{\tau}{R} \big(1-e^{-\Delta^S}\big) \right)^2 -\frac{1}{2} \, \hat Q
  \left( -\Delta ^W + \frac{\tau}{R} \big(e^{\Delta^W} -1\big)
  \right)^2\bigg]
\end{eqnarray}
\end{itemize}
The subscript `R' reminds us that the above expressions hold for data
collected from $R$ unzippings of the experiment.  Let us stress that the
field distributions $Q$ (and their cumulative functions $\hat Q$)
appearing in the expressions of $\epsilon$ and $\chi ^{dis}_1$ above depend on
the sequence through the ratio $x$, see eqns
(\ref{expressq},\ref{defab},\ref{tauiter}).

\begin{figure}
\begin{center}
\psfig{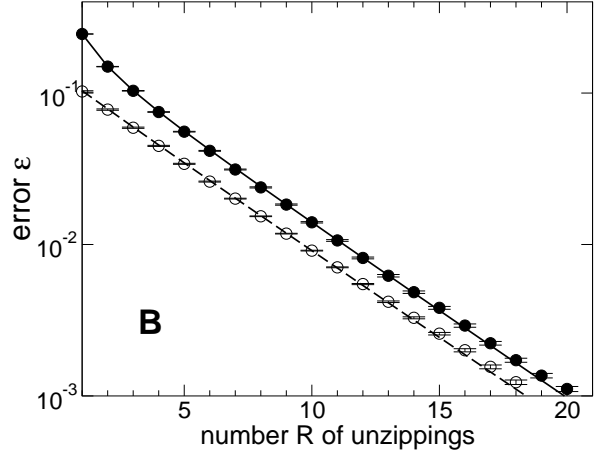}
\caption{Probability of misprediction for repeated $WW$ (full line) and
$SS$ (dashed line) sequences as a function of the number $R$ of unzippings
in the infinite force limit and in presence of stacking interactions. 
Here, $g_0(W,W)=-1.5, g_0(S,W)=
g_0(W,S)=-2.5, g_0(S,S)=-3.5$. The strong and weak sequences are repeated
$SS$ and $WW$ sequences respectively. Simulation results are shown with 
the error bars. Remark that the slope of $\ln \epsilon$ is about twice
the one for the non-stacking case with $\Delta =1$ (Fig.~\ref{omteo}), see
eqn (\ref{rcstack}) and attached discussion.}
\label{omteostack}
\end{center}
\end{figure}

\begin{figure}
\begin{center}
\psfig{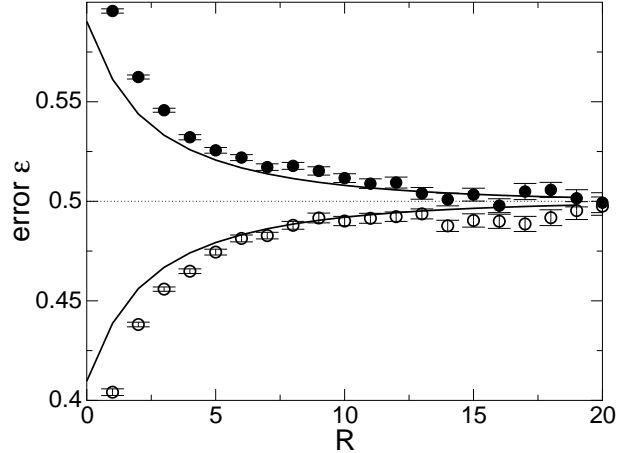}
\caption{Probabilities $\epsilon ^{SW,S}_R$ and $\epsilon ^{SW,W}_R$ 
of mispredicting, respectively, a $S$ (black dots, full curve) 
and $W$ (empty dots, dashed curve) base in a repeated $SW$ sequence
as a function of the number $R$ of unzippings
in the infinite force limit.  The  stacking interactions are
$g_0(W,W)=-1.5, g_0(S,W)=
g_0(W,S)=-2.5, g_0(S,S)=-3.5$. Simulation results are shown with 
the error bars, while continuous curves correspond to the theoretical
expression (\ref{omegasw}). As $R$ grows the prediction on a single
base becomes essentially random  ($\epsilon \to \frac{1}{2}$) since
$SWSW...$ and $WSWS...$ sequences cannot be distinguished from
one another.  }
\label{omseqalt}
\end{center}
\end{figure}

\begin{figure}
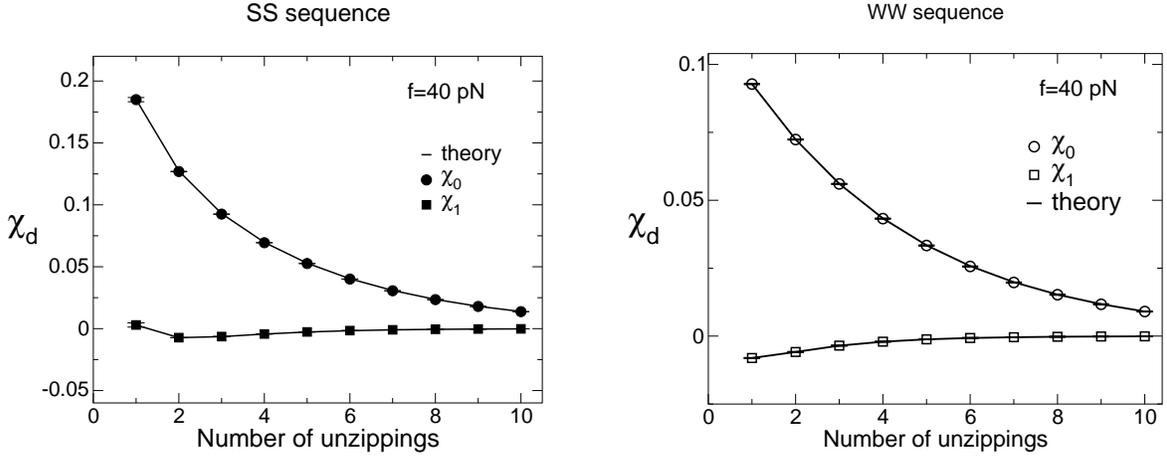

\begin{center}
\psfig{figure=fig16a.eps,height=6cm,angle=0} \hskip 1cm
\psfig{figure=fig16b.eps,height=6cm,angle=0} 
\caption{Connected correlation function $\chi _1$ at distances $d=1$ 
for, respectively, repeated $SS$ (left panel) and $WW$ (right panel) 
sequences as a function of the number $R$ of unzippings
in the infinite force limit ($f=40$ pN in simulations). For
comparison we show the $d=0$ correlation function, $\chi _0=\epsilon
(1-\epsilon)$. }
\label{figcorrssww}
\end{center}
\end{figure}

The above theoretical predictions are shown in Fig.~\ref{omteostack}
and Fig.~\ref{omseqalt} for
the three sequences, and perfectly agree with numerical experiments.
For $SS$ and $WW$ sequences, we find
that the quality of predictions tends to 100\% accuracy as the number
$R$ of unzippings increases. It is shown in Appendix \ref{app2c} that the
asymptotic scaling of $\epsilon _R$ is given by
\begin{equation} \label{epsilonwwss}
\epsilon _R \simeq \frac{\tau^2 \; e^{-2 \, R \, (\tau-1-\ln \tau)}}
{\sqrt{4 \pi R}\; (\tau -1)}
\end{equation}
where $\tau$ equals
\begin{equation} \label{deftwwtss}
\tau ^{WW} = \frac{\Delta ^W}{1- e^{-\Delta ^W}} \quad \hbox{\rm and}
\quad \tau ^{SS} = \frac{\Delta ^S}{e^{\Delta ^S}-1}
\end{equation}
for $WW$ and $SS$ sequences respectively. The above formula shows that
the number of unzippings must exceed
\begin{equation} \label{rcstack}
R_c = \frac{1}{2\, (\tau -1 -\ln \tau)}
\end{equation}
in order to achieve good recognition; we find $R_c \simeq 4.3$ and
$R_c\simeq 3.3$ for $WW$ and $SS$ sequences respectively.
 The nearest-neighbor correlation function $\chi _1$ 
in Fig.~\ref{figcorrssww} is very small, even for $R=1$ unzipping. 
The quasi-independence of predictions can be understood from
the analytical calculation of Appendix \ref{app2b}, and is essentially
due to the fact that the sums of the diagonal and off-diagonal
elements of  the $g_0$ matrix are equal. We have numerically checked that the correlation
function is very small at all distances $d$, not only at high forces, but for all forces above criticality.

The above findings can be easily understood from the findings of Section
\ref{nostackingtheory}. Consider for instance the $WW$ sequence.  When
$R$ gets very large, very few bases $S$ are (wrongly) predicted to be
in the sequence. Call $\epsilon$ the probability that a single base
$S$ is predicted. The predicted event $WSW$ violates two stacking
interactions (bonds) with respect to the correct event $WWW$. Let us
make the simplifying hypothesis that these two violations are
independent: $\epsilon = \mu ^2$, where the probability $\mu$ of one
bond violation depends on the free energy excess $\Delta ^W$
(\ref{fediff}) of the erroneous bond $WS$ (or $SW$) with respect to
the true bond $WW$. We estimate the value of $\mu$ from the theory of
Section \ref{nostackingtheory}: $\mu = \epsilon _R$
(\ref{nostackomega}) with $\tau = \tau ^{WW}$, see
(\ref{fr},\ref{deftwwtss}). This simple argument explains why the
quality of predictions is much closer to 100\% success in presence
than in absence of stacking (for the same number of unzippings). In
particular, the cross-over number of unzippings $R_c$ required to achieve
good recognition is twice smaller in the former case (\ref{rcstack})
than in the latter case (\ref{rcnostack}).

The behavior of the error $\epsilon$ for the alternate $SW$ sequence 
is slightly more subtle to interpret, see Fig.~\ref{omseqalt}. 
From expressions (\ref{omegasw},\ref{chisw}), we find (see
Appendix \ref{app2c}), in the infinite $R$ limit,
\begin{equation} \label{omegaswasym}
\epsilon _R^{SW,S}\ \&\ \epsilon _R ^{SW,W}\to \epsilon ^{SW}
_{\infty} = \frac{1}{2} \qquad \hbox{and} \qquad (\chi _1)_R ^{SW}\to
(\chi _1)_\infty ^{SW} = \frac{1}{2} \ .
\end{equation} 
The limit value of $\epsilon$ is at, first sight, disappointing. There
is 50\% probability that a $S$ or $W$ is predicted at a given position
$i$ along the sequence, showing that our prediction is not better than
a purely random guess! However, the nearest-neighbor correlation
function $\chi$ is much higher than the value $(1-\epsilon)^2$ it
would have if there were no correlation. Indeed, we find that the
probability that base $i+1$ is correctly predicted provided its
neighbor at position $i$ is equals
\begin{equation}
\frac{ \langle n_i n_{i+1} \rangle}{\langle n_i \rangle} \to \frac{
\chi _\infty ^{SW}}{1-\epsilon _\infty ^{SW}} = 1
\end{equation}
as the number of unzippings increases. In other words, only two sequences
can be predicted, either the correct one $SWSWSW...$ or its mirror
sequence $WSWSWS...$. Actually, both sequences produce identical
unzipping signals since the pairing matrix $g_0$ is symmetric, which
is not the case for the true matrix (Table \ref{tableg0}).

\begin{table}
\begin{center}
\begin{tabular}{|c|c|c|c|c|}
\hline &A&T&C&G \\ \hline A&18&75&72&51\\ \hline T&8&14&14&13\\
\hline C&13&51&50&39\\ \hline G&14&72&69&50\\ \hline
\end{tabular}
\hskip 1cm
\begin{tabular}{|c|c|c|c|c|}
\hline &A&T&C&G \\ \hline A&51&44&12&13\\ \hline T&59&51&13&14\\
\hline C&14&13&11&8\\ \hline G&12&12&7&7\\ \hline
\end{tabular}
\\ \vskip .1cm $b=A$  \hskip 3cm  $b=C$   
\end{center}
\caption{Single base mutation decay constant $R_c ^{sm} (xby)$, that 
is, value of the number of unzippings necessary for a good prediction 
at high force of a base $b$ as a function of the 
contiguous bases $x$ (row) and $y$ (column).  See equation (\ref{id})
for a precise definition.  Left: the central base is $b=A$; the
most dangerous mutation is $b=A\to b'=T$ for all contiguous bases,
except for $xy=AA$ where $b'=G$ . Right: the central base is
$b=C$; the most dangerous mutation is $b=C\to b'=G$ for all contiguous
bases, except for $xy=CC$ where $b'=A$. }
\label{tab1}\label{tab2}
\end{table}

\begin{table}
\begin{center}
\begin{tabular}{|c|c|c|c|c|}
\hline &A&T&C&G \\ \hline A &151 &151 &89 & 89\\ 
                   \hline T &15 &32 &118 &118 \\
                   \hline C& 78 &78 &22& 16  \\
                   \hline G & 139& 139 & 14&21\\ \hline
\end{tabular}
\end{center}
\caption{Decay constant $R_c(xb \to xb')$, that is, number of 
unzippings necessary for a good prediction, at high force, of a bond 
between base $x$ (fixed as in the sequence, value indicated in the leftmost
column) and base $b$ (value reported in the top line), potentially
predicted to be of $b'$ type. The most dangerous (requiring the 
largest number of unzippings) mutation $b \to b'$ are given by $b'$ equal to
  the complementary base of $b$, except  for the 
cases $TT \to TC,\; CC \to CA,\; GG \to GT.$  } 
\label{tab00}
\end{table}

\subsection{High force theory: decay constants $R_c$ 
for heterogeneous sequences}
\label{stacksec2}

Let us turn to the realistic case of a non-repeated sequence with four 
base types, and stacking interactions between neighbouring bases.
From the numerical findings of Section \ref{mlp} and the theoretical
analysis of repeated sequences of Section \ref{stacksec} we expect the
error on a base to decay exponentially with the number $R$ of unzippings.
In a first step we estimate the decay constant within a single mutation
assumption: all bases are assumed to be correctly predicted but the one
under study \cite{corto}. 
However this single mutation assumption is not always 
correct. We will show that the decay of the error in predicting one base
is often due to the difficulty in predicting a whole block of co-mutated
bases, and give the corresponding expression of the decay constant $R_c$. 

\subsubsection{Decay constant in the single base mutation assumption}
\label{yyy}

Consider a triplet of contiguous bases along the
sequence, $xby$ and let us start by calculating the error
due to a predicted sequence with a single base
mutation e.g. $b\to b'$ when  keeping  bases $x$ and $y$ to the correct
values. In this case 
the argument following eqn ({\ref{rcstack}) and obtained in the 
case of repeated sequences is still valid.  
As a result of stacking interactions the probability 
$\epsilon^{b\to b'}$ of this mistake
is the product of the probabilities $\epsilon^{xb\to xb'}$ and
$\epsilon^{by\to b'y}$ of either bond violation.
The large $R$ behavior of the error probability
\begin{equation}
\label{errex2}
\epsilon^b_R \sim e^{-R/R_c^{sm} (xby)}
\end{equation}
on base $b$ is then obtained by selecting 
the worst value for the mutation $b'$,
\begin{equation} \label{id}
\frac{1}{R_c^{sm} (xby)} = \min _{b' (\ne b)} \left[ \frac{1}{R_c (xb \to
xb')} + \frac{1}{R_c (by \to b'y )} \right]
\end{equation}
where $R_c (xb\to xb')$ is the decay constant of the error 
obtained  in the no-stacking theory of Section \ref{nostackingtheory} 
(applied here to a bond and not to a base violation); it is given
by formula (\ref{fr2}) with $\Delta= g_0 (x,b)-g_0 (x,b')$
and $\tau=\Delta/(e^\Delta-1).$  
The values of $R_c$ obtained from formula (\ref{id}) are given in
Table~\ref{tab1} (after rounding to the closest integer) for base
triplets $xby$ with central base $b=A$ and $b=C$ respectively. The
values of $R_c$ for triplets with central bases $b=T$ and $b= G$ can
be deduced from the decay constants of the complementary triplets, expressed
in reversed order, due to the symmetry of the interaction matrix $g_0$
of Table \ref{tableg0} {\em e.g.}
$R_c^{sm}(ATT)=R_c^{sm}(AAT)$. The value $b'$ of the most difficult base to
distinguish from $b$, see (\ref{id}), is $T$ when the central base is
$A$ and $G$ when the central base is $C$, except in the $AAA$, $CCC$
cases where $b'=G$, $b'=A $ respectively.

\subsubsection{Propagation of errors, and blocks of locked-in bases}

The above single base mutation offers only a lower bound to the true value of 
the decay constant $R_c(i)$ of the error $\epsilon _i$ in predicting
base pair $i$. Strictly speaking, to calculate $R_c(i)$, one
must consider all the $3\times 4^{N-1}$ sequences where base $i$
differ from its value in the true sequence, and find among those sequences
the one which requires the largest number of unzippings to be discarded. 
In other words errors on bp $i$ may result from the difficulty of 
correctly predicting a block of more than one bp located around bp $i$ 
rather than this bp alone.
 
We start by defining the decay constant for the large $R$ behavior 
of the single bond  misprediction probability $\epsilon^{xy\to x'y'}$
 for two contiguous mutations  $(xy\to x'y')$,
\begin{equation}
\epsilon^{xy\to x'y'}_R \sim e^{-R/R_c (xy\to x'y')}
\label{errex}
\end{equation}
where $R_c (xy\to x'y')$ is given by eqn~(\ref{fr}) with $\Delta
= g_0 (x,y)-g_0 (x',y')$ and $\tau =\Delta/(e^\Delta-1)$ (\ref{fr2}).  
We then define, in the maximum likelihood framework, 
the probabilities (with respect to the random variables $t_i$) 
$\mu^\rightarrow_i (b)$ and $\mu^\leftarrow_i (b)$
of predicting base pair $i$ to be of $b$-type when,
respectively, the bases located to the right and the left of $i$ are
ignored. We assume that 
\begin{equation}
\mu^\rightarrow_i (b)=e^{-R \pi^\rightarrow _i (b)} \quad
\mbox{and}\quad 
\mu^\leftarrow_i (b)=e^{-R \pi^\leftarrow _i(b)} 
\end{equation}
for a large number $R$ of unzippings, with boundary conditions 
$ \pi^\rightarrow _1(b)=0$ and
$ \pi^\leftarrow _N(b)=0$ for all $b$. These probabilities can be 
evaluated from the probabilities of the most dangerous subsequence
to the left and right of base pair $i$, according to the
recurrence equations
\begin{eqnarray}
\label{rcprop}
\pi^\rightarrow _i(b')&=&\min_{b}\left( \pi^\rightarrow _{i-1}(b)+
\frac 1{R_c (b_{i-1}^Lb_{i}^L\to b\,b')} \right)   \\ 
\pi^\leftarrow _i(b')&=&\min_{b}\left( \pi^\leftarrow _{i+1}(b)+
\frac 1{R_c (b_{i}^Lb_{i+1}^L\to b'\,b)} \right)  \ , \nonumber
\end{eqnarray}
remember $b_i^L$ denotes the true type of bp $i$. These recurrence
equations have a simple meaning. The probability that bp $i$
is of $b'$ type, when there is no base to the right of $i$, is 
simply given by the sum over $b$ of the probability that bp $i-1$
is of $b$ type times the probability of predicting the bond 
$b\,b'$ instead of $b_{i-1}^Lb_i^L$. Notice that
recurrence eqns~(\ref{rcprop}) are simply the asymptotic
counterpart of eqn~(\ref{recurq}) in the large $R$ limit (for four 
and not two base types). They can be obtained from eqn (\ref{recurpi})
and (\ref{recurqpim}) by choosing for $t_i$  the time
having equal probabilities with the true bond $b_{i-1}^Lb_i^L$ and
the erroneous bond $b\,b'$  distributions \cite{altro}. 

The decay constant $R_c(i)$ of the error on bp $i$ is obtained by selecting 
the most dangerous value for the type $b$,
\begin{equation}
\frac 1{R_c(i)}= \min_{b\neq b_i^L} \;\left ( \pi^\leftarrow _i(b)+ 
\pi^\rightarrow _i(b) \right ) \ .
\label{rcprop1}
\end{equation}
In general $R_c(i)$ differs from the single mutation value,
$R^{sm}_c(i)$.  The latter depends only on the base and its two
neighbors while the former depends on the whole sequence.
Equations~(\ref{rcprop}) and~(\ref{rcprop1}) can be interpreted by
considering $\pi^\leftarrow _i(b)+
\pi^\rightarrow _i(b)$ as the free energy for the lowest 
excited state (sequence) with the base $i$ fixed to a value, $b$, 
distinct from the one, $b_i^L$, in the ground state (real sequence). 
If the base $i$ has a very large value for $R^{sm}_c(i)$, because both 
the bonds on the right and on the left of the base have a large $R_c$ 
(see eqn~\ref{id}),  the
most dangereous sequence is exactly this 'single mutation' sequence.
In this case the minimum over $b$ in (\ref{rcprop}) is exactly obtained
for $b=b_{i-1}^L$ and $b=b_{i+1}^L$, and the recursion halts after
the nearest neighbors. 
However, when the bond constant $R_c^{sm}$ is small, we can expect that 
it is less costly, in terms of free energy, 
to propagate the excitation at site $i$  in a configuration where the base  
and its neighboring base are both mutated into their
complementary values. The decay constant $R_c$ for such a bond 
is indeed large because it is difficult to
distinguish two bases from the complementary ones (Table \ref{tab00}).
This 'defect' propagates,
in the recurrence eqn~(\ref{rcprop}), until an interface with a 
large value for $R_c$ is found. Obviously this propagation mechanism
takes place on both sides of bp $i$.
The most dangerous excitations are thus blocks of complementary bases
of the real sequence. The bases in a block have then roughly the same 
$R_c$ and are locked-in together (Fig~\ref{epsilonif40}). 

The high force behaviour of the errors $\epsilon _i$  (for $R=1,50,200$),
obtained by the numerical inference and shown  in Fig~\ref{epsilonif40}
agree with these theoretical results. The theoretical values for the decay 
constants $R_c(f\ge 40 pN,i)$ obtained from (\ref{rcprop},\ref{rcprop1}) 
are shown  in Fig.~\ref{Rpd} (dotted line). By solving eqn~(\ref{rcprop}) 
we find that bp $i=6$ belongs to a block extending from  bp 1 to 9.
The boundary bp 1 has  $R_c$ on the left equal to $\infty$ and   bp
8 has $R_c(GA\to CA)=139$. From eqn~(\ref{rcprop})  we obtain  
$R_c=114$ for the whole block 1-9.  
This value coincide with the  decay of the error at large $R$ found from
simulations and shown in Fig~\ref{omrun}. We  obtain $R_c^{num}=113 
\pm 2$ from a fit of $\log \epsilon _i$ vs. $R$ at $f=40$ pN.
\footnote{A fit of the slope of the curve in figure \ref{omrun}  
gives $R=100 \pm 1$ while a better fit $R=113 \pm 2$ is obtained by 
taking into account
 the multiplicative $1/\sqrt R$ term in the error in formula
 ~(\ref{nostackomega}).}
Base pair 27  belongs to a block on the right
spreading over the whole sequence down to base 1, while the block on the left 
stops on the base itself.
The number of unzippings needed for a good prediction
of bp 27 is smaller: we obtain from theory $R_c= 24$, and from simulation
$R_c^{num}= 25 \pm 1$. 
Note that the propagation of the error by blocks of complementary bases
in this section go beyond the single mutation approximation
reported in~\cite{corto}.

\subsection{Moderate force theory}

\begin{figure}
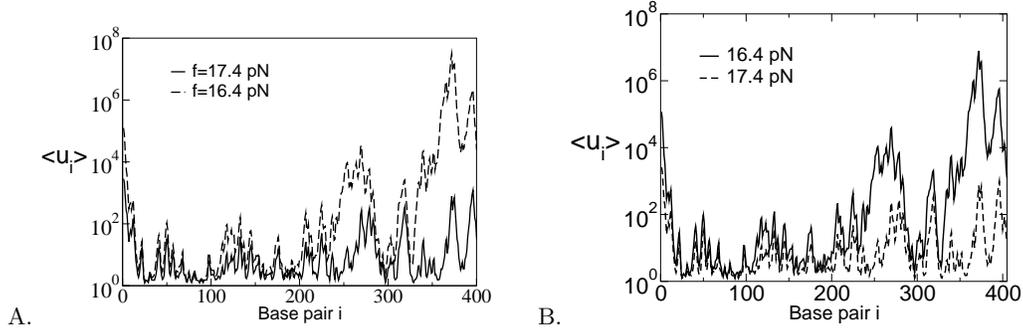

\begin{center}
A. \epsfig{file=fig17a.eps,width=6.cm,angle=0} \hskip .5cm 
B. \epsfig{file=fig17b.eps,width=6.cm,angle=0}
\caption{Average number $\langle u_i \rangle$ of openings of bp $i$
for the $\lambda$--phage sequence during one unzipping for forces
16.4 and 17.4 pN.  {\bf A}. theoretical values in the
limit of infinite time.  {\bf B}. numerical values from MC simulations
with $ M=10^7$ steps.  Note that the
infinite time theoretical values coincide with the numerical values
up to some base index $i_{max}$ such that $\sum _{i< i_{max}} 
\langle u_i \rangle\ll M$
{\em e.g.}  $i_{max}\simeq200$ for $f=16.4$ pN and $M=10^8$ steps. }
\label{visit}
\end{center}
\end{figure}

\subsubsection{On the number of single-base openings}
\label{avuisec}

We now investigate the case of unzipping under a finite force. 
The opening fork may go backward, closing a previously 
open base pair, and reach this base pair later. Therefore the number 
$u_i$ of opening transitions $i\to i+1$, $u_i$, is not always equal to 
unity but is stochastic and varies from experiment to experiment, and base to
base. 
To calculate the distribution of  $u_i$ it is convenient to think of 
the opening and closing process as an unidimensional random walk
where, at each move, the probability to go backward and forward
(closing and opening transitions respectively) are equal to $q_i$ and
$1-q_i$ respectively, with
\begin{equation} \label{defq}
q_i=\frac{e^{g_{s}(f)}}{e^{g_{s}(f)}+e^{g_0(b_i,b_{i+1})}} \ .
\end{equation}
For forces larger than the critical force, we have $q_i < \frac{1}{2}$: the
random walk is submitted to a forward drift and is transient.  
We define the probability of escape, $E_i$, as the probability of
never reaching back position $i$ starting from position
$i+1$. The case of infinite force corresponds to $E_i=1$.
For a homogeneous sequence the free energy landscape $G(n,f)$ in which the 
random walk takes place is
simply a tilted line; $E=(1-2q)/(1-q)$ 
depends on the force and on the sequence type.
For  a heterogeneous sequence the free energy landscape $G(n,f)$ is
more complex (Fig.~\ref{ombaserun2}), $E_i$ depends not only on the force  
and  on the base  type $b_i$ (and on its neighbor $b_{i+1}$) but also 
on its environment e.g. whether
base $i$ is located in a local minimum or in a local maximum of the free-energy
landscape. We show how to calculate $E_i$ in Appendix~\ref{sec93} for any
given sequence. 

The distribution $\rho_1$ of the number $u_i$ of opening transitions 
$i\to i+1$ during a single unzipping is simply obtained from $E_i$ and reads
\begin{equation} \label{defrho1}
\rho_1(u_i)= \left(1-E_i\right)^{u_i-1}\;E_i
\end{equation}
From equation~(\ref{defrho1}) we have that the average number of openings
of bp $i$ is
\begin{equation}
\langle u_i \rangle = \frac{1}{E_i} \ .
\label{un}
\end{equation}
$\langle u_i\rangle$ is shown  in Fig.~\ref{visit} for
forces $f=16.4,17.4$~pN for the first
400 bases of the $\lambda$ phage DNA sequence.
Theoretical values for $\langle u_i\rangle$ are obtained 
in the limit of infinite time while MC simulations (or experiments)
duration is finite. Call $t_i ^{last}$ the expectation
value of the last-passage time of the fork at site $i$; $t_i^{last}$ is finite
since the random walk is transient. Clearly theoretical and 
MC values for $\langle u_i\rangle$ will coincide
for bases of indices $i<i_{max}$ where $t_{i_{max}}^{last}$ is equal to
the duration of the simulation. In practice we estimate $i_{max}$ through 
the condition
$\sum _{i<i_{max}} \langle u_i\rangle \simeq M$, where $M$ is the number of
MC moves. The outcome for $i_{max}$ is plotted in the
inset of Fig~\ref{basi1run}. For instance, as shown in 
Fig.~\ref{visit}, $i_{max}\simeq 200$ for $f=16.4$~pN and $M=10^8$. 
 $\langle u_i\rangle$ varies a lot from
 base to base, and reaches values up to $10^8$ (for the considered force).

The generalization of the calculation of the distribution $\rho
_R(u_i)$ of the number of openings of base pair $i$ to the case of  
$R$ unzippings is immediate (Appendix \ref{app1}). The result is the
$R^{th}$ convolution power of $\rho_1$, and reads
\begin{equation} \label{rhor}
\rho_R(u_i) = 
{u_i-1 \choose R-1} \,(1-E_i)^{u_i-R}\;E_i^R \ .
\label{pru}
\end{equation}

\subsubsection{Error in predicting a base in the absence of stacking}
\label{rcfiniteforcesec}

The number of opening transitions of a base at finite force, $u_i$,
plays the same role as the number $R$ of repetitions of the unzippings
at large force.  As the fork visits again and again the same base pair
more and more data are collected on the sojourn time $t_i$ on this
base and the prediction error becomes smaller and smaller.  However,
contrary to $R$, $u_i$ is a stochastic variable.  The error in
predicting base pair $i$ of type $b_i=W,S$, in the absence of stacking
is then obtained by averaging the error on this bond at large force
and after $u_i$ unzippings, $\epsilon_{u_i}^{b_i}$ (\ref{pnostack4}),
over the distribution $\rho_R$ (\ref{rhor}),
\begin{equation} 
\label{errf} \label{formpred}
\epsilon _{f,R} ^{b_i}= \sum_{u_i\ge 1} \rho_R(u_i) \;
\epsilon_{u_i}^{b_i} \ ,
\end{equation}
where the $f$ subscript indicate that the above formula holds for a
finite force.
A detailed derivation of eqn (\ref{errf}) is given in
Appendix~\ref{phadia}. In the limit of large force
$E_i\to 1$ from (\ref{defq}), $\rho_R(u_i)\to \delta_{u_i,R}$ from
(\ref{rhor}), and $\epsilon_{f,R}^{b_i} \to \epsilon_R^{b_i}$ as expected.

Error~(\ref{formpred}) can be easily computed when the error 
$\epsilon_{u_i}^{b_i}$ is replaced with  asymptotic expression 
(\ref{nostackomega}).  Using the expression for the generating
function of the probability $\rho_R$ with argument $\exp(-1/R_c)$ given 
in Appendix~\ref{app1} we obtain
\begin{equation}
\label{err}
\epsilon^{b_i}_{f,R} \simeq e^{-R/{R_c(f,i)}} \quad \mbox{with} \quad
R_c (f,i)= \left[\ln \left(1+\langle
    u_i\rangle(e^{1/{R_c}}-1)\right)\right]^{-1}
\label{epsmedio}
\end{equation}
The above decay constants $R_c$ can be approximated with the simpler expression
\begin{equation} \label{rcforce} 
R_c(f,i) \simeq \frac{R_c }{\langle u_i\rangle} 
\end{equation}
which are quantitatively accurate unless the number of required
unzipping at large force, $R_c$, becomes much smaller than $\langle
u_i\rangle$ {\em i.e.} close to the critical force. This formula
simply expresses that the effective number of unzippings to correctly
predict base $i$ at finite force is $R \times \langle u_i\rangle$
rather than $R$.  Recall that the value of the decay constant of the
error at high force, $R_c$, depends only on the free energy difference
between $W$ and $S$ bases. At finite force this decay constant is
roughly divided by $\langle u_i\rangle$.  The latter depends on the
whole free energy landscape around the base. Therefore at finite
force, even in the absence of stacking interaction, the error on a
base depends on the whole sequence of bases. Moreover bases with a large $R_c$ 
that are in a valley of the free energy landscape can be better
predicted than bases with a small $R_c$ located on the
top of barriers in the  landscape. 

Let us apply the above result to the case of a homogeneous sequence,
with two base types, $b=W,S$. 
The decay constant $R_c$ (\ref{fr2}) at high force
depends only on the free energy difference $\Delta$ between W and S bases. 
For a homogeneous sequence the average number of openings of
each base is simply  $\langle u\rangle =\frac{1-q}{1-2q}$, where $q$ is 
obtained  from formula~(\ref{defq}) with  $g_0(b_i,b_{i+1})=g_0 (b)$.
In Fig~\ref{omteoforce} we plot the error for $W$ bases for 
$\Delta=2.8$ (to distinguish a sequence of bases $A$ or $T$ 
from a sequence of bases $G$ or $C$) and $\Delta=0.5$ (to distinguish  
a sequence of $A$ bases from one of $T$ bases, or a sequence of $C$ bases 
from one of $G$ bases). The plot for a repeated sequence of $S$ bases is 
similar.
As shown in Fig~\ref{omteoforce} the error sharply decreases
when the force reaches its critical value from above e.g.
$f_c=9.25$~pN for  $g_0(W)=-1.1$~k$_B$T.
As shown in Fig~\ref{omteoforce} the decay constant (\ref{epsmedio}) 
\begin{equation}
\label{epsmedioo}
R_c (f)=\left[ \ln\left( \frac {(1-q)e^{1/R_c} -q} 
{1-2q} \right)\right]^{-1}
\end{equation}
obtained by approximating $\epsilon _{u} ^b$ with a pure exponential
is in perfect agreement
with the numerical calculation of formula $\epsilon_{f,R}^W$.
The simplified expression (\ref{rcforce}) 
\begin{equation} \label{rcf}
R_c (f) =  R_c   \times \frac{1-2q}{1-q}  \,.
\end{equation}
is in very
good agreement with $R_c (f)$, except in the case 
$\Delta=2.8$, $f=f_c+2$ pN for which the decay constant is very small. 

The value of $R_c(f)$ is plotted as a function of the force in
Fig \ref{phdi} for various sequences, and allows us to
draw the phase diagram for the prediction 
in the force vs. number of unzippings plane.
The prediction becomes perfect, $\epsilon _{f,R} ^{b} \ll 1$, if the
number $R$ of unzippings is (much) larger than some crossover value
$R_c$ (\ref{epsmedioo}). It appears that $R_c(f)$ is always
smaller than its infinite force value $R_c$, and vanishes when the
force reaches the critical unzipping force from above, $f\to f_c ^+$.
In this limit, $q\to \frac{1}{2}$: the motion of the opening fork
becomes purely diffusive, and each base is visited a very large number
of times going to infinity for an infinite duration of the experiment.
Predictions made from a single unzipping are reliable provided
$R_c(f)< 1$ {\em i.e.}  the force $f$ does not exceed by a large
amount its critical value $f_c$,
\begin{equation}
f -f _c \leq \frac{\Delta ^2}{8 \, d_c}
\end{equation}
where $d_c = |dg_{s}/df(f_c)|$ is twice the extension of a DNA
single strand monomer at the
critical force, and we have used expression (\ref{fr2}) for $R_c$.
Typically, $d_c\sim 1~$nm $\simeq 0.25$ k$_B$T/pN, leading to $f-f_c <
\frac{1}{2}\Delta ^2$ pN with $\Delta$ expressed in units of
k$_B$T. Notice that this theoretical result does not consider the
actual number of open base pairs, which decreases as the force is
lowered to its critical value, but only the quality of their
prediction.

\begin{figure}
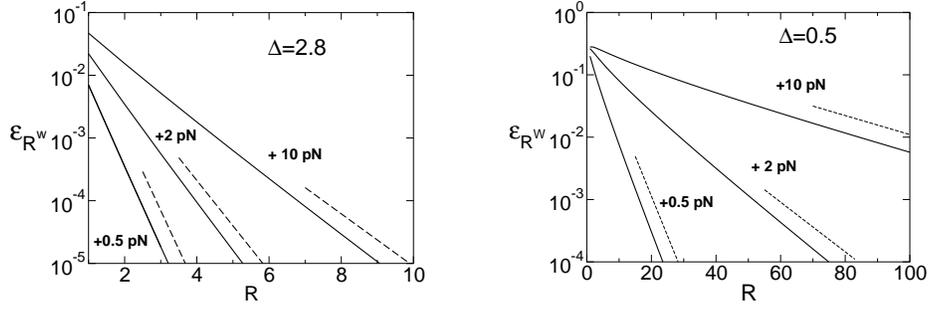

\begin{center}
\psfig{figure=fig18a.eps,height=4cm,angle=0} \hskip 1cm
\psfig{figure=fig18b.eps,height=4cm,angle=0}
\caption{Probability $\epsilon$ of misprediction on repeated sequences
of $W$ (empty dots, dashed lines) and $S$ (black dots, full lines)
bases for pairing free--energy differences $\Delta =2.8$ ({\bf A}) and
$\Delta=0.5$ ({\bf B}) in the absence of stacking. For each case we
show the error as a function of the number $R$ of unzippings for
forces above the critical force by $0.5$, 2 and 10~pN.
The decay constants have for $\Delta=2.8$ the following values: 
$R_c(f=\infty)=32$; for $f=f_c+10$ pN, 
 $R_c=28.5$; for $f=f_c+2$ pN,  $R_c=10.9$; for $f=f_c+0.5$ pN, $R_c=3.4$.}
\label{omteoforce}
\end{center}
\end{figure} 

\begin{figure}
\begin{center}
\epsfig{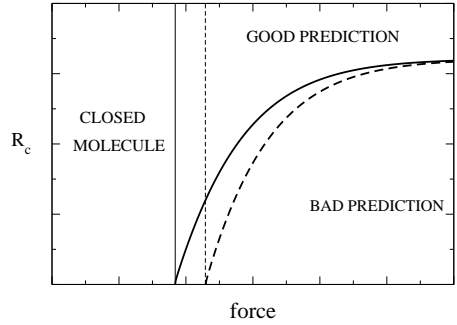} 
\caption{Phase diagram in the number of unzippings vs. force plane.
Efficient prediction is possible above the critical line $R_c(f)$
(\ref{epsmedioo}). Here $g_0(W)=-1.06,g_0(S)=-1.55$. 
The full line indicates the repeated $W$ sequence,
the dashed line corresponds to the repeated $S$ sequence. For forces
smaller than the critical value $f_c\simeq 9$~pN for the W sequence,
$f_c\simeq 12$~pN for the S sequence (vertical lines) 
the molecule remains closed. At large force the number of required
unzippings reaches a common value $R_c\simeq 30$.}
\label{phdi}
\end{center}
\end{figure}

\subsubsection{Results for heterogeneous sequence in presence
of stacking interactions}\label{sechs}

The above theory tells us how many unzippings are necessary to
recognize a base type from another at moderate force, 
when the pairing free energies of these
two base types differ by $\Delta$ and when the fork opens the base 
$\langle u_i\rangle$ times in each unzipping. 
It can be applied to the case of bond and not base recognition as we 
have done at large force in Section \ref{yyy}. The number of
unzippings $R_c(f,i,b_i^Lb_{i+1}^L\to b\,b')$ necessary to
recognize that the bond between  base pairs $i$ and $i+1$ is not  
$b\,b'$ is given 
by expression (\ref{epsmedio}) or (\ref{rcforce}) with $R_c$ substitued
with $R_c(b_i^Lb_{i+1}^L\to b\,b')$, see Section \ref{yyy}, 
which depends on the biochemical parameters $g_0(b_i^L,b_{i+1}^L)-
g_0(b,b')$ given in Table~\ref{tab1}. 

The decay constant of the error on base $i$ at finite force,
 $R_c(f,i)$, is calculated by applying
the recursive formula~(\ref{rcprop}) and  the minimization 
formula (\ref{rcprop1}) after replacing
the  bond decay constants at infinite force with the 
ones at finite force, 
\begin{eqnarray}
\label{rcprope}
\pi^\rightarrow _{i,f}(b')&=&\min_{b}\left( \pi^\rightarrow _{i-1,f}(b)+
\frac 1{R_c (i,f,b_{i-1}^Lb_{i}^L\to b\,b')} \right)   \\ 
\pi^\leftarrow _{i,f}(b')&=&\min_{b}\left( \pi^\leftarrow _{i+1,f}(b)+
\frac 1{R_c (i,f,b_{i}^Lb_{i+1}^L\to b'\,b)} \right)  \ , \nonumber
\end{eqnarray}
with bondary condition $ \pi^\rightarrow _{1,f}(b)=\pi^\leftarrow _{N,f}(b)=0$.
The minimization condition then reads
\begin{equation}
\frac 1{R_c(f,i)}= \min_{b\neq b_i^L} \;\left ( \pi^\leftarrow _{i,f}(b)+ 
\pi^\rightarrow _{i,f}(b) \right )\ .
\label{rcprope1}
\end{equation}
Figure \ref{Rpd} shows the values of $R_c(f,i)$ at $f=17.4$
pN  (full line)
for the first 400 base pairs of the $\lambda$--phage derived from
(\ref{epsmedio}). 
$R_c(f,i)$ is in very good agreement with the decay constant
 of the error $\epsilon _i$ obtained through the numerical inference
procedure and shown in Fig~\ref{sigmabase}A. Indeed, roughly,
for all bases with $R_c(f,i) \leq 15$ the numerical inference errors 
goes to zero with $R=40$ unzippings. For a more precise comparison
we have focused on two specific bases (Fig~\ref{omrun}).

Base pair 6 is located in a valley of the landscape $G$ at force of 17.4 pN,
hence the number of openings of the base, $\langle u_i\rangle$, and
of its neighbors, $\langle u_j\rangle$ with $j$ close to $i$, are large 
{\em e.g.} $\langle u_{1} \rangle = 28000$,
 $\langle u_{6} \rangle = 60$ as shown in Fig~\ref{visit}. 
The decay constant of the error quickly decreases 
with the force from $R_c(f\geq 40\mbox{ pN},i=6)=114$ 
to  $R_c(f=17.4\mbox{ pN},i=6)=2$; these theoretical values are in  
very good agreement with the numerical findings of Fig~\ref{omrun}.
Moreover the connected correlation function $\chi_{i,6}$ 
at $f=17.4$~pN has non-zero value up to the base $i=20$.  
Solving the recursive eqns~(\ref{rcprope},\ref{rcprope1}) we found
that the decay of the prediction error on $i=6$ originates 
from a 20 defect--sequence where bases 1-20 are locked-in into 
their complementary values with respect to the true sequence. 

Base pair 27 lies, on the contrary, on a barrier of the free energy 
landscape and the numbers of openings (at a force of $17.4$ pN)
of this base (and its neighbors) is smaller:  $\langle u_{27} \rangle 
=1.5 $ as shown in Fig.~\ref{visit}. The decay constant decreases 
slightly when the force diminishes, from $R_c(f\geq 40\mbox{
pN},i=27)= 24$ to  $R_c(f=17\mbox{ pN},i=27)=15$.
These theoretical values agree very well with the fit of the
numerical simulations in Fig \ref{omrun}. 
Moreover the decay of the prediction  error on base  27 at $f=17.4$ pN came
from a two-defect excitation of bases 26-27.
Note that numerical results are limited by the
finite number of samples from which the error $\epsilon _i$ is
calculated. The number of samples $M_p$ necessary to estimate accurately
the error must be much larger than the inverse of the probability of 
misprediction.
With $M_p = 2\,10^{4}$ (Fig~\ref{omrun}) errors smaller than 
$\epsilon=10^{-3}$ cannot be measured. As $\epsilon$ decreases
exponentially with $R$, $M_p$ must scale as $\exp(R\,\mu)$ with 
$\mu > R_c$ to reach a good estimate of $R_c$. 
Finite sampling could also lead to
statistical bias due to the large deviation fluctuations of
$u_i$. We show that these effects are negligible in Appendix \ref{appi}. 

\begin{figure}
\begin{center}
\epsfig{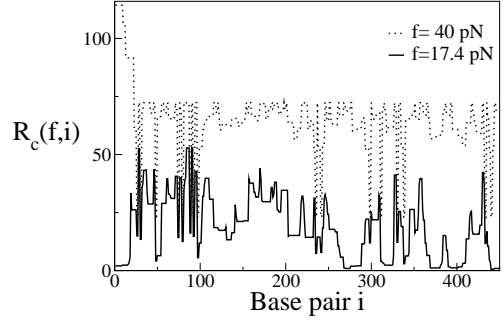}
\caption{Theoretical values for the number $R_c(f,i)$ of unzippings
necessary for a good prediction of base $i$ at force $f=17.4$ (full
line) and $f\geq 40$ pN (dashed line)  for the first 400 bases of the
$\lambda$ phage sequence obtained from formula (\ref{epsmedio}) }
\label{Rpd}
\end{center}
\end{figure}

\subsection{Inference from two-way unzippings}
\label{twowaysec}

We hereafter consider that the molecule can be unzipped from both
extremities (two-way opening) and want to infer its sequence from the
data collected in both directions. This investigation is motivated by
the observation  that the 
free energy landscape is flipped {\em i.e.} multiplied by $-1$ when
the molecule is opened from the other extremity. Bases that were 
located in local maxima in the landscape, hence poorly
predicted, become local minima in the new landscape, and  are much
better predicted.   

Let us denote $+$ the normal direction of unzipping of the
molecule: the $i^{th}$ base (along the $5'\rightarrow 3'$ strand of
molecule) in this direction is simply $b_i$.
The free energy to open the first $n$ bases of the molecule is
$G^+(n,f;B)$, equal to $G$ defined in (\ref{p}). 
In the reverse direction, denoted by $-$, we denote by $b^-_i$ the
$i^{th}$ base along the $5'\rightarrow 3'$ direction:
$b_i^-=\mbox{compl}(b_{N+1-i})$ where $\mbox{compl}(b)$ denotes the
complementary base of $b$. The free energy to open the 
first $n\ge 0$ bases of the molecule in the $-$ direction is
\begin{equation} \label{grev}
G^-(n,f;B)=\sum_{i=0}^{n-1} g_0(b^-_i,b^-_{i+1})- n\; g_s(f) = 
\sum_{i=N-n+1}^{N} g_0(b_{N-i},b_{N-i+1})- n\; g_s(f) = -
G^+(N-n,f;B) + G(N,f;B)
\end{equation}
where we have used the symmetry
$g_0(b,b')=g_0\big(\mbox{compl}(b'),\mbox{compl}(b)\big)$
of the $g_0$ interaction matrix (Table \ref{tableg0}) \footnote{To define 
properly the change in the free energy $G$ (\ref{p}) of the molecule when
its last base $i=N$ is opened we have added a
fictitious $i=N+1$ base; the contribution to the free energy is symbolized
by $\Delta g =g_0(b_N,b_{N+1})$. In practice $\Delta g$ is not given
by Table \ref{tableg0} but may have a more complicated origin. For instance
the molecule may end with a loop, $\Delta g$ will then be equal to the
gain in entropy when the loop opens.}. Therefore, up to an irrelevant
additive constant, the free energy to open $n$ bp in the $-$ direction is 
simply the opposite of the free energy to open $N-n$ bp in the $+$ direction.

If we unzip $R$ times the molecule in the $+$ direction the
error in predicting base $i$ will decay exponentially with $R$ with a
decay constant equal to $R_c ^+(f,i)$ given by eqn (\ref{rcf}).
We may instead open $R$ times the molecule in the $-$ direction, and 
infer the value of base $i$ (labeled $N+1-i$ in the $-$ nomenclature).
The probability of a mistake is again an exponentially decreasing function
of $R$ with decay constant $R_c ^-(f,i)$ (\ref{rcf}), calculated 
from the number of openings of base $i$ in the $-$ direction 
(Appendix~\ref{appdouble}). 

Assume now that the unzip $R/2$ times the molecule in the $+$ direction
and $R/2$ times in the $-$ direction. We show in Appendix~\ref{appdouble} 
that the probability of predicting that the bases attached to the 
bond $i,i+1$ are $b,b'$ 
decays exponentially with $R$  with a decay constant equal to
\begin{eqnarray}\label{rcavin}
R_c ^{+ \& -} (f,i,b_i^Lb_{i+1}^L \rightarrow b\,b') &=&
\left[\ln \left(1+\langle
    u_i^+\rangle (e^{1/2\,R_c(b_i^Lb_{i+1}^L \rightarrow
    b\,b')}-1)\right)+ \ln \left(1+\langle u_{i+1}^-\rangle (e^{1/2\,
 R_c(b_i^Lb_{i+1}^L \rightarrow b\,b'
)}-1)\right)\right]^{-1} \nonumber \\
&\simeq & 2\;R_c(b_i^Lb_{i+1}^L \rightarrow b\,b'
)/\left( \langle u_i^+\rangle+\langle u_{i+1}^-
  \rangle\right)
\end{eqnarray}
We have taken into account the effects of stacking interactions
between nearest neighbor base pairs as done in
Section~\ref{stacksec2}.  The decay constant of the error $\epsilon
_i$ in the two-way unzipping at force $f$, $R_c ^{+ \& -}(f,i)$, is 
obtained using recurrence eqn~(\ref{rcprope1}) upon substitution 
of $R_c (f,i,b_i^Lb_{i+1}^L \rightarrow b\,b')$ with $R_c ^{+ \& -}
(f,i,b_i^Lb_{i+1}^L \rightarrow b\,b')$. The results for 
$R_c ^{+ \& -}(f,i)$ are shown in Fig~\ref{rcbf}. 
A comparison with Fig \ref{Rpd} shows that the number of unzippings
necessary for a good prediction greatly decreases with the 
two-way unzipping procedure with respect to the one-way unzipping (for the
same amount of collected data). 

\begin{figure}
\epsfig{file=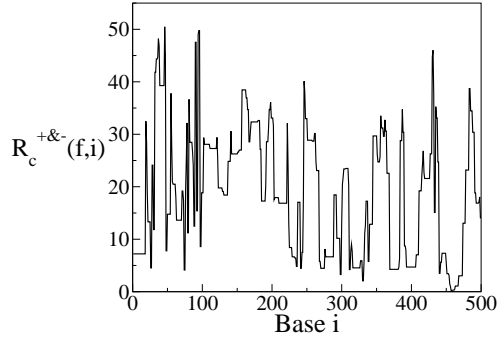,width=6.5cm,angle=0}
\caption{Decay constant $R_c^{+\&-}(f,i)$ of the prediction error on
  base $i$ for
the first 450 base pairs of the $\lambda$ phage DNA, at force $f=17.4$
pN from the two-way unzipping numerical unzipping.}
\label{rcbf}
\end{figure}
\section{Towards more realistic data modeling}

\subsection{Finite-bandwidth inference}
\label{mstepsec}

So far we have assumed that the temporal resolution was infinite. A time-trace
contains a perfect information on the opening dynamics {\em i.e.} 
on the motion of the fork (set of numbers $u_i$ ,$d_i$) and on the
sojourn times $t_i$ for every base $i$ of the chain.
Real experiments obviously do not have such a perfect sensitivity:
actual feedback systems and detectors are limited to delays between
measures of about $\Delta t\sim 0.1-1$ ms. This temporal resolution is
a major limitation: during the delay $\Delta t$ the fork can explore up 
to 100-1000 bases around the starting position, depending on the local
structure of the free energy landscape.  The true dynamics of the fork
is therefore unknown  and the prediction algorithm has to consider
all the trajectories of the fork (in a $\sim$100 bp window). This problem is
studied in detail in \cite{altro}. Hereafter we limit ourselves to the
case of a finite but very large bandwidth where the delay 
$\Delta t$ between two
measures is of the order of the opening time of a bp (and not much
smaller as considered so far).

\subsubsection{Typical jump between two measures}

Rates (\ref{ratemd}) define the non zero (off diagonal) elements
of the elementary transitions matrix 
\begin{equation}
\hat H_{i',i} = r_o(i)\cdot \delta_{i',i+1} + r_c(f) \cdot
\delta_{i',i-1}- (r_o(i)+r_c(f)) \cdot \delta_{i',i}
\end{equation}
The evolution operator after a time $\Delta t$ is given by the
matrix exponential
\begin{equation}\label{matrix_exp}
\hat U = \exp \big[\Delta t\; \hat { H}\big]
\end{equation}
The entry $\hat U_{i',i}$ represents the probability of going from base $i$ to
base $i'$ in the time interval $\Delta t$.
In principle all transitions are allowed and $\hat U$ is therefore a
$N\times N$ matrix. In practice jumps $j=i'-i$ are unlikely to exceed
(in absolute value) the ratio $\Delta t/\tau$ where $\tau$ is the 
typical time to open a bp. The probability distribution of jumps $j$, 
averaged over the starting base $i$, is shown in Fig
\ref{p_jump} for $f=16.4$ and 17.4 pN, and $\Delta t$ ranging 
between $10^{-5}$s and $10^{-3}$s.
As the force and the sampling interval increases the distribution
gradually spreads over larger jump values, and long tails appear.
Nevertheless, long jumps  seem to be rare events, restricted to
particular regions of the landscape.   Most of the
information on the opening dynamics can therefore be kept
when discarding displacements larger than some threshold $J$ e.g. $J=10$
in Fig \ref{p_jump}. To do so, given the starting base $i$, we
construct a reduced $(2J+1)\times (2J+1)$ matrix $\hat H^{(J,i)}$ as follows,
\begin{equation} 
\label{reduced_H}
\hat H^{(J,i)}=\left(
\begin{array}{ c c c c c}
- r_o(i-J) - r_c& r_c& 0 & \ldots & 0\\  r_o(i-J) & - r_o(i-J+1) -r_c &
r_c & \ldots & 0\\  0 &  r_o(i-J+1)  &  - r_o(i-J+2) -r_c
 & \ldots & 0 \\\vdots & \vdots& \vdots & \vdots& \vdots\\
0& \ldots & 0& r_o(i+J-1) & - r_o(i+J)-r_c\\
\end{array} \right)
\end{equation}
and the associated evolution operator 
$\hat U^{(J,i)} =\exp\big[ \Delta t\; \hat  H^{(J,i)}\big]$, which encodes
all the jumps from base $i$ of amplitude less or equal to $J$. 
There are $4^{(2J+1)}$ different $\hat U^{(J)}$ matrices, one for each 
possible choice of the $2J+1$ bases involved.

\begin{figure}
\begin{center}
\psfig{figure=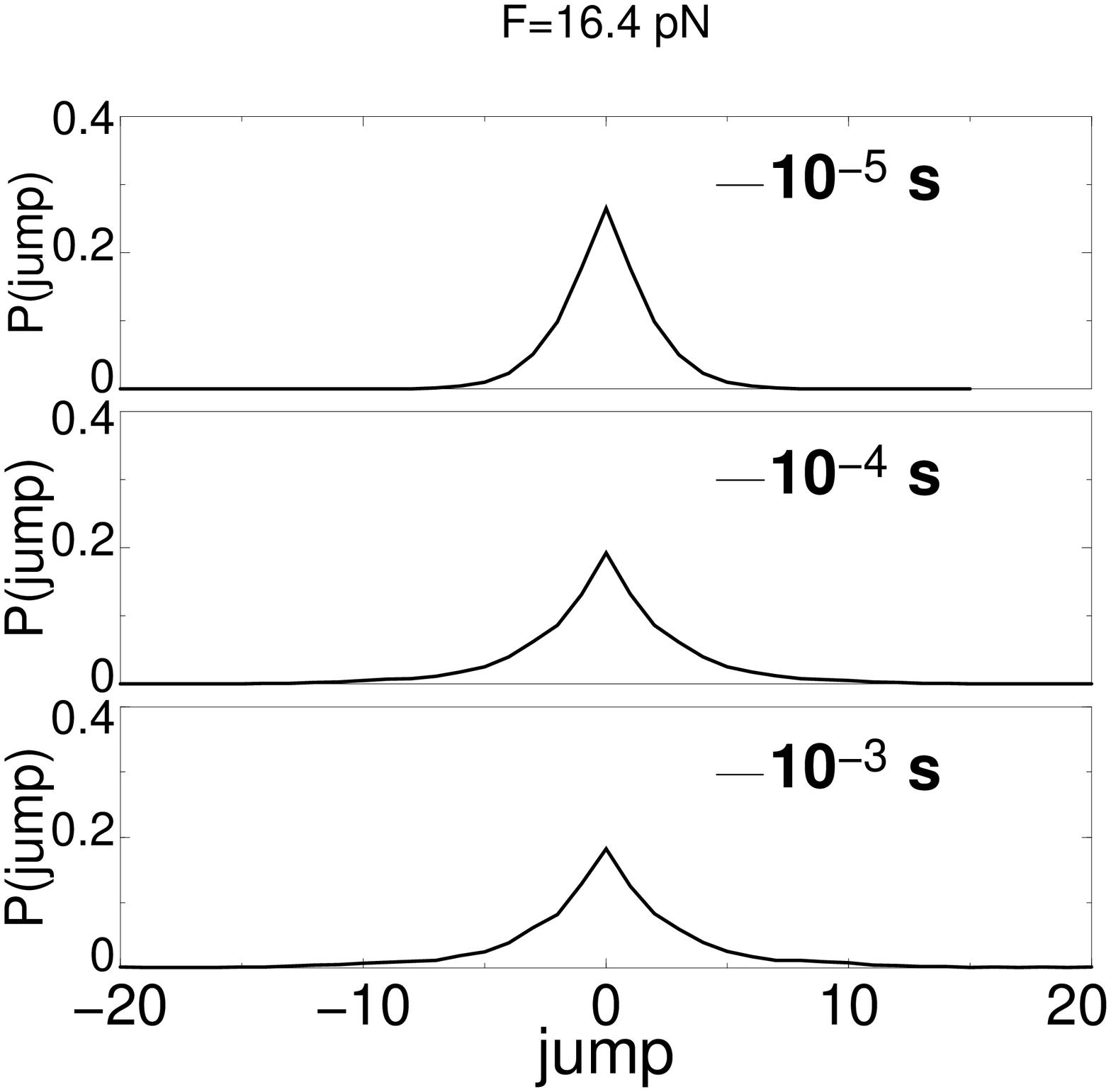,height=6cm,angle=-90}
\psfig{figure=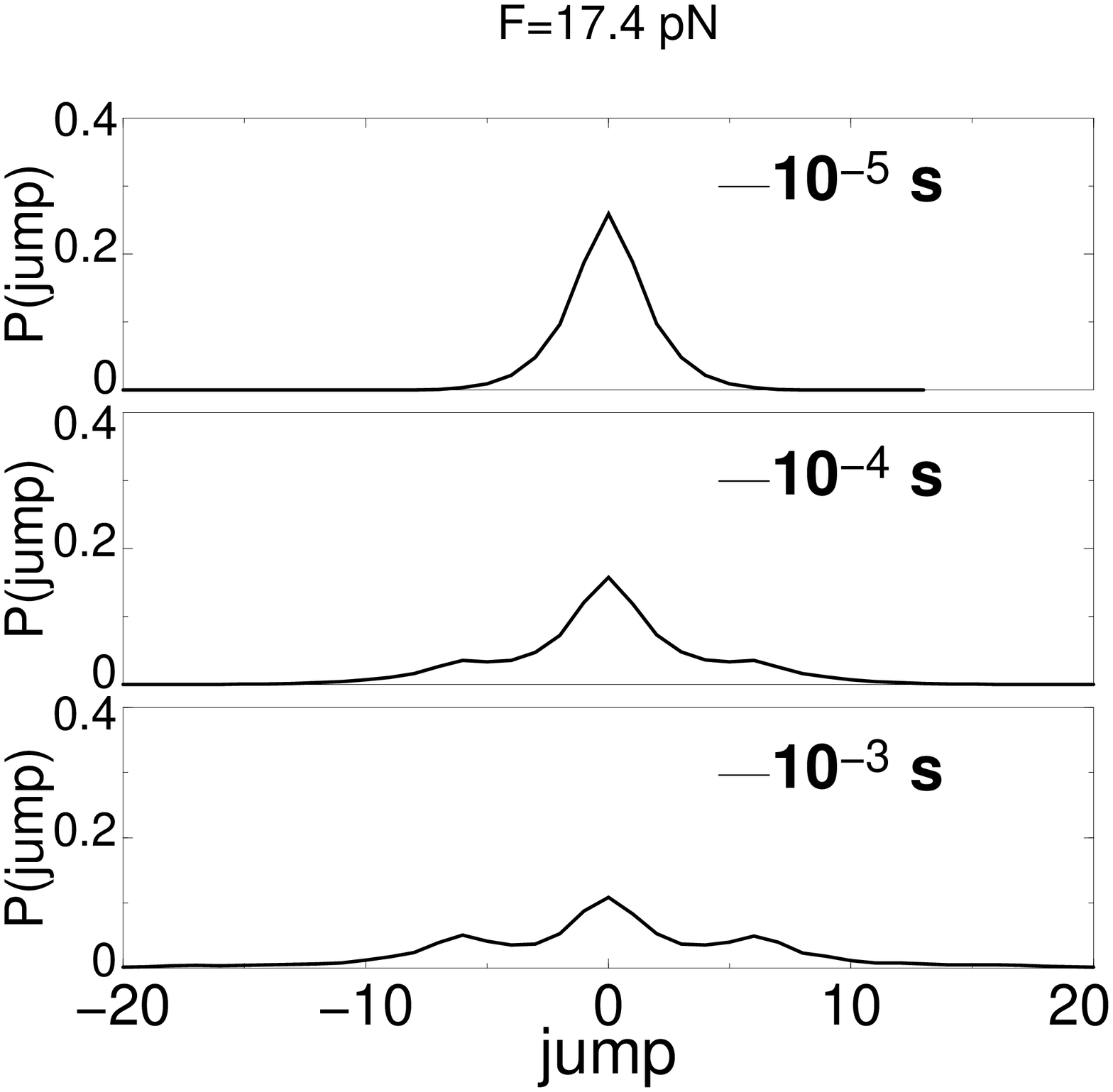,height=6cm,angle=-90}
\caption[]{ Probability distribution of a $j$-base jump for $\Delta t$
between $10^{-5}$s and $10^{-3 }$s and for forces $f= 16.4$ and 17.4 pN. Notice that the probability is
not necessarily a monotonously decreasing function of $|j|$, see extra humps
in the right column, due to sequence effects.}
\label{p_jump}
\end{center}
\end{figure}

\subsubsection{Extended Viterbi algorithm}

Given a sequence $B$ for the molecule the probability of a
time-trace $T$ (where the number of open bp is measured at times
multiple of $\Delta t$) is given by a product of $4^J\times
4^J$ transfer matrices
\begin{equation}
{\cal P}^{(J)}(T|B)= \prod_i M^{(J,i)}(b_i, \dots b_{i+J})
\end{equation}
with
\begin{equation}\label{tmatrix_mstep}
M^{(J,i)}(b_i, \dots, b_{i+J})= (\hat U^{(J,i)}_{i,i})^{k^{(0)}_i} 
\prod_{j=1}^{J}(\hat U^{(J,i)}_{i+j,i})^{k^{(j)}_i} (\hat
U^{(J,i+j)}_{i,i+j})^{k^{(-j)}_{i+j}}
\end{equation}
and $k_{i}^{(j)}$ is the number of transitions $i\to i+j$, with
$j=-J,-J+1,\ldots,J-1,J$ in $T$. Notice that $k_i^{(0)}$,
$k_i^{(1)}$ and $k_i^{(-1)}$ coincide with $t_i/\Delta t$, $u_i$ and $d_i$
respectively.

An extended Viterbi algorithm allows us to find the most probable
sequence. We now have to consider the probability of a sequence of $J$ 
contiguous base, starting from $i$, and write a recursion equation for
this probability,
\begin{equation} \label{recurj}
P^{(J)} _{i+1}(b_{i+1}, \dots ,b_{i+J-1}) = \max_{b_i} \big[
M^{(J,i)}(b_i, \dots , b_{i+J}) \times P^{(J)}_i(b_{i}, \dots ,
b_{i+J-1}) \big] \ ,
\end{equation}
which extends eqn (\ref{recur}) to $J\ge 2$. 
For the first base $i=1$ the optimization is simply
\begin{equation}
P^{(J)}_2(b_2, \dots , b_{J+1}) = \max_{b_1} M^{(J,2)}(b_1,b_2, \dots ,b_{J+1})
\end{equation}
The optimal choice for $b_1$ depends on the $J$ next base values, 
$b_1^*= b_1^{max}(b_2, \dots, b_{J+1})$. Then we find the next base,
$b_2^*$ as a function of $b_3,\ldots, b_{J+2}$ through (\ref{recurj}),
and so on, until the last base of the chain is reached. 
Its most probable value is selected and the whole optimal sequence is
recursively  reconstructed from the $b_i^{max}$ functions.

\subsubsection{Numerical study}

We first generate a set of numerical data by recording the
MC output (fork position) at discrete times multiple of a
sampling interval $\Delta t$; intermediate states are simply ignored 
as the instrument does not have the resolution to appreciate them.
Then we preprocess this partial time-trace
to obtain the transition number $k_{i}^{(j)}$, and make a prediction 
for the sequence using the above extended Viterbi algorithm. 

Figure \ref{om_jumps}A shows the quality of prediction
as a function of the delay $\Delta t$ at fixed range $J=2,3,4,6$
and for a single unzipping ($R=1$). 
Data shows that, for a given range $J$, there exists a threshold value for
$\Delta t$ above which the maximum displacement permitted becomes too
small to properly describe the unzipping dynamics.  The information
collected is no longer sufficient for a reliable prediction and the error
$\epsilon$ rapidly increases (see Fig \ref{om_jumps}A).  As
expected the threshold $\Delta t$ increases with the range, meaning
that larger ranges are better suited to deal with longer sampling
intervals.  When $\Delta t$ is small, comparable with the elementary
sojourn time on a base ($\tau \simeq 1\mu$s for a weak base), 
the performances are equivalent to the one of the $J=1$ case. 

The relationship between the range $J$
and the largest delay $\Delta t$ it can sustain is better seen on the
case of uniform sequences. The characteristic sojourn time on a base,
$\langle t\rangle $ (\ref{taui}), 
is then uniform throughout the sequence e.g. $\langle t\rangle \simeq
1~\mu$s for a repeated sequence of $W$ bases.  
Fig \ref{om_jumps}B shows that the prediction 
is perfect up to a temporal resolution $\Delta t \simeq 
J \times \langle t\rangle$, where $\langle t\rangle $ is the
characteristic sojourn time on a base pair, and $J$ is the range of
the algorithm. The existence of a threshold for the delay is clearer
at high $R$ than for $R=1$ (Fig  \ref{om_jumps}A) due to the presence
of larger fluctuations in the sojourn time in the latter case.

Figure \ref{om_run} (left) shows that the quality of the prediction betters
when the information from several opening experiments is collected.
As long as the typical jump associated to a delay $\Delta t$ is smaller
than the range $J$ (Fig \ref{p_jump}) the error $\epsilon$ can be 
reduced and values of order $10^{-2}$ are reached after 50 unzippings
for the $\lambda$-phage sequence at force $f=16.4$ pN .  Once the threshold
$\Delta t$ is crossed, however, the loss of information can not be
`repaired' and repetitions of the experiment appear to be useless.
The fork has moved too far away during the delay $\Delta t$ and a lot of
information falls out the window of size $J$ our algorithm is based
on, an effect which cannot be compensated with multiple experiments.
The effect is qualitatively similar for the weak/strong (AT/GC)
distinction shown in Figure \ref{om_run}, but is somewhat less
dramatic from a quantitative point of view. 

\begin{figure}
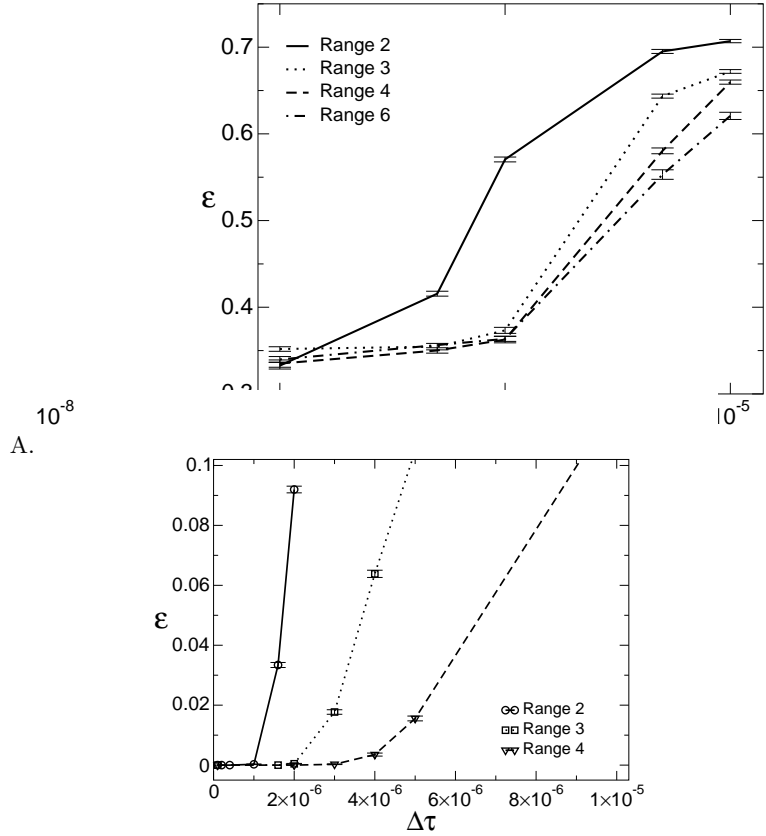

\begin{center}
A.\psfig{figure=fig23a.eps,height=6cm, angle=0}
\hskip 1cm
B.\psfig{figure=fig23b.eps,height=5cm,angle=0}
\caption[]{Error $\epsilon$ as a function of the delay $\Delta t$
between measures for various ranges (shown on Figure).
{\bf A.} Case of one unzipping ($R=1$) of a $\lambda$-phage DNA
molecule at $f=16.4$ pN.
{\bf B.} Case of $R=20$ unzippings of a uniform sequence of weak
bases at $f=11.8$ pN. 
Results are averaged over 50 samples in both panels. }
\label{om_jumps}
\end{center}
\end{figure}

\begin{figure}
\begin{center}
\psfig{figure=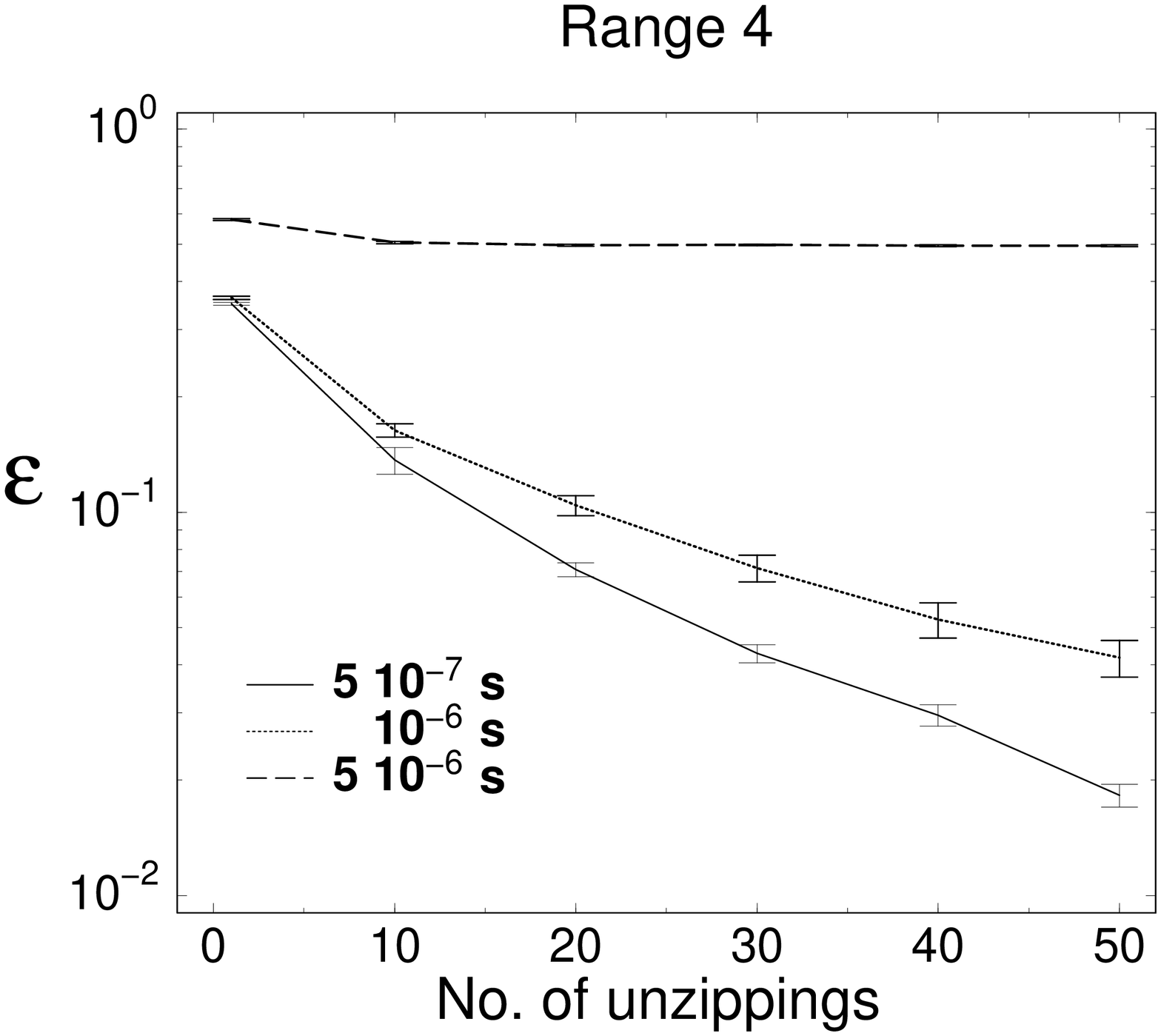,height=6cm, angle=-90 }
\psfig{figure=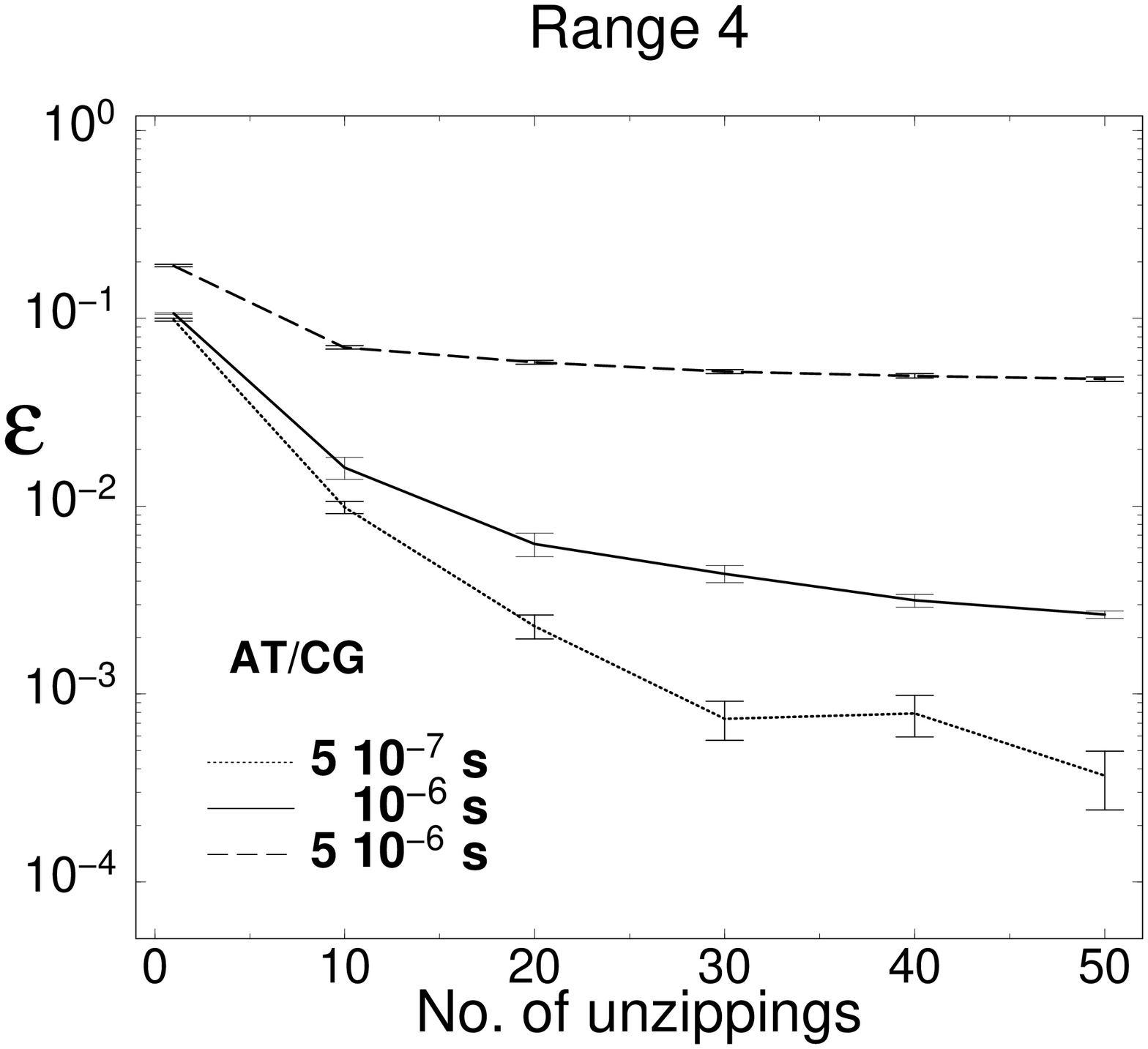,height=6cm, angle=-90 }
\caption[]{Left: Fraction of mispredicted bases $\epsilon$ as
a function of the number of unzippings for different temporal
resolutions $\Delta t$. 
The value of the range is $J=4$. Right: same as right but we only
discriminate among strong and weak bases.  Data refer to the opening
of a $\lambda$-phage sequence at f=16.4 pN and they are averaged over
50 samples.}
\label{om_run}
\end{center}
\end{figure}

\subsection{Fluctuations of the unzipped DNA strands}
\label{flucsec}

Real experiments give access to the extension $x$ of the open DNA
(ssDNA) strands, and not to the number $i$ of open bp (Fig.~\ref{fig1}). 
Due to the intrinsic elasticity of the strands $x$ fluctuates even at
fixed $i$, and these fluctuations grow with $i$. Indeed a strand is 
made of $i$ monomers, each acting as a spring with stiffness constant 
$K\simeq 170$~pN/nm at $f= 16$~pN and room temperature \cite{Coc3}.  The
distribution $A(x|i)$ of the extension $x$ for a given $i$ is roughly
Gaussian, with mean $i \, x_0$ where $x_0 = dg_{ss}/df\simeq .9$~nm is
twice the average extension of a ssDNA monomer, and standard deviation
$\sqrt i\, \delta x$ where $\delta x = \sqrt {2\,k_B\,T/K}\simeq
.2$~nm (Fig~\ref{fig-kernela}). Distribution $A$ could be precisely measured
through a combination of optical trap and single-molecule fluorescence
techniques \cite{Lan03}.

\begin{figure}
\epsfig{file=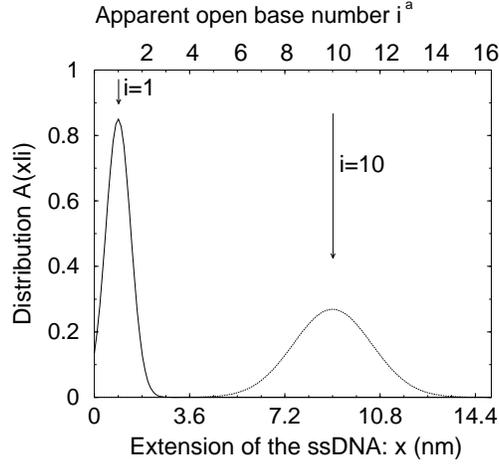,width=6cm,angle=-90}
\caption{Distribution $A(x|i)$ of the extension $x$ of the open ssDNA at fixed
position of the opening fork, $i=1$ and $i=10$. The r.m.s.  of the
distribution (at a force of 16 pN) increases as $\sqrt i$. 
The apparent value of the number of opened bases corresponding to a 
given $x$, $i^a$ (\ref{apparent}), 
is shown on the top  axis. }
\label{fig-kernela}
\end{figure}

\subsubsection{Effect of ssDNA fluctuations on the Bayesian inference}

We hereafter study the effects of these fluctuations
on the inference problem in the absence of stacking interactions and
at high force. We start by making more precise the notion of the
time spent on a base:
\begin{itemize}
\item {\em the real time $t^r_i$}: this is the time really
spent by the fork on bp $i$, simply denoted by $t_i$ so far. This number 
is stochastic since the fork undergoes a random walk motion, 
with a distribution depending
on the nature of base $i$ (\ref{pnostack1}). The absence of stacking
ensures that real times attached to distinct bases are uncorrelated;
the probability of the set of real times $T^r=\{t_i^t\}$ given a
sequence $B$ is, up to a sequence-independent multiplicative factor,
\begin{equation} \label{pdetbis}
{\cal P}(T^r|B) \propto \prod _{i}
\exp \big[ g_0(b_i)  -  r \,  
 e^{g_0(b_i)}\; t_i ^r \big]
\end{equation}
which corresponds to (\ref{pdetb},\ref{defmatrixmbayes}) in the
limiting case of high force and no stacking. Given a set of real times
the  best sequence $B^*(T^r)$ is the one 
maximizing ${\cal P} (T^r |B)$.  The probability of predicting
sequence $B$ is, given the true sequence $B^L$, 
\begin{equation} \label{qb1}
{\cal Q}^r(B) = \int dT^r \;  {\cal P}(T^r|B^L) \; \prod _{B' (\ne B)}
\theta \big(  {\cal P}(T^r|B) -  {\cal P}(T^r|B') \big) \, 
\end{equation}
where $\theta$ is the Heaviside function, $\theta(x)=1$ if $x>0$,
0 otherwise. In practice, however, one has no access to the real times. 

\item {\em the apparent time $t_i^a$}: Given a measure for the 
extension $x$ of the ssDNA we define the apparent position of the fork
through
\begin{equation} \label{apparent}
i^a = \hbox{\rm Closest integer to} \ \frac{x}{x_0} \ .
\end{equation}
The value of $i^a$ is stochastic, with a probability $A$ depending
on the real position of the fork, $i^r$. 
Considering Rouse dynamics for the monomers \cite{sebas} the longest relaxation
time of a strand is, denoting the viscosity of the solvent by $\zeta$, $t_r (n)
\sim \zeta/(K\pi^2)\times (2n)^2\sim 100 \, n^2$~ps. For
molecules with $< 100$~bp ssDNA reaches equilibrium faster than the fork
moves. The probability to observe
$i^a\ge 1$ at some instant thus depends only on the true position $i^r$ 
of the fork at the same time, and reads, when $i^r\ge 1$, 
\begin{equation} \label{defflua}
A_{ i^a , i^r } = \int _{i^a -\frac 12} ^{i^a +\frac 12}
\frac{d\nu}{\sqrt{2\pi \, i^r \, \sigma^2 }}\; \exp \left[ 
- \frac{(\nu - i^r)^2}{2 \, i^r\, \sigma^2} \right]
\end{equation}
with $\sigma ^2 = \delta x /x_0$; 
the expression for $i^a=0$ is obtained from (\ref{defflua})
upon replacement of the lower integration limit with $-\infty$. When
the molecule is entirely closed ($i^r=0$) all values of $i^a$ have zero 
probability except $i^a=0$ $(A(0|0)=1$); this choice amounts to
neglect the fluctuations in the extension of the DNA linkers.   

We call $t_i^a$ the time apparently spent by the fork on bp $i$, that
is, the number of measures in a time-trace in which the fork appears
to be at location $i$ according to (\ref{apparent}), divided by the
delay $\Delta t$ between two measures. Matrix $A$ (\ref{defflua})
implicitly define the probability distribution of a set of apparent
times $T^a=\{t_i^a\}$ given a set $T^r$ of real times, see Appendix
\ref{app3} for more details. Multiplicating by (\ref{pdetbis}) and
integrating over the real times formally defines the probability
${\cal P} ^a (T^a|B)$ of a set $T^a$ of apparent times given a sequence
$B$.
Given an apparent signal $T^a$ the best sequence $B^*(T^a)$ is the one 
maximizing ${\cal P}^a (T^a |B)$.  The probability of predicting
sequence $B$ is, given the true sequence $B^L$, 
\begin{equation} \label{qb2}
{\cal Q}^a(B) = \int dT^a \;  {\cal P}^a(T^a|B^L) \; \prod _{B' (\ne B)}
\theta \big(  {\cal P}^a(T^a|B) -  {\cal P}^a(T^a|B') \big)\ .
\end{equation}
\end{itemize}

Consider first the ideal case where the delay $\Delta t$ between 
successive measures is vanishingly small. In this limit, given the
set of real times, the apparent times $t^a_i$ are not stochastic 
but simply obtained through the convolution of the $t_i^r$'s with 
matrix $A$ (\ref{defflua}): $T^a=A\cdot T^r$ in vectorial notation. 
Starting from the probability (\ref{qb2}) of predicting a sequence from the 
apparent times and performing the change of variable $T^r=A^{-1}\cdot T^a$
we obtain ${\cal Q}^a(B)={\cal Q}^r(B)$ (\ref{qb1}). 
The probability, within Bayes framework,
of predicting the true sequence $B^L$ is the same as in the absence of 
fluctuations. In particular the values for $R_c$ calculated in the
previous Section are unaffected by the presence of ssDNA elasticity.

This result does not hold for finite delays $\Delta t$ where, given a
set $T^r$ of real times, the apparent times $t^a_i$ are stochastic due
to the finite number of samplings during the sojourn time 
on each base. Let us assume that the delay $\Delta t$ between 
successive measures is small with respect to the sojourn time 
$\langle t\rangle$ on a base pair but non zero. 
The Bayesian probability ${\cal Q}^a(B)$ of a sequence  now 
depends on the fluctuation matrix $A$. 
For the sake of simplicity we consider only
the case of a large number of unzippings, and a repeated sequence of
bases $S$ with a unique $W$ base at location $i$. Let
\begin{equation} \label{defrho}
\rho = \frac{ \Delta t} {\langle t\rangle ^S} = r \, e^{g_o(S)}\, \Delta t
\end{equation}
denote the ratio of the delay over the average time spent on a $S$ base;
by hypothesis $\rho \ll 1$. The probability
that the $W$ base is not correctly predicted reads (Appendix \ref{app3}), 
\begin{equation}\label{ludeau2}
\epsilon _{R,i} = {\cal Q}^a(B^S) \sim e^{-R/R_c(i)} \quad \hbox{\rm where}
\quad R_c(i) \simeq \frac{8}{\Delta^2\; \big( A^T\, \beta^{-1}\, A\big)_{i,i}  
}\ , \quad
\beta_{j,k} = (1-\rho)\,  \big( A\, A^T)_{j,k} + \rho \, Id_{j,k}
\ .
\end{equation} 
and $A^T$ denotes the transposed matrix of $A$. 
The above formula holds for a small difference $\Delta$ of free energies 
between the weak and strong bases, see (\ref{rcnostack}). The outcome
for $R_c(i)$ is shown in Fig \ref{ssdnafig2}A for $\rho=0.1$ and grows 
as the square root of $i$ \cite{corto}. More precisely we find $R_c(i)
\propto \sigma \; \sqrt{i}$ where $\sigma=\sqrt{\delta x/x_0}$, and
the proportionality factor depends
on $\rho$. Perfect prediction is still
possible, but at the price of a number of unzippings growing with the base 
index.

\begin{figure}
\begin{center}
A.\psfig{figure=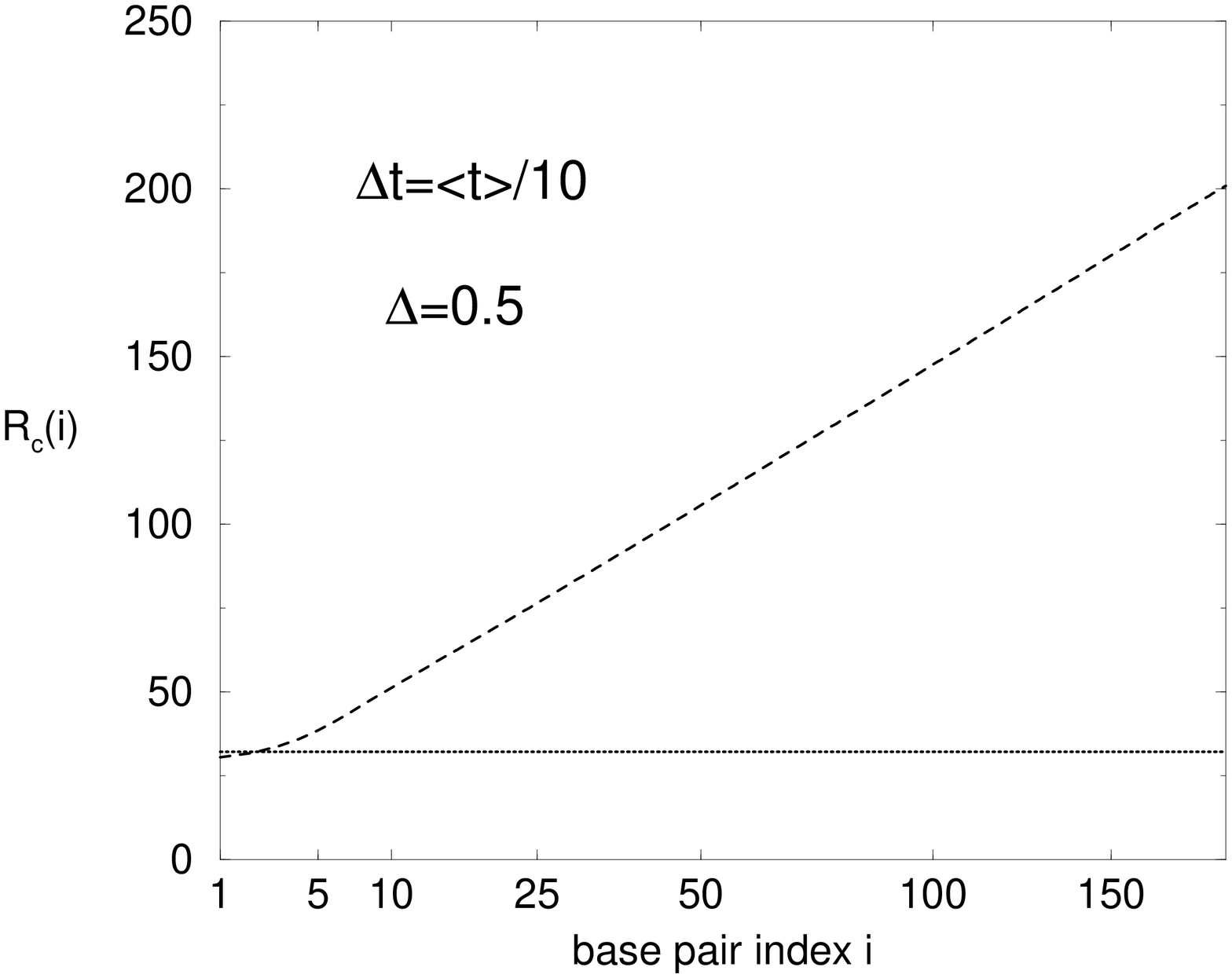,height=7cm, angle=-90}
\hskip 1cm
B.\psfig{figure=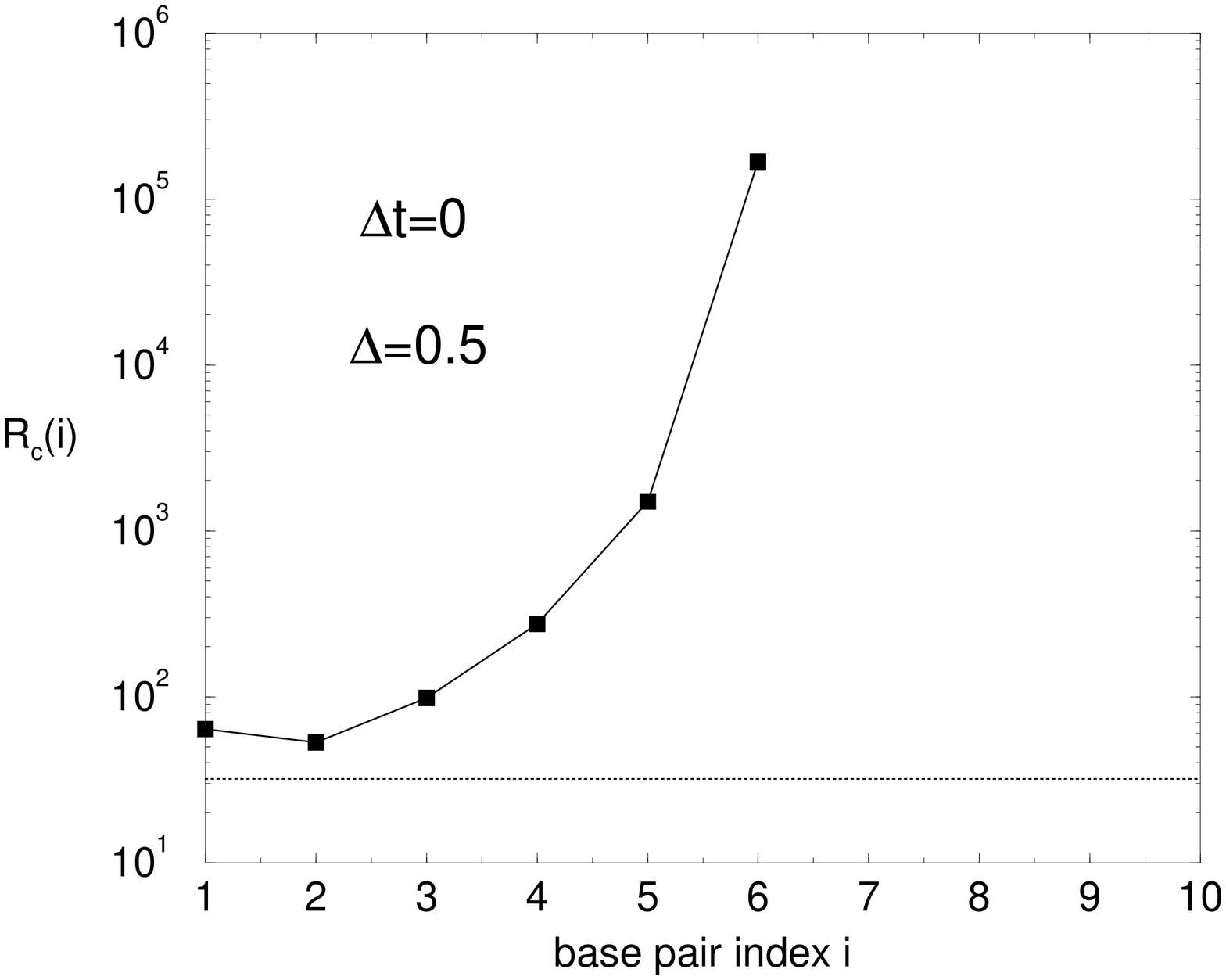,height=7cm, angle=-90}
\caption{Value of the number of unzippings controlling the decay of the error
in predicting a base, $R_c(i)$, as a function of the base index $i$.
The sequence is made of bases $S$ with a single $W$ base at position $i$.
The dotted line shows the value of $R_c$ in
the absence of ssDNA fluctuation, for a difference of free energy 
between $S$ and $W$ bases equal to $\Delta =0.5$.
{\bf A. Case $\Delta t =\langle t\rangle/10$}. The decay constant $R_c(i)$ 
for the Bayesian error (\ref{ludeau2}) grows as $\sqrt i$ (dashed line).
{\bf B. Case $\Delta t\to 0$}. The full line shows $R_c(i)$
for the Viterbi procedure without deconvolution;
for $i\ge 7$ $R_c(i)$ is infinite, meaning that the $W$ base is almost surely
predicted to be of $S$ type. With appropriate deconvolution the dotted line
value for $R_c$ is recovered. }
\label{ssdnafig2}
\end{center}
\end{figure}

\subsubsection{Sequence prediction through deconvolution}

The above results do not tell us how to make a prediction
for the sequence given an apparent signal $T^a$. The expression for
${\cal P}^a$ is highly non local: the probability of the time $t_i^a$
does not depend on the type $b_i$ of base at location $i$ but also on its 
neighbors. A practical procedure consists in calculating, once 
the apparent times $T^a$ are measured, the set of deconvoluted times 
$T^d=\{t_i^d\}$ through the formula
\begin{equation} \label{deconvol}
t_i^ {d} = \sum _j D_{i,j}\; t_j^a
\end{equation}
where $D$ is an appropriate deconvolution kernel to be specified
later. Ideally, after deconvolution, the probability of $T^d$ given
the sequence $B$  should coincide with the local probability (\ref{pdetbis}).
The prediction for the sequence is then done through the maximization of ${\cal
P}$ (\ref{pdetbis}) over $B$, given the set $T^d$ of deconvoluted times.

We start by showing how the performances of the inference procedure 
are dramatically worsened by fluctuations if no deconvolution 
is performed ($D=Id$), and then show how the effects of
fluctuations are cured when deconvolution is performed.
We focus here on the cases $R=1$ and $R\gg 1$
only, and concentrate on the case $\Delta t\to 0$ first. 
Consider the base at location $i$, which we suppose to be, say, of
type $W$. The error in predicting this base
reads, see Appendix \ref{app3}, 
\begin{equation} \label{error1}
\epsilon ^W _{1,i} = \sum _i \prod _{j (\ne k)} \left( 1-
\frac{C_{i,j}}{C_{i,k}} \right) ^{-1} \ e^{ - \tau^W/
C_{i,k}} \ ,
\end{equation}
where 
\begin{equation} \label{defcni}
C_{i,j} = \exp ( g_o(b_i) - g_o(b_j)) \; (D\, A) _{i,j} \ ,
\end{equation}
and $\tau^W,\tau^S$ are defined in (\ref{fr}). The subscript 1 refers
to the value $R=1$ of the number of unzippings. 
Figure \ref{ssdnafig} shows $1-\epsilon ^W _{1,i}$ as a function of
$\sqrt i$ for a repeated sequence $SSSS\ldots$, and for an alternate
sequence $SWSW\ldots$ in the absence of deconvolution ($D=Id$).
The error increases from a value for $i=1$ essentially equal to
its counterpart $\epsilon _1^W$ (\ref{pnostack3}) in the absence of strand
fluctuation, to reach unity at large $i$.
This behavior is easily interpreted: in the absence of deconvolution
the apparent time $t_i^a$ (more precisely, the reduced time
$\tau _i^a$ (\ref{deftaui})) on base $i$ is the sum of the real times $t_j^r$ 
spent on each base $j$, weighted with the probability $C_{i,j}$ 
(\ref{defcni}). As $i$ grows more and more bases $j$ contribute to the
sum with smaller and smaller weights, with a number of contributing terms
scaling as $\sqrt i$. The law of large numbers tells us that the 
distribution of $\tau _i^a$ is asymptotically concentrated around a single
value, equal to $\tau_\infty ^a = e^\Delta$ and to $\tau_\infty ^a = 
\frac 12 (1+e^\Delta)$ for the $SSSWSSS\ldots$ (where the unique $W$
base is located at position $i$) and  $SWSW\ldots$
sequences respectively. As these values exceed $\tau^W$ (\ref{fr}) the
base is almost never correctly predicted\footnote{The same argument 
indicate that
the probability to mispredict base $b_i=W$ base among a repeated
$WWWW\ldots$ sequence vanishes when $i$ tends to infinity. The reason
is that the apparent time on base $i$ converges to the average time on
the neighbors which are all of the right type $W$.}. The very tiny
probability of success is due to the tail of the times below $\tau^W$,
which decreases exponentially with $\sqrt i$ (Fig  \ref{ssdnafig}).

In the limit of a large number $R$ of unzippings the error decreases
as (Appendix \ref{app3})
\begin{equation}\label{ludeau}
\epsilon _{R,i}\sim e^{-R/R_c(i)} \quad \hbox{\rm where}
\quad R_c(i) = \left\{ \begin{array}{c c c}
\frac{2\, \sum _j C_{i,j}^2}{\left(1+\frac \Delta2 -
\sum _j C_{i,j} \right)^2} & \hbox{\rm if} & \sum _j C_{i,j} < 1+\frac
\Delta2\\
+\infty  & \hbox{\rm if} & \sum _j C_{i,j}  \ge 1+\frac \Delta2
\end{array} \right .
 \ .
\end{equation} 
The above expression was derived when the free energy difference
$\Delta$ between $W$ and $S$ bases is small, the hardest case from
the inference point of view. In the absence of fluctuation $A=D=Id$
we find back result (\ref{fr2}) as expected. Notice $R_c=\infty$
simply means that the error does not converge to zero when $R$
increases. An illustration of this situation is given in 
Fig \ref{ssdnafig2}A. The number $R_c(i)$ of unzippings necessary to 
correctly predict a unique $W$ base located at position $i$ inside a
repeated $SSSS\ldots$ sequence increases with $i$, and diverges 
for $i\ge 7$ in the absence of deconvolution. The reason for this
failure is the same as in the above $R=1$ case: the apparent time on
base $i$ is corrupted by too many $S$ bases and the true nature of
the base cannot be recognized.

Fortunately the situation drastically improves when the signal is 
deconvoluted with the kernel
\begin{equation}
D= A^\dagger \label{kera}
\end{equation}
equal to the pseudo-inverse of matrix $A$. We have not encountered any
numerical problem to calculate this pseudo-inverse from the inverse of
$A^TA$ for sequences with a few hundred bases. The matrix $C$ in
(\ref{defcni}) then reduces to the identity matrix, and the errors for a single
(\ref{error1}) and a large number (\ref{ludeau}) of unzippings decrease to
their respective values in the absence of fluctuations. In
particular the number of unzippings necessary to correctly predict a
base is simply $R_c\sim 8/\Delta^2$, independently of $i$. As a
conclusion, through an adequate and sequence-independent 
deconvolution procedure, we have been able to completely remove the effect of
ssDNA fluctuations.

\begin{figure}
\begin{center}
\psfig{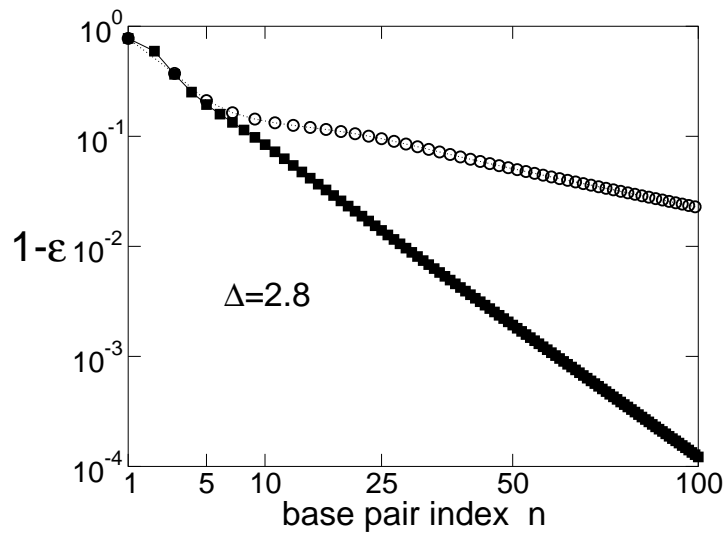}
\caption{Probability that a base is correctly predicted, $1-\epsilon_i^W$,
as a function of its location $i$ in the case of: a repeated sequence of 
$S$ bases with a single $W$ base at position $i$ (black dots), an alternate
sequence $SWSW\ldots$ (empty dots). In both cases the rate of success decreases
exponentially with the square root of $i$. The difference of free energies 
between $S$ and $W$ bases is $\Delta =2.8$.
}
\label{ssdnafig}
\end{center}
\end{figure}

In the case of a finite delay $\Delta t$ we expect that an appropriate
deconvolution with the kernel (\ref{kera}) is sufficient to correctly 
infer the sequence with the extended Viterbi algorithm of Section
\ref{mstepsec} \cite{altro}.

\section{Summary and Conclusion}

In this paper we have studied the inference of a DNA sequence from 
Monte-Carlo generated unzipping signals. Inference is made uneasy by
the fact that unzipping signals are  largely affected by thermal 
noise, due to the fact that the free energy to open a base pair (the 
loss in binding free energy plus the work to stretch the unpaired DNA strands)
are of the order of k$_B$T. The main goal of the present work was
precisely to reach a theoretical understanding of how to cope with 
thermal noise in the inference process. 

The present study is in part numerical and in part analytical.  From
the numerical side we have first generated, from a given sequence,
unzipping data by a Monte Carlo algorithm based on a previously
introduced dynamical model of the unzipping \cite{Coc4}.  We have then
implemented algorithms to reconstruct the most probable sequence
from the unzipping signal.  The prediction error on each base can be
simply evaluated through the comparison between the true and the
predicted sequences.  From a theoretical side we have calculated the
error (probability of misprediction) with the aim to understand its
dependence on the sequence, the intrinsic parameters {\em i.e.} the
biochemical base pair free energies, and the extrinsic parameters {\em
i.e.} the unzipping force, the number of repetitions of the unzipping,
the collection of unzippings from both sides of the molecule, ....
Numerical results compare very well with analytical calculations. Our main
analytical finding is that the average prediction error on a base $i$
decreases exponentially with the number $R$ of unzippings. The decay
constant $R_c(i)$ gives the number of unzippings required to achieve an excellent
prediction of the base. We have analytically calculated the value of
$R_c$ in the following cases: (high force) repeated sequences without (\ref{nostackomega},\ref{fr2}) and with (\ref{rcstack}) stacking interactions, heterogeneous sequences (\ref{rcprop1}); (moderate force) with (\ref{epsmedio},\ref{rcforce}) and without stacking interactions (\ref{rcprope1}), for two-way unzippings (\ref{rcavin}), and taking into account the fluctuations of the extension of the unzipped strands (\ref{ludeau2},\ref{ludeau}).
 
We have first considered the ideal case in
which it is possible to follow directly the dynamics of the opening
fork with a perfect temporal resolution; in this limit all base pair
opening and closing events are detected.  The only source for stochasticity
is the thermal motion of the fork. In the absence of stacking
interaction the decay constant
$R_c(f,i)$ for the base $i$ and at a force $f$ can be obtained, in
this case, as the ratio of the decay constant at large force,
$R_c(f=\infty,i)$, over the average number of openings of the base
during a single unzipping, $\langle u_i\rangle$.  
The average number of openings of a base,
$\langle u_i\rangle$, depends on the free energy landscape of the
molecule, determined by the force and the sequence content, and was
computed in Appendix \ref{sec93}. In the presence of
stacking interactions $R_c(f,i)$ depends on the whole sequence and
was calculated through an asymptotic version of the Viterbi
algorithm (Section \ref{sechs}). Base pairs exhibit a
lock-in phenomenon : there exist blocks of neighbouring bases with the
same decay constant $R_c(f,i)$, while bases in different blocks have
much weaker correlations. We also show that much better predictions
on the value of a base can be obtained from the same amount of
collected data if the  molecule is unzipped from both extremities rather than
from one extremity (as done so far).

The assumption of infinite temporal bandwidth and precise knowledge of
the fork position dynamics allows us to start
from the simplest case for the sequence prediction analysis.
The advantage is that Bayesian inference can be done exactly with a fast
procedure, the so-called  Viterbi algorithm. The most likely sequence,
given a measured unzipping signal, is found in a time scaling linearly
with the number of the bases. 
The existence of a fast, exact algorithm allowed us to check 
analytical results; the latter are indeed always obtained for the 
optimal sequence, irrespectively of the existence of a practical
algorithm capable of finding this sequence.

In the second part of the paper we have made a step forward toward the
analysis of real experimental data and have included in the inference 
analysis two major sources of instrumental limitations: the finite
data acquisition bandwidth, and the elastic fluctuations of the
unzipped DNA strands. 

The finite resolution in time is such that
during the time interval between two data acquisitions the opening fork can 
move by (much) more than one base.  The exact Viterbi algorithm has 
been generalized to the case of a large but finite 
bandwidth, by considering all the forward and backward transitions of the
opening fork which can take place, within a range $J$, during the time
 interval $\Delta t$ between two measures. This new algorithm 
is able to reconstruct the sequence when the range $J$ is of the order
of the ratio between  $\Delta t$ and the typical sojourn time 
$\langle t\rangle$
on a base pair. Though our extended Viterbi algorithms still runs in
a time growing linearly with the number of bases, it is exponential
in the range $J$, and is limited in practice to $J \le 10$.
This algorithm is thus implementable for
 $\Delta t \sim 10\; \langle t\rangle$,
{\em i.e.} up to about 10 $\mu$s. In other word the bandwidth frequency
should be larger than 100 KHz, a larger value than the current value
for the bandwidth in  real experiments of the order of 1-10 KHz. 
Other algorithms presumably  not guaranteed to reach the most likely
sequence, but with a running time polynomial in the range $J$,
should be implemented. 

In addition  we have considered the effects of the
fluctuations in the extension of the DNA strands. Indeed, even if the 
distance between the extremities of the unzipped strands is typically 
known within $< 1$ nm accuracy \cite{Boc02,Bus03}, thermal 
fluctuations in the strand
length (and possibly in the linkers) are responsible for a larger
uncertainty over the position of the opening fork.  We have, in
particular, extended our theoretical formalism to calculate the decay
constant of the error with the number of unzippings $R_c$ at high
force, without stacking, in presence of DNA strand fluctuations and with an
interval $\Delta t$ between two measures finite but small with respect
to the sojourn time $\langle t\rangle$.  We have obtained that the
decay constant $R_c$ for the error on base $i$ is multiplied 
by $\sqrt i$ with respect to its counterpart in the absence of DNA
fluctuations. The further from the beginning of the
sequence a base is, the larger is the number of unzipping to
reach a good prediction.

The theoretical formalism for  $\Delta t \to 0$  suggests a way
to  preprocess the signal by deconvoluting it with the pseudo-inverse
of the (sequence--independent) DNA fluctuation  matrix
(\ref{defflua}). This signal can then be
processed with the usual Viterbi algorithm, and the quality of the prediction
is the same as in the absence of strand fluctuations.
A natural question is whether the same deconvolution procedure could
be applied to the realistic case of a finite bandwidth or not.
We are currently working on this problem, and are developing a formalism 
for the calculation of $R_c$ in the presence of DNA strand
fluctuations  and for experimental value of $\Delta t \sim 0.1$~ms
\cite{altro}. The design of efficient inference algorithms in this
realistic case is a challenging issue. 

An implicit but not well justified assumption we have so far is to
have a perfect knowledge of the pairing free energies and dynamics of 
unzipping {\em i.e.} of the conditional probability  $P(T|B)$. 
In practice, however, modeling cannot be perfect and any functional 
form for $P(T|B)$ will be only approximate for a given
experimental setup. Numerical investigations show, not surprisingly, 
that the quality of prediction 
deteriorate when the rates used by the Viterbi procedure differ too
much from their values in the data generating Monte Carlo procedure. 
A possible way out should be based on a learning principle: in
a first stage unzipping data corresponding to a known sequence 
($\lambda$-phage) are collected to caliber the rates, in a second stage
predictions are made for new sequences.
Last of all we have here considered unzipping at constant force.
Investigation of the constant velocity case \cite{Boc02} would be very
interesting. Local minima are well predicted and remarkably the force 
signal may be affected by the substitution of one base pair \cite{Boc02}.

Let us finally mention a related albeit more complex problem, the 
analysis of RNA unzipping data. The non complementarity of single 
strands in RNA molecules  give rise to complex folded secondary
structures with multiple helices. Gerland and collaborators
have suggested a way to reconstruct RNA secondary structure by combining
the recording of the force-extension curve and the passage through a
nanopore~\cite{Ger04}. 
The passage through the nanopore would indeed force to the
helices to open one after the other with a sequence-specific order. In 
this respect, thanks to the nanopore geometry, the RNA unzipping
problem is reduced to a unidimensional  problem for which the
inference methods presented here could be of interest.

\vskip .5cm
{\bf Acknowledgments.} 
We thank U. Bockelmann for repeated and useful discussions, and
F. Zamponi for a critical reading of the manuscript. 
We are grateful to H. Isambert for his suggestion of two-way
unzipping at the origin of Section \ref{twowaysec}.
This work has been partially sponsored by the European EVERGROW
(IST-001935) and STIPCO (HPRN-CT-2002-00319) programs, and the
French  ACI-DRAB \& PPF Biophysique-ENS actions.

\appendix
\section{Implementation of the Extended Viterbi Algorithm}
\label{app_vit}

A time trace $T$ of the unzipping signal, produced by the Monte Carlo
procedure, is first encoded in a vector $\mathcal{K}=
\{k_{i}^{(-J)},k_{i}^{(-J+1)} \dots k_i^{(0)}, k_{i}^{(+1)}, \dots
k_{i}^{(J)} \} $ where $k_i^{(j)}$ is the number of transitions $ i
\rightarrow i+j$.  $J$ fixes a cutoff on the displacement taken into
account: only jumps by $|j| <J $ bases are considered.  The
information on the opening dynamics $i.e.$ the vector $\mathcal{K}$,
the applied force $f$ and the temporal resolution $\Delta t$ is used
to construct the transfer matrix $M^{(J,i)}$ (\ref{tmatrix_mstep}) for
the $i^{th}$ base.


The matrix exponentiation, needed to compute $\hat U$
(\ref{matrix_exp}), is carried out by solving the set of $2J+1$
coupled differential equations
\begin{equation}\label{diff_eq}
\frac{dy _j }{dt}= \sum _{j'} \hat H^{(J,i)} _{j,j'} \; y _{j'}
\end{equation}
where $j=-J,\ldots, J$, and $ \hat{H^{(J,i)}}$ is defined in
(\ref{reduced_H}). The initial conditions read
\begin{eqnarray}\label{in_cond}
y^0_i&=&1 \nonumber \\ y^0_j&=&0 \quad j\neq i \ .
\end{eqnarray} 
The value of $y_j$ at time $\Delta t$ is the matrix element 
$\hat U^{(J,i)}_{i+j,i}$ of the truncated evolution operator.
The operation is repeated for the various values of the starting base
index $i$ to obtain the whole operator.
From a numerical point of view we solve (\ref{diff_eq}) 
using a classical $4^{th}$ order 
Runge-Kutta method for integration of ordinary differential equations.  

Once the matrix
$\hat U^{(J,i)}$ is computed, the transfer matrix $\hat M^{(J,i)}$
can be easily evaluated knowing the unzipping dynamics $i.e.$ the
vector $\mathcal{K}$.  The probability of a sequence $B$
given the unzipping signal $T$ is then maximized via a transfer-matrix-like
algorithm.  To avoid errors due to small numbers we apply the
recursive procedure (\ref{recurj}) to the logarithm of the probability
instead of the probability itself.  The general
optimization step is therefore
\begin{displaymath}
\ln P^{(J)}_{i+1} (b_{i+1},b_{i+2}, \dots ,b_{i+J-1}) = 
\max_{b_i}\big[ \ln P^{(J)}_{i} (b_{i},
\dots ,b_{i+J-1})+ \ln M ^{(J,i)} (b_{i}, \dots , b_{i+J}) \big]
\end{displaymath}
At each step, the type of the $i ^{th}$ base that maximizes $\ln P^{(J,i)}_i$,
$b^*_i$,  is stored for each
of the $4^J$ possible choices of following $J$ bases
$(b_{i+1},b_{i+2}, \dots ,b_{i+J})$.  $4^J$ possible sequences are
thus constructed and kept in memory.  When the algorithm reaches the
end of the sequence the maximization over the last base type selects 
the best sequence and all previous
bases can be simply reconstructed from the $b_i^*$, going backwards
from the last base to the first one.

Some problems in memory allocation and state labeling must be faced.
The dimension of each vector $b_i^*(b_{i+1},b_{i+2}, \dots ,b_{i+J})$ 
grows as $4^J$ and there are $N$
(up to 48,502 for a $\lambda$-phage DNA) different vectors.
When the range $J$ is large, the memory space required to store this
information becomes huge.  To circumvent this problem we have reduced the
space complexity of the algorithm by increasing its time complexity.
To do so we apply the algorithm more times, memorizing, and
reconstructing, only a portion of length $D$ of the sequence during each
execution.  During the first execution only the last $D$ bases
of the sequence are reconstructed.
In the second execution the algorithm stops at base $N-D$, where $N$
is the total number of open base pairs, and another set of $D$ bases
are predicted. This procedure goes on until the first base of the
molecule  is reached. 
$D$ is of course an adjustable parameter and the number of times the
algorithm is repeated is chosen consequently.

Our code is written in a range independent way.  The user 
simply sets $J$ at the beginning of the program, without changing
anything else.
The $4^{(2J +1)}$ choices of the variables $( b_{i-J }, \dots
b_{i+1},b_{i+2}, \dots ,b_{i+J})$ that define a specific
reconstruction `state' are represented by a bit string whose length
depends on the fixed range $J$.
The string is assigned in the following way: 2 bits identify the type
selected for a base, the lower bits referring to  the base
with the lower index along the chain, see 
Table \ref{variable_label}.  The binary number $s$ encoding a string
of $2J+1$ bases is called its label. 

\begin{table}
\label{variable_label}
\begin{center}
\begin{tabular}{|c|c|c|c|c||c||c|}
\hline \rule{0pt}{20pt} \bf{$i+J$} & \bf{$i+J-1$} & &\bf{$i-J+1$} &
\bf{$i-J$} & \bf{s} &\bf{ Sequence}\\[2pt] \hline \hline
\rule{0pt}{20pt} 00 & 00 & \dots & 00 & 00 & \bf{0} & AA \dots
AA\\[2pt] \hline \rule{0pt}{20pt} 00 & 00 & \dots & 00 & 01 & \bf{1} &
AA \dots AT\\[2pt] \hline \rule{0pt}{20pt} 00 & 00 & \dots & 00 & 10 &
\bf{2} & AA \dots AC\\[2pt] \hline \rule{0pt}{20pt} 00 & 00 & \dots &
00 & 11 & \bf{3} & AA \dots AG\\[2pt] \hline \rule{0pt}{20pt} 00 & 00
& \dots & 01 & 00 & \bf{4} & AA \dots TA\\[2pt] \hline
\rule{0pt}{20pt} 00 & 00 & \dots & 01 & 01 & \bf{5} & AA \dots
TT\\[2pt] \hline \rule{0pt}{20pt} 00 & 00 & \dots & 01 & 10 & \bf{6} &
AA \dots TC\\[2pt] \hline \rule{0pt}{20pt} \dots & \dots& \dots &
\dots& \dots & \vdots& \dots \\[2pt]
\end{tabular}
\caption[]{Table of variable labeling for a set of $(2J+1)$
bases. Each sequence is identified by a label $s$ in its binary
writing: 2 bits identify the type assigned to each base, the lower bit
being corresponding to the base with the lower index along the
sequence. }
\end{center}
\end{table}

The largest range we could test is $J\sim 10$.  Like the memory cost, the
execution time of the program scales linearly with $N$ but
exponentially with the range $J$.  The time needed to perform a single
unzipping (without considering the statistics over samples) 
increases as $n_{RK}
\times 4^J \times (2J+1)^3$, where $n_{RK}$ is the number of
integration steps in the Runge-Kutta subroutine.

\section{Convolution products for $R$ unzippings}

\subsection{Distribution of the sojourn time}
\label{appa}

The distribution $P_R$ of the total sojourn time $\tau$ (\ref{totaltau}) 
spent on a base for $R$ unzippings is defined as
\begin{equation}
P_R(\tau) = \int _0 ^\infty d\tau ^{(1)} P_1 (\tau ^{(1)}) 
\int _0 ^\infty d\tau ^{(2)} P_1 (\tau ^{(2)}) \ldots
\int _0 ^\infty d\tau ^{(R)} P_1 (\tau ^{(R)}) \; \delta \bigg( \tau -
\big( \tau _i^{(1)} + \tau _i^{(2)} + \ldots + 
\tau _i^{(R)} \big) \bigg)
\end{equation}
where $P_1$ is defined in eqn (\ref{dist1}). Taking the Laplace
transform, we obtain
\begin{equation}
\int _0 ^\infty d\tau P_R(\tau) \, e^{-s\, \tau}
= \left( \int _0 ^\infty d\tau P_1(\tau) \, e^{-s\, \tau}
\right) ^R = \left( 1 + s\right)^{-R} \ .
\end{equation}
It is a simple check that this expression coincides with 
the Laplace transform of the right hand side of eqn (\ref{dist2}), hence
proving identity (\ref{dist2}) for $P_R$. 

\subsection{Distribution of the number of fictitious unzippings}
\label{app1}
 
To calculate the $R^{th}$ power (for the convolution product) of $\rho _1$
(\ref{defrho1}) we introduce the generating function
\begin{equation} 
\label{gx}
g(x)=\sum_{u=1}^{\infty} \rho _R (u)\;x^u= 
\left( \sum_{u=1}^{\infty} \rho _1 (u)\;x^u\right) ^R =
 \left(\frac{E_i\;x}{1-(1-E_i)x}\right)^R \ .
\end{equation}
Thus $\rho _R(u)$ is the coefficient of $x^u$ in the above rightmost
expression. It is convenient to define
\begin{equation} 
\tilde g(x)=\sum_{i=1}^{\infty} \rho _R (u)\;x^{u-R}= 
\left(\frac{E_i}{1-x(1-E_i)}\right)^R
\end{equation}
We then obtain expression (\ref{pru}) from the identity
\begin{equation}
\rho _R(u)=\frac 1{(u-R)!} \left. \frac{\partial^{u-R} \tilde g}
{\partial x^{u-R}} \right|_{x=0} \ .
\end{equation}

\section{Stationary distribution of loglikelihood fields}
\label{appstack}

Assume that the sequence is repeated; hence we can drop the base index $i$ in
the definition of function $F_i$ (\ref{deffuncf}) and in the distribution $Q_i$
of the loglikelihood. We rewrite eqn (\ref{recurq}) 
under the form
\begin{equation} \label{recurq2}
Q (h') = \int _{-\infty}^\infty dh\, T_R( h',h) \, Q ( h) \ ,
\end{equation} 
where the kernel $T_R$ is defined through
\begin{equation} \label{recurq3}
T_R (h ',h) = \int _0^{\infty} d\tau \,P_R(\tau ) \; 
\delta \big( h ' - F(h, \tau ) \big)\ .
\end{equation}
In addition we define 
\begin{equation}
\tau_1 (h)= \frac{h + \Delta ^W}{x\, (e^{\Delta ^W} - 1)}  \quad , \quad
\tau_2 (h) = \frac{h + \Delta ^S}{x\, (1- e^{-\Delta ^S}) } \ .
\end{equation}
where we have used definition (\ref{tauiter}) for parameter $x$.
We now rewrite  
\begin{equation} \label{b4}
F(h, \tau ) = -h + x\, (e^{\Delta ^W} -1) \; \max 
( \tau_1 (h)- \frac {\tau}R, 0)  + x\, (1- e^{-\Delta ^S}) \; 
\min( \tau_2 (h)- \frac{\tau}R,0)
\end{equation}
The value of above function of $\tau$ depends on the relative values of
$\tau_1$ and $\tau_2$.
Let us make the hypothesis $(H1): \ e^{\Delta ^W} + e^{-\Delta ^S} > 2$.
Then, $\tau_1 < \tau_2$ if and only if $h > h_0$ with 
\begin{equation}
h_0 = \frac{ \Delta ^W\, ( 1- e^{-\Delta ^S}) - \Delta ^S \,
(e^{\Delta ^W} - 1)}{e^{\Delta ^W} + e^{-\Delta ^S} - 2}
\ .
\end{equation}
Assume in addition that $(H2): \ \Delta ^W \le \Delta ^S$. Then
\begin{equation}
h_0 \le \frac{ \Delta ^S\, ( 1- e^{-\Delta ^S}) - \Delta ^S \,
(e^{\Delta ^W} - 1)}{e^{\Delta ^W} + e^{-\Delta ^S} - 2} = - \Delta ^S
\ .
\end{equation}
We obtain from (\ref{b4}),
\begin{equation}
F(h, \tau )  = \left\{ \begin{array}{c c c}
\Delta ^W - \frac{\tau}R\, x\, (e^{\Delta ^W}-1) & \hbox{\rm if} & 
h> -\Delta^W \ \hbox{\rm and} \ \tau < \tau_1(h) \\
-h & \hbox{\rm if} & h> -\Delta^S \ \hbox{\rm and} \ \tau_1(h) < \tau < 
\tau_2(h) \\
\Delta ^S - \frac{\tau}R
\, x \, (1-e^{-\Delta ^S}) &\hbox{\rm if}& \tau > \tau_2(h)
\end{array} \right.
\end{equation} 
and the following expression for the kernel $T_R$ (\ref{recurq3}),
\begin{equation} \label{recurq4}
T_R (h ',h) = \left\{ \begin{array}{c c c}
P_R\big( - R\,\tau _1 (-h') \big) /(R\, x)/ (e^{\Delta^W}-1)& \hbox{\rm if} & 
h> - \min (h',\Delta ^W) \\
\delta (h'+h) \times \big[ \gamma ( R,R \,\max (0,\tau _1(h))) -
\gamma (R,R \,\tau_2(h)) \big] & \hbox{\rm if} & h> -\Delta ^S\\
P_R\big( - R\,\tau _2 (-h') \big) /(R\, x)/
 (1-e^{-\Delta ^S}) &\hbox{\rm if}& h'<\min(-h,\Delta ^S) \\
0 & \hbox{\rm if} & h' > \Delta ^S\ \hbox{\rm or} \ 
\Delta ^W \le -h<h'\le \Delta ^S
\end{array} \right.
\end{equation}
where $\gamma$ is the incomplete Gamma function (\ref{defgammaincomplete})
and distribution $P_R$ is defined in (\ref{dist2}). We then inject expression
(\ref{recurq4}) for $T_R$ in the fixed point eqn (\ref{recurq2}),
and integrate both sides over $h'$ over the interval $H \le h' \le \infty$.
As a result we obtain the remarkably simple identity
\begin{equation} \label{recur5a}
\hat Q (H) = A(H) - B(H) \; \hat Q (-H)
\end{equation} 
where the cumulative distribution $\hat Q$ is defined in (\ref{defqhat}),
and functions $A,B$ in (\ref{defab}). 

From (\ref{recurq4}) (fourth line) $Q(h')$ vanishes when $h'>\Delta ^S$.
Hence $Q(H)=0$ for $H > \Delta ^S$ (third line of (\ref{expressq})). 
Choose now $H< -\Delta ^S$; then $\hat Q(-H)=0$ and, from (\ref{recur5a}), 
$Q(H)=A(H)$ (first line of  (\ref{expressq})). Then we iterate 
(\ref{recur5a}) to obtain
\begin{equation} \label{recur5b}
\hat Q (H) = A(H) - B(H) \; \big[ A(-H) - B(-H) \; \hat Q (H) \big] \ ,
\end{equation} 
from which we extract the expression of $\hat Q(H)$ in the range
$-\Delta ^S \le H \le \Delta ^S$ (second line of  (\ref{expressq})).
It is easy to check that $\hat Q$ is a continuous function of
its argument both in $-\Delta^S$ and $+\Delta^S$. Notice that hypothesis
(H1,H2) hold for typical values of the binding free-energies.

It is quite remarkable that an exact analytical expression for $Q(h)$ 
is available for our model. Indeed the recurrence
equation for the field distribution of most disordered one-dimensional
cannot be solved in a closed form \cite{Luck}. Dyson noticed in his
original paper \cite{Dyson} that a case of solvable model is obtained
when the site disorder (here, the time $t_i$ spent on each base)
is exponentially distributed. The present study
generalizes this observation to the case of the convolution of exponentials.

\section{Calculation of the error $\epsilon$ and the correlation function
$\chi ^{dis}$} \label{app2b}

Assume the sequence is very long ($N\gg 1$), and consider the base at 
location $i$ far away from the extremities ($1\ll j\ll N$). 
Base $i$ can be predicted to be $b$ ($=W$ or $S$), with probability
\begin{equation} \label{pdag}
P_i^\dagger (b_i) = \exp(-R\,\pi _i^\dagger (b) )
\end{equation}
 depending on the stochastic set of times 
$\{t_i\}$ spent on the bases.  We look for the distribution of the 
loglikelihoods of base $i$,
\begin{equation}
Q^\dagger (h^\dagger) = \hbox{\rm Probability} \big[ 
h^\dagger = \pi _i^\dagger (S) - \pi_i^\dagger (W)  \big]
\end{equation}
where the probability is calculated over the sets of sojourn
times $\{t_i\}$. Notice that we do not expect $Q^\dagger$ to vary with $j$ 
in the bulk of the repeated sequence (see calculation of the correlation 
function below).  

$Q^\dagger$ does not coincide with the distribution $Q$ of fields 
used in the iteration equation (\ref{recurq}). Indeed the latter merely
expresses the dependence of the loglikelihood over base $i+1$ upon 
the choice for base $i$, independently of what happens at site $i+2$.
In other words, eqn (\ref{recurpi}) is a propagation equation
for the left-to-right likelihoods
$\pi^\to_{i}$; the $\to$ subscript has been omitted so far to lighten
notations. The direction of propagation is arbitrary: it corresponds
to the choice of running the Viterbi algorithm from the first to
the last base, determining the value of this last base, and then
deducing the values of all bases from the last one to the first one. 
Clearly, we could have decided to run the Viterbi procedure in the 
opposite direction. The recurrence equation for 
the right-to-left likelihoods $\pi^\leftarrow _i$ is straightforwardly 
established, and reads
\begin{equation} \label{recurqpim}
\pi_{i} ^\leftarrow(b _i) = \min _{b_{i+1}} \big( \pi_{i+1}^\leftarrow 
(b_{i+1}) - g_0(b_i,b_{i+1})
+r\, e^{g_0(b_i,b_{i+1})}\, t_i/R \big) \ .
\end{equation}
When the binding energy matrix is symmetric, the above recursion can be
rewritten as 
\begin{equation} \label{recurqpim2}
\pi_{i} ^\leftarrow(b _{i+1}) = \min _{b_{i}} \big( \pi_{i+1}^\leftarrow 
(b_{i}) - g_0(b_i,b_{i+1}) +r\, e^{g_0(b_i,b_{i+1})}\, t_i /R\big) \ ,
\end{equation}
and is identical to the recurrence equation (\ref{recurpi}) for $\pi^\to$.
We deduce that the stationary probability distribution of right-to-left
fields, $h^\leftarrow _i = \pi_{i}  ^\leftarrow (S) - 
\pi_{i}  ^\leftarrow (W)$, is  equal to the left-to-right
field distribution $Q$. 

Obviously, the actual prediction for base $i$ is the base $b_i$ maximizing
$P_i^\dagger$ (\ref{pdag}) and  depend on
the bases located on both left and right sides, that is, on 
left-to-right and right-to-left likelihoods, 
\begin{equation}
P _i ^\dagger (b_i) = P _i ^\to (b_i) \times P _i ^\leftarrow (b_i) 
\quad i.e. \quad
\pi _i ^\dagger (b_i) = \pi _i ^\to (b_i) + \pi _i ^\leftarrow (b_i) 
\quad ,
\end{equation}
when taking the logarithm.
Translating the above equation in terms of fields we obtain
\begin{equation}
h ^\dagger_i =  h_i^\to + h_i^\leftarrow 
\quad .
\end{equation}
A symbolic representation of the above equality is proposed in 
Fig.~\ref{dess}A. The distribution of 'true' likelihoods is thus
given by
\begin{equation}
Q^\dagger (h^\dagger ) = \int dh^\to Q(h^\to)  \int dh^\leftarrow 
Q(h^\leftarrow) \; \delta \big ( h ^\dagger - h^\to - h^\leftarrow \big)
\quad .
\end{equation}
The error in predicting base $i$ is therefore, 
\begin{equation}
\epsilon ^W = \int _{-\infty} ^ 0 dh^\dagger Q^\dagger (h^\dagger) 
= 1 - \int _{-\Delta ^S} ^{\Delta ^S} dh\, Q(h)\, \hat Q(h) 
\qquad , \quad
\epsilon ^S = \int _0 ^{\infty}  dh^\dagger Q^\dagger (h^\dagger) 
= \int _{-\Delta ^S} ^{\Delta ^S} dh\, Q(h)\, \hat Q(h) 
\end{equation} 
for repeated sequences of $W$ or $S$ bases respectively, see formulae
(\ref{omegaww},\ref{omegass},\ref{omegasw}). 
We have here used definition (\ref{defqhat}) for the cumulative distribution 
$\hat Q$ of fields. 

A similar approach can be used to calculate the 
disconnected nearest neighbor correlation function $\chi ^{dis}$ 
(\ref{ccdis}). Assume for simplicity that the true sequence 
is a repeated sequence of $S$ bases, and consider the two bases at 
locations $i$ and $i+1$. Call $h_i^\to$ and $h_{i+1}^\leftarrow$ the
left-to-right and right-to-left likelihoods incoming onto bases $i$
and $i+1$ respectively, see Fig.~\ref{dess}B.
Let $n_i=1$ if base $i$ is (correctly) predicted 
to be $S$, 0 if the prediction is $W$. We define a similar variable, 
$n_{i+1}$, attached to site $i+1$. Finally call $\tau$ the 
normalized sojourn time on base $i$ with distribution (\ref{dist2}).
Given a pair of incoming 
likelihoods $(h_{i}^\to, h_{i+1}^\leftarrow)$ and the sojourn time
$\tau$, the Bayesian prediction for $(n_i,n_{i+1})$ is 
\begin{equation}
\big(n_i(h_{i}^\to, h_{i+1}^\leftarrow,\tau),n_{i+1}(h_{i}^\to, h_{i+1}^
\leftarrow,\tau)\big) = \hbox{\rm arg}\!\!\max _{(n,n')} 
\Xi^{SS} (n,n',h_{i}^\to, h_{i+1}^\leftarrow,\tau) 
\end{equation}
where 
\begin{equation} \label{defpiapp2b}
\Xi ^{SS} (n,n',h,h',\tau) = h \, (1-n) +  h'\, (1-n') +
\bar g^{SS}(n,n') - \frac {\tau}R \exp \; \bar g^{SS}(n,n')   
\end{equation}
and
\begin{eqnarray} \label{defgbarapp2b}
\bar g ^{SS}(n,n') &=& g_0 (S,S) (\,n\,n'-1) + g_0 (W,W) \,(1-n)\,(1-n') + 
 g_0 (W,S) \,\big[ n\, (1-n') + n'\, (1-n)\big] \nonumber \\
&=&  n\,n'\, ( \Delta ^W - \Delta ^S ) +
(n+n') \, \Delta ^W + \Delta ^W + \Delta ^S \ .   
\end{eqnarray}
The correlation function between $n_i,n_{i+1}$ is
\begin{equation} \label{chi89}
\langle n_i \, n_{i+1} \rangle  = \int d\tau \, P_R(\tau) 
\int dh_{i}^\to\, Q(h_{i}^\to) \,
dh_{i+1}^\leftarrow \, Q(h_{i+1}^\leftarrow )\; 
\delta _{n_i(h_i^\to , h_{i+1}^\leftarrow ,\tau),1} \;
\delta _{n_{i+1}(h_i^\to , h_{i+1}^\leftarrow ,\tau),1}
\end{equation}
where $\delta_{a,b}=1$ if $a=b$, 0 otherwise. An inspection of 
(\ref{defpiapp2b}) shows that both bases are correctly predicted 
to $S$ when  $h_i^\to$ and $h_{i+1}^\leftarrow$ are both 
smaller than $-\Delta^S + \frac {\tau}R (e^{\Delta ^S}-1)$. Hence 
formula (\ref{omegass}). Formulae (\ref{omegaww}) and (\ref{chisw}) for
repeated WW and SW sequences can be obtained along the same lines
through substitution of (\ref{defpiapp2b}) and (\ref{defgbarapp2b}) with, 
respectively,
\begin{eqnarray} 
\Xi ^{WW} (n,n',h,h',\tau) &=& h \, n +  h'\, n' +
\bar g^{WW}(n,n') - \frac {\tau}R \exp \; \bar g^{WW}(n,n')   
\\
\bar g ^{WW}(n,n') &=& n\,n'\, ( \Delta ^W - \Delta ^S) +
(n+n') \, \Delta ^S - \Delta ^W - \Delta ^S  \ ,    \nonumber 
\end{eqnarray}
and
\begin{eqnarray} 
\Xi ^{SW} (n,n',h,h',\tau) &=& h \,(1- n) +  h'\, n' +
\bar g^{SW}(n,n') - \frac {\tau}R \exp \; \bar g^{SW}(n,n')   
\\
\bar g ^{SW}(n,n') &=& n\,n'\, ( \Delta ^S - \Delta ^W) -
n \,  \Delta^S + n' \, \Delta ^W
\ .    \nonumber 
\end{eqnarray}

\begin{figure}
\begin{center}
\psfig{figure=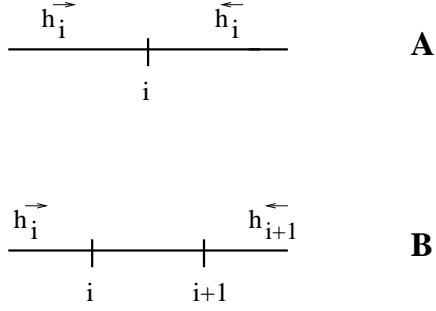,height=4cm,angle=0}
\end{center}
\caption[]{Symbolic representation of the calculation of the distribution of 
likelihood at site $i$ ({\bf A}), and the joint distribution at sites
$i$ and $i+1$ ({\bf B}). The left part of the sequence induces a
left-to-right likelihood $h_i^\to$ on base $i$, while the right part 
contribution is $h_i^\leftarrow$ ({\bf A}) or $h_{i+1}^\leftarrow$ ({\bf B}).
}
\label{dess}
\end{figure}

\section{Large $R$ asymptotic} \label{app2c}

A saddle-point calculation of the incomplete Gamma function 
(\ref{defgammaincomplete}) gives the following large $R$ asymptotic
for $z\ne 1$,  
\begin{equation} \label{gammasym}
\gamma ( R, R \, z) \simeq \theta (1-z) + \frac{\exp\big[ -R ( z-1-\ln z)\big]}
{\sqrt{2\pi R}\; (z-1)} 
\end{equation} 
where $\theta$ is the Heaviside function: $\theta(1-z)=1$ if
$z<1$, 0 if $z>1$. Application of this formula to the error (\ref{pnostack4})
in the no-stacking case  yields the large $R$ scaling of $\epsilon$ 
in (\ref{nostackomega}).

Consider now the case of stacking interactions between neighboring bases.
We first calculate the cumulative distribution $\hat Q$ (\ref{expressq}) 
of likelihoods in the $R\to\infty$ limit, then derive finite $R$ corrections.
With definitions (\ref{defab},\ref{tauiter}) we obtain, in the infinite $R$ 
limit,
\begin{equation}\label{asymab}
A(h) \to \theta ( h^S - h)
\ , \quad
B(h) \to \theta ( h - h^W) - \theta (h-h^S)
\end{equation}
where 
\begin{equation}\label{hwhs}
h^W =  \Delta ^W - x\, \big(e^{\Delta^W} -1\big)
\ , \quad 
h^S = \Delta ^S - x\, \big(1- e^{-\Delta^S} \big) \ .
\end{equation}
For repeated sequences of, respectively, bases $W$ and $S$, we have 
$x=e^{-\Delta ^W}$ and $x=e^{\Delta ^S}$. 
It is a simple check that, whatever the value of $b$, $h^W$ and $h^S$ have 
the same sign (positive for the $WW$ sequence, negative
for the $SS$ sequence). Thus the product $B(h)\, B(-h)$ in (\ref{expressq}) 
vanishes. We find that the cumulative distribution $\hat Q(h)$ of fields is 
a step function. More precisely, 
\begin{equation}
Q(h) \to \delta (h - h_\infty ^b) \quad \hbox{\rm where}
\quad h_\infty ^W = \Delta ^W - 1 + e^{-\Delta^W} \ , \
h_\infty ^S = -\Delta ^S +1 + e^{-\Delta^S}\ ,
\end{equation}
from which we deduce that the error in predicting a base vanishes in the
large $R$ limit.
The case of the alternate sequence $SW$ is more complicated. Setting $x=1$
in (\ref{hwhs}) we have $h^W <0$ and $h^S>0$. Using (\ref{asymab}) and 
(\ref{expressq}) we merely obtain $\hat Q(h)=1$ for $h < h^W$, 
$\hat Q(h)=0$ for $h > - h^W$ and 
\begin{equation} \label{mixed}
\hat Q(h) = 1 - \hat Q(-h) \qquad (h^W < h < -h^W) \ .
\end{equation}
Though (\ref{mixed}) is not sufficient to characterize the
likelihood distribution it allows us to calculate the error from 
(\ref{omegasw}), with the result (\ref{omegaswasym}).

Let us now calculate the corrections to the infinite $R$ limit. The
calculation of the error $\epsilon$ is similar for $WW$ and $SS$ sequences,
and is thus reproduced below in the $WW$ case only. Let us introduce
\begin{equation}
\alpha (h) =  \frac{\Delta ^S -h}{x ( 1- e^{-\Delta ^S})} \quad ,
\quad \beta (h) =  \max \bigg( 0, \frac{\Delta ^W \pm h}{x ( 
e^{\Delta ^W}-1)} \bigg) \qquad ( x = e^{-\Delta ^W}) \ .
\end{equation}
Using the large $R$ expansion (\ref{gammasym}) for the functions $A$ and $B$ in
(\ref{defab}) we obtain from (\ref{expressq}) the asymptotic expression for
the cumulative distribution of loglikelihoods
\begin{equation}
\hat Q (h) = 1 - \frac{\exp\big[ -R \big( \beta (h) -1-\ln \beta (h)
\big)\big]} {\sqrt{2\pi R}\; (\beta (h)-1)} 
\end{equation}
and, through differentiation with respect to $h$,
\begin{equation}
Q (h) = \sqrt{\frac R{2\pi}}\, \frac{\beta(h)}{1- e^{-\Delta ^W}}\,  
\exp\big[ -R \big( \beta (h) -1-\ln \beta (h) \big)\big] \ .
\end{equation}
These expressions hold when $\beta (h) < \alpha (h)$. This condition
happens to be fulfilled for the choice of parameters of Section 
\ref{stacksec}, and in the vicinity of $h=0$. From (\ref{omegaww}) we
have
\begin{eqnarray}
\epsilon ^{WW} &=& \int _{-\Delta ^S} ^{\Delta^S} dh\, Q(h)\, 
\big[ 1- \hat Q(-h) \big] \nonumber \\
&=&  \frac{ \exp[- R\, ( \beta(0)-1-\ln \beta(0))]}
{2\pi \, (1- e^{-\Delta ^W})} \; \int _{-\Delta ^S} ^{\Delta^S} dh\,
\frac{\beta(h)}{\beta(-h)-1}\, \exp\bigg[ R \, \ln \bigg( 1 -
\bigg(\frac h{\Delta ^W} \bigg) ^2 \bigg) \bigg] \ . 
\end{eqnarray}  
The dominant contribution to the integral comes from the $h\simeq 0$
region. Expanding the integrand to the second order in $h$ and
calculating the Gaussian integral we obtain expression (\ref{epsilonwwss}).

Finally we consider the case of finite temperature prediction of Section 
(\ref{finitetemp}) for a base $b$ ($S$ or $W$), in the absence of
stacking. Let $\Delta $ be the difference of free-energy between the 
two base types, and $\tau$ given by (\ref{deftau84}).
Integrating (\ref{omegat=1}) by part and 
performing the change of variable $\tau = R(x+\Delta)/(e^{\Delta}-1)$,
we obtain the  following expression for the error,
\begin{eqnarray}
\epsilon _R(T=1) &=&  \int _ 0 ^\infty dx \bigg[ 1- \gamma\left(R, 
\frac{R(x+\Delta^S)}{e^{\Delta^S}-1} \right) \bigg] \;
\frac{R \, e^{-R\,x}}{(1+e^{-R\,x })^2} \label{er23} \\
&=&\sqrt{\frac{R}{2\pi}} \int _ 0 ^\infty dx\;   \frac{ \exp[- R\, G(x)]}
{1 - (\Delta^S +x)/(e^{\Delta ^S}-1)} \label{er24}
\end{eqnarray}
where we have made use of (\ref{gammasym}) to obtain (\ref{er24}) from
(\ref{er23}), and have defined 
\begin{equation}
G(x) = \frac{\Delta^S +x}{e^{\Delta^S}-1} -1 - \ln \left( 
 \frac{\Delta^S +x}{e^{\Delta^S}-1} \right) + |x| \ .
\end{equation}
The maximal contribution to the integral comes from the $x=0$ region,
with $G(0)=\tau-1-\ln \tau$.
Defining $\tilde x= R \, x$ and expanding $G$ around $x=0$ to the first
order, we obtain
\begin{equation}
\epsilon _R(T=1) =  \frac {e ^{- R \, (\tau -1 -\ln \tau )}}
{\sqrt {2\pi R } \; (1-\tau) } \; \int _ {-\infty} ^\infty d\tilde x \;
\frac{e^{-\tilde x (1- \sigma) }}{(1+e^{-\tilde x })^2} =
\frac {e ^{- R \, (\tau -1 -\ln \tau )}}
{\sqrt {2\pi R } \; (1-\tau) } \; \frac{\pi\, \sigma}{\sin (
\pi \, \sigma)}
\end{equation}
where $\sigma = |G'(0)|$ is given by (\ref{deftau84}).

\section{Calculation of the escape probability $E_i$}
\label{sec93}

In this appendix we calculate the escape
probability $E_i$ that the fork moves away 
from base pair $i$ (never reaches it back) after its first visit. 
Assume the fork starts its motion from base $j$. We define $p_{j}^{(i)}$ 
as the probability that
the fork will never reach position $i$ at any future instant.
This probability is larger than
zero when $i<j$ since the walk is transient. 
Given the bp index $i$ the probabilities $p_{j}^{(i)}$s fulfill the recursion
relation
\begin{equation}
p_{j} ^{(i)} = q_{j}\, p_{j-1}^{(i)} + (1-q_j)\, p_{j+1}^{(i)}
\label{recrel}
\end{equation}
where, in analogy with definition (\ref{defq}) for a repeated
sequence,
\begin{equation}
q_j = \frac{e^{g_s(f)}}{ e^{g_s(f)}+ e^{g_0(b_j,b_{j+1})}}
\label{defq_i}
\end{equation}
is the probability that the next base visited by the fork in $j$ is
$j-1$. Equation
(\ref{recrel}) is complemented by the boundary $p_i^{(i)}=0$ and
$p_N^{(i)} = 1$. Mathematically speaking the probability of not
reaching $i$ from $N$ is not equal to unity since a random walk on a
finite sequence is recurrent. However this approximation is
quantitatively excellent for the long sequences considered here.
Defining
\begin{equation} \label{ric0} 
E^{(i)}_j= \frac{p _{j} ^{(i)} }{p _{j+1} ^{(i)} }
\end{equation}  
we obtain the Riccati recursion relation
\begin{equation}
E^{(i)}_j=0\,; \hspace{2cm}
E^{(i)}_{j+1}=\frac{1-q_{j+1}}{1-q_{j+1}\;E^{(i)}_j}
\hspace{1cm} {\mbox {for}\ j\geq i} \ . 
\label{ric}
\end{equation}
We have solved equation (\ref{ric}) numerically for the
$\lambda$--phage sequence. The escape probability from $i$ is then
obtained from (\ref{ric0}) and (\ref{un}),
\begin{equation}
E _i = \frac 1{p_{i+1}^{(i)}}= \prod_{j\geq i+1} {E^{(i)}_j} \ .
\label{pj}
\end{equation}

\section{Average error $\epsilon$ at finite force}

\subsection{Case of one-way unzippings}
\label{phadia}

\begin{figure}
\begin{center}
\epsfig{file=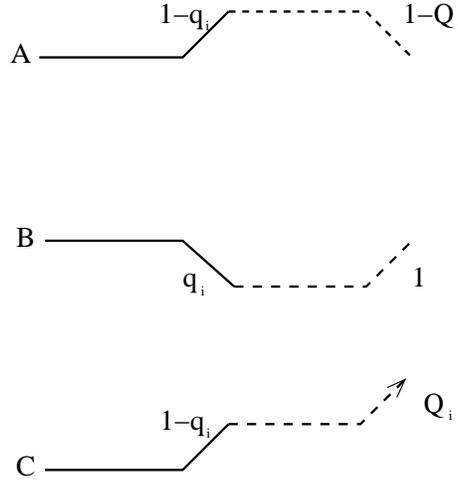,width=6cm}
\caption{Patterns $A$, $B$, and $C$ present in the transition trace
around base pair $i$. See text for definition.}
\label{path}
\end{center}
\end{figure}
In this appendix we calculate the average prediction error after
one unzipping over the
distribution of the unzipping
time traces. For a given time-trace, the
 prediction error depends only on the observed
 set $\{t_i,u_i,d_i\}$ 
of times $t_i$ spent on each base, and numbers of  opening
($u_i$) or closing ($d_i$) of each base. 
To make the average  we have  to calculate  the distribution
$P_1(\{t_i,u_i,d_i\})$   of such sets on all the time traces. 
$P_1(\{t_i,u_i,d_i\})$  is therefore the product of the probability to observe
a set of $\{t_i,u_i,d_i\}$ in a given time trace 
(given in equation~ \ref{pdetb}) time the
multiplicity of such a set $\{t_i,u_i,d_i\}$ on all the possible time
traces.

 Let us start by calculate the
 distribution $P_1(u_i,d_i)$ 
 ignoring for a while the time $t_i$ spent
on this base. 
Let us focus on base $i$; the sequence of opening and closing
transitions around this base, hereafter referred to as transition
trace, can be decomposed into three kinds of elementary patterns
schematized in Fig~\ref{path}, and labeled with letters A, B and C:
\begin{itemize}
\item Pattern A (Fig~\ref{path}A) corresponds to staying on base $i$
for some time, moving forward ($i\to i+1$, probability $1-q_i$), then
coming back to $i$ after a random walk throughout the upper part of
the sequence ($i+1 \ldots N$) with probability $1-E_i$. The probability
of pattern A is thus $P_A= (1-q_i)\times (1-E_i)$.
\item Pattern B (Fig~\ref{path}B) corresponds to staying on base $i$
for some time, moving backward ($i\to i-1$, probability $q_i$), then
coming back to $i$ after a random walk throughout the lower part of
the sequence ($i+1 \ldots N$) with probability $1$. The probability
of pattern A is thus $P_B=q_i$.
\item Finally, pattern C (Fig~\ref{path}C) corresponds to staying on
base $i$ for some time, moving forward ($i\to i+1$, probability
$1-q_i$), without ever coming back to this base later on (probability
$E_i$). This final pattern has probability $P_C = (1-q_i)\times
E_i$.
\end{itemize}
The number of closing transitions in a transition trace, $d_i$, is simply equal to the
number of B patterns  around base $i$.
Similarly, the number of opening transitions, $u_i$, is the sum of the
numbers $N_A$ and $N_C$ of A and C patterns respectively.  As $N_C=1$
by definition, we have $N_A=u_i-1$. A and B patterns can be randomly
located in the transition trace and are followed by one C pattern, the
distribution  $P_1(u_i,d_i)$ on the ensemble of transition trace is therefore:
\begin{equation}
P_1(u_i,d_i)= {u_i-1+d_i \choose d_i} \;q_i^{\,d_i}\,\left[(1-q_i)\,(1-E_i)
\right]^{\,u_i-1}\,E_i\,(1-q_i)\;. \
\qquad (u_i\ge1,d_i\ge0) \ .
\end{equation}
Let us now focus on the total time $t_i$ spent on base $i$. It is the
sum of $u_i+d_i$ times each exponentially distributed with average
sojourn time
\begin{equation} \label{sojbis}
\langle t_i\rangle = \frac{1}{r\, (e^{g_0(b_i^L,b_{i+1}^L)} + e^{g_{s}(f)})}
\end{equation}
Thus, $\tau_i = t_i/\langle t_i \rangle$ is a stochastic variable
obeying distribution $P_R$ (\ref{dist2}) where $R=u_i+d_i$ plays the
role of a fictitious number of unzippings.  We obtain the joint probability
of time $\tau_i$, opening and closing moves $u_i$ and $d_i$,
\begin{equation}
P_1(\tau _i ,u_i,d_i)= \frac{q_i^{\,d_i}\left[(1-q_i)\, 
(1-E_i) \right]^{\,u_i-1}\,E_i\,(1-q_i) \;e^{-\tau_i}\,
\tau_i^{\,d_i+u_i-1}}{d_i!\;(u_i-1)!}\,.
\end{equation} 
Summation over all values for $d_i$ lead to the (single base)
probability for unzipping data
\begin{equation}
P_1(\tau _i,u_i)=\frac{E_i\,(1-q_i)}{(u_i-1)!}\,\left[(1-q_i)\,
(1-E_i)\, \tau_i \right]^{\,u_i-1} \;e^{-\tau_i(1-q_i)}\ .
\label{ptu}
\end{equation}
Neglecting stacking effects between bases, the content $b_i$ of base
$i$ is chosen to maximize the probability
\begin{equation}\label{pbisenzadi}
P(b_i | \tau _i,u _i) = \frac{ \exp \bigg( g_0(b_i)\,u_i - r \,
 e^{g_0(b_i)} \, \langle t_i \rangle\, \tau _i \bigg)}{\exp \bigg(
 g_0(W)\,u_i - r \, e^{g_0(W)} \, \langle t_i \rangle\, \tau _i
 \bigg)+ \exp \bigg( g_0(S)\,u_i - r \, e^{g_0(S)} \, \langle t_i
 \rangle\, \tau _i \bigg)} \ ,
\end{equation}
where the average sojourn time is given by eqn (\ref{sojbis}).  This
maximization can be done along the lines of Section
\ref{nostackingtheory} devoted to the case of infinite force. We find
the average fraction of mispredicted bases at force $f$,
\begin{equation} \label{pnostack3bis}
\epsilon _{f,1} ^W = \sum _{u_i\ge 1}\int_0 ^{\frac{u_i \tau_i^W}{1-q_i}}
d\tau_i \, P_1(\tau_i,u_i) \quad , \quad
\epsilon _{f,1} ^S = \sum _{u_i\ge 1} \int_{\frac{u_i
\tau_i^S}{1-q_i}} ^{\infty} d\tau_i \, P_1(\tau_i,u_i) \ ,
\end{equation}
with definition (\ref{ptu}) for $P_1$. Hence eqn (\ref{errf}). 

\subsection{Case of two-way unzippings}
\label{appdouble}

We now suppose that the sequence is opened in both ways, and denote
by $\sigma=+$ the left-to-right and $\sigma=-$ the right-to-left
openings respectively. Let $u_i ^\sigma, \tau _i^\sigma$ 
denote the number of openings of bp $i$ and the 
time spent by the fork on $i$ for both directions ($\sigma =\pm$). 
The joint distribution of $u_i^\sigma,\tau_i^\sigma$ is $P_1$ (\ref{ptu}) 
with $q_i,E_i$ replaced with, respectively, $q_i^\sigma$, the probability to
close back bp $i$ when the fork is in $i$, and $E_i^\sigma$, the escape 
probability from base $i$ in the $\sigma$ direction. $q_i^+$ and $E_i^+$ are
simply given by (\ref{defq_i}) and (\ref{pj}) respectively. In addition
$q_i^-=q_{N-i+1}^+$, and $E_i^-$ can be obtained along the lines of
Appendix \ref{sec93}.

As the unzippings in both directions produce statistically
uncorrelated data the joint distributions of $u_i^+,\tau_i^+$ and
$u_i^-,\tau_i^-$ factorize.  The Bayesian probability that base $i$ is
of type $b_i$ is simply given by (\ref{pbisenzadi}) with $u_i=u_i^+ +
u_i^-$, $\tau_i = \tau_i^+ + \tau_i^-$.  In the framework of Maximum
Likelihood Prediction we maximize this quantity to obtain the error on
base $i$,
\begin{equation}
\epsilon_{f,1}^{b_i} = \sum_{u_i^+,u_i^-\geq 1} \;\rho_1(u_i^+) \;
\rho _1(u_i^-)\; \epsilon_{u_i^+,u_i^-}^{b_i} \ ,
\end{equation}
where
\begin{eqnarray}
\epsilon_{u_i^+,u_i^-}^W = \int_0^{+\infty} dx\, dy\; \theta\left(
x+y-\tau^w\left(\hat{u}_i^++\hat{u}_i^-\right) \right)\,\frac{e^{-x}\,
x^{\,(u^+-1)}}{(u_i^+-1)!}\;\frac{e^{-y}\;
y^{\,(u_i^--1)}}{(u_i^--1)!}\nonumber\\
\epsilon_{u_i^+,u_i^-}^S=\int_0^{+\infty} dx\,
dy\;\theta\left(\tau^s\left(\hat{u}_i^++\hat{u}_i^-\right)
-x-y\right)\, \frac{e^{-x}\;
x^{(u^+-1)}}{\,(u_i^+-1)!}\;\frac{e^{-y}\;
y^{\,(u_i^--1)}}{(u_i^--1)!}
\end{eqnarray}
and $\rho _1$ is defined in (\ref{defrho1}) (beware of the dependence
of $E_i^\sigma$ on the unzipping direction $\sigma$).

The generalization to the case of $R/2$ unzippings in each direction
is done along the lines of Section~\ref{phadia}, with the immediate
result
\begin{equation}
\epsilon_{f,R}^{b_i}= \sum_{u_i^+,u_i^-} \rho_{R/2}(u_i^+) \;
\rho_{R/2}(u_i^-) \;\epsilon_{u_i^+,u_i^-}^{b_i}\ ,
\end{equation}
where $\rho_{R/2}$ is the $(R/2)^{th}$ convolution power of the
probability $\rho_1$, see eqn (\ref{rhor}).

\section{Calculation of $R_c$ in presence of DNA strands fluctuations}
\label{app3}

Let  $\hat T_i ^r$ be the number of measures 
where the fork is really at location $i=0,1,\ldots,N$. 
These integer numbers are stochastic and distributed according to,
given the sequence $B$,
\begin{equation}
\hbox{\rm Proba}[\{ \hat T_i^r\}|B] = \prod _{i }
e^{ -\Delta t \,r_o(b_i)\, \hat T_i^r} \; \bigg( 1 - 
e^{ -\Delta t \,r_o(b_i)} \bigg) \ .
\end{equation} 
The number of times the fork is apparently at position $j$, $\hat T_j^a$,
given the set of $\hat T_i^r$,  is stochastic too. Their probability is 
given by
\begin{equation}
\hbox{\rm Proba}[\{\hat  T_j^a\}|\{\hat  T_i^r\}] = 
\sum _{\{f_{ij} =0,1,2,\ldots\}} \prod _{i} 
\left\{ \frac {\hat T_i ^r!}{\prod _j f_{ij}!} 
\prod _j  [A_{j,i} ]^{f_{ij}} \;
\delta (\hat T_j^a, \sum _i f_{ij}) \right\}
\delta (\hat T_i^r, \sum _j f_{ij})
\end{equation}   
where $\delta (a,b)=1$ if $a=b$, 0 otherwise is the Kronecker
function, and the fluctuation matrix $A$ is defined in
(\ref{defflua}). It is convenient to work with the generating function
of the $\{\hat  T_i^a\}$, 
\begin{equation}
G_1( \{y_j\}|B) = \sum _{\{\hat T_j^a\},\{\hat T_i^r\}} 
\hbox{\rm Proba}[\{ \hat T_j^a\}|\{ \hat T_i^r\}] \;\hbox{\rm 
Proba}[\{ \hat T_i^r\}|B] \; \prod _j e^{y_j \, {\hat T_j^a}}
= \prod _{i } \left( \frac{  1 - e^{ -\Delta t \,r_o(b_i)}}
{1-  e^{ -\Delta t \,r_o(b_i)}  \sum _{j}  A_{j,i}\; e^{y_j }}\right)\ .
\end{equation}  
The generating function of the probability distribution of the
apparent times $t^a_j=\hat T^a_j\times\Delta t$ is simply $G_1( \{y_j\, 
\Delta t\}|B)$.

The above expression for $G_1$ holds for one unzipping. For $R$ 
unzippings the generating function $G_R$ is simply given by the
$R^{th}$ power of $G_1$. In the large $R$ limit we obtain 
\begin{equation}
G_R( \{y_j\, \Delta t\}|B) = \exp\big[ -R \; \sum _i \ln (1 +
  \chi_i(\{y_j\})) \big]
\end{equation}
where, to the first order in $\Delta t$,
\begin{equation}
\chi_i(\{y_j\}) = \sum _{j} A_{j,i} \, y_j \big(\frac{\Delta t}2 - 
r_o(b_i)^{-1} \big)- 
\frac{\Delta t}2 \sum _j A_{j,i} \, y_j^2 \, r_o(b_i)^{-1} \ .
\end{equation}
Assume now that the true sequence $B^L$ is a repeated sequence of $S$
bases with a $W$ base at location $n$; we call $B^S$ the sequence made
of $S$ bases only. We furthermore assume that the free energy
difference $\Delta$ is small which makes inference harder. Using
$\rho$ defined in (\ref{defrho}) and introducing $s_j=y_j/r_o(S)$, we obtain
\begin{equation}
G_R( \{s_j\}|B^L) = \exp\big[ R \; \gamma (\{s_j\}|B^L) \big]
\end{equation}
where 
\begin{equation}\label{expressgamma}
\gamma (\{s_j\}|B^L) = -\sum _j s_j\, h_j(B^L) -\frac 12 \sum _{j,k}
s_j \, \beta_{j,k}\, s_k +O(s_j^3)\quad \hbox{\rm with} \quad
h_j (B^L)= 1-\frac \rho2 +\Delta \; A_{j,n}
\end{equation}
and matrix $\beta$ defined in (\ref{ludeau2}). Notice that
the expressions for $h$ and $\beta$ were obtained using the
approximation $\sum _i A_{i,k}=1$ for any $k$, and in the limit of
small $\rho,\Delta$.
The expression for $\gamma (\{s_j\}|B^S)$ is 
obtained from (\ref{expressgamma}) when $\Delta\to 0$.

We obtain the large deviation expression for the distribution $P_R$ of the
apparent times through the Legendre transform of $\gamma$,
\begin{equation} \label{legtra}
P_R  (\{ t_j^a = \tau _j^a /r_o(S)\}|B ) = \exp( - R \; \omega 
(\{\tau_i^a\}|B)\quad \hbox{\rm with} \quad  \omega (\{\tau_i^a\}|B)
= - \max _{\{s_j\}} \bigg[ \gamma (\{s_j\}|B) + \sum _j s_j\,\tau_j
\bigg]
\end{equation}
for the two sequences $B=B^L,B^S$. When $\Delta$ is small we expect
the distribution of apparent times for the two sequences to be very close
and thus the set of times $\{\tau _j^a\}^*$ for which they are equal will
be close to the most likely apparent times with both
distribution. This justifies the second order expansion in $s$ in 
(\ref{expressgamma}). The exponent $\omega^* =
\omega (\{\tau_i^a\}^*|B^L)= \omega (\{\tau_i^a\}^*|B^S)$ of the 
probability of this crossing time $\{\tau _j^a\}^*$ is equal, in the large 
$R$ limit, to the inverse of $R_c(n)$. This statement can be
    graphically understood from the  Figure 2 in
\cite{corto}. A more detailed explanation will be
given in \cite{altro}. The calculation of $\omega^*$ is immediate from
(\ref{legtra}) and leads to (\ref{ludeau2}). For $\rho=0$ the value
for $R_c(n)$ coincide with its expression (\ref{rcnostack}) in the
absence of ssDNA fluctuation.

We now turn to the analysis of the Viterbi algorithm in the limit 
$\Delta t=0$. The Laplace transform of the probability distribution 
$P_R^{(i)}$ of the deconvoluted time $\tau_i^d=t_i^d \, r_o(W)$ on
base $i$ is obtained from $G_R$ by applying the deconvolution kernel
$D$, with the result
\begin{equation}
\int _0^\infty dt _i^d\, P_R ^{(i)} (t _i^d ) \, e^{-y_i\,t _i^d} =
\prod _{j=0}^N \frac 1{(1+y_i \; C_{ij})^R}
\end{equation}
where $C_{ij}$ is defined in (\ref{defcni}).
The error in predicting base $i$ is then given by the integral of
$P_R^{(i)}$ over $\tau _i^d > \tau^W$ since $b^L_i=W$, see
(\ref{fr},\ref{pnostack4}),
\begin{equation}
\epsilon ^W _{R,n} = \int _{\frac{R\,\Delta}{1-e^{-\Delta}}}
 ^\infty dt _i^d\, P_R ^{(i)} (t _i^d ) = \int _o^\infty \frac{dx}R
\int _{-i\infty} ^{+i\infty} \frac{ds}{2i\pi}\; e^{R\,f(x,s)}
\end{equation}
where
\begin{equation} \label{deff}
f(x,s) = \left(x + \frac{\Delta}{1-e^{-\Delta}} \right)\, s -
\sum _i\ln ( 1+s\, C_{n,i}) \ .
\end{equation}
The result for $R=1$ unzipping is given by (\ref{error1}). In the
large $R$ limit we obtain expression (\ref{ludeau}) through a saddle-point 
calculation and a small $s$ expansion (valid for small $\Delta$). 
The saddle-point value for $x$ can be located in 0, or in
a strictly positive real value. This corresponds to the two cases
listed in (\ref{ludeau}).

\section{On sampling and large deviations of the error $\epsilon$}
\label{appi}

\begin{figure}
\begin{center}
A.\epsfig{file=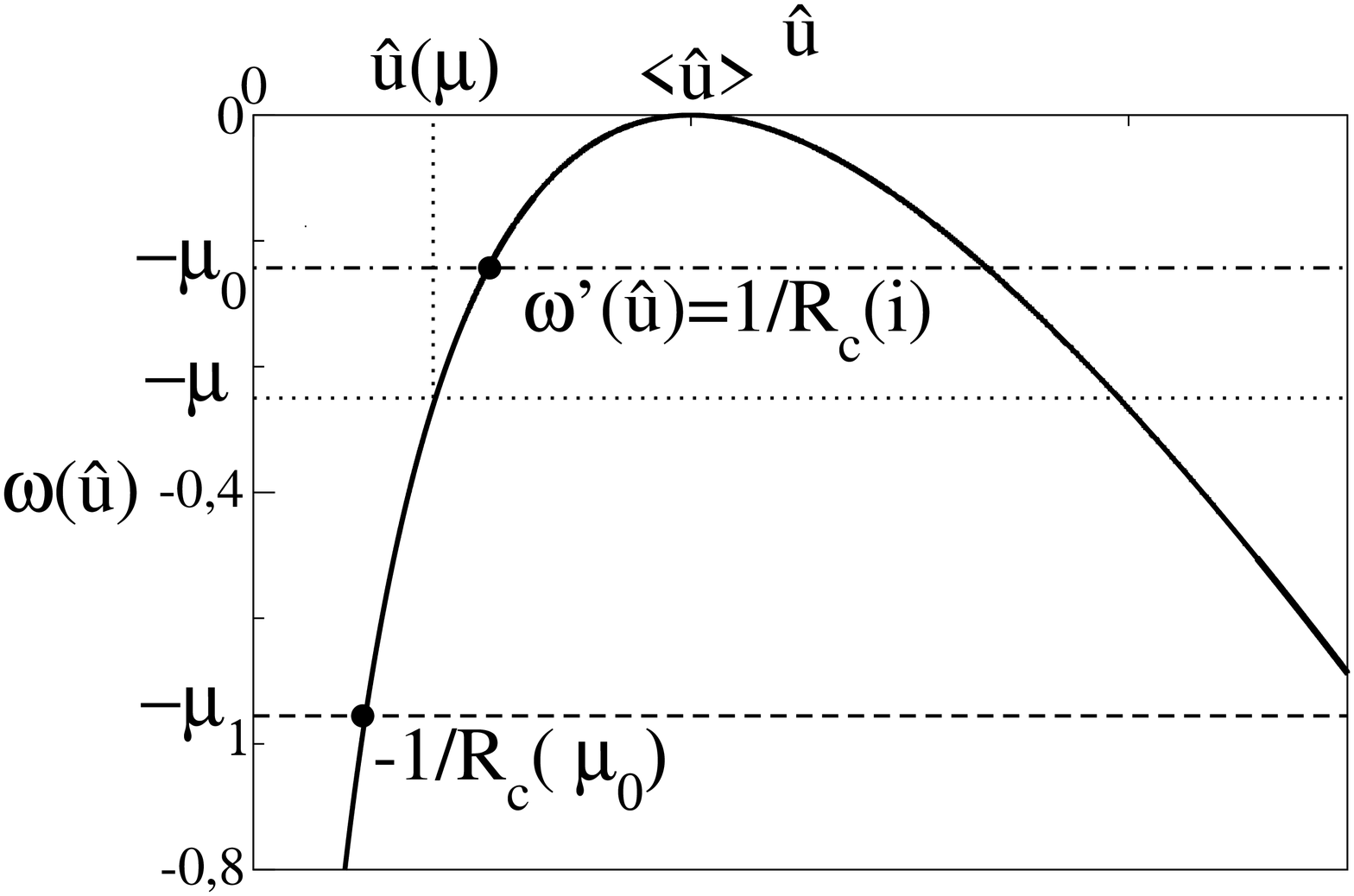,width=7cm}\hskip1cm
B.\epsfig{file=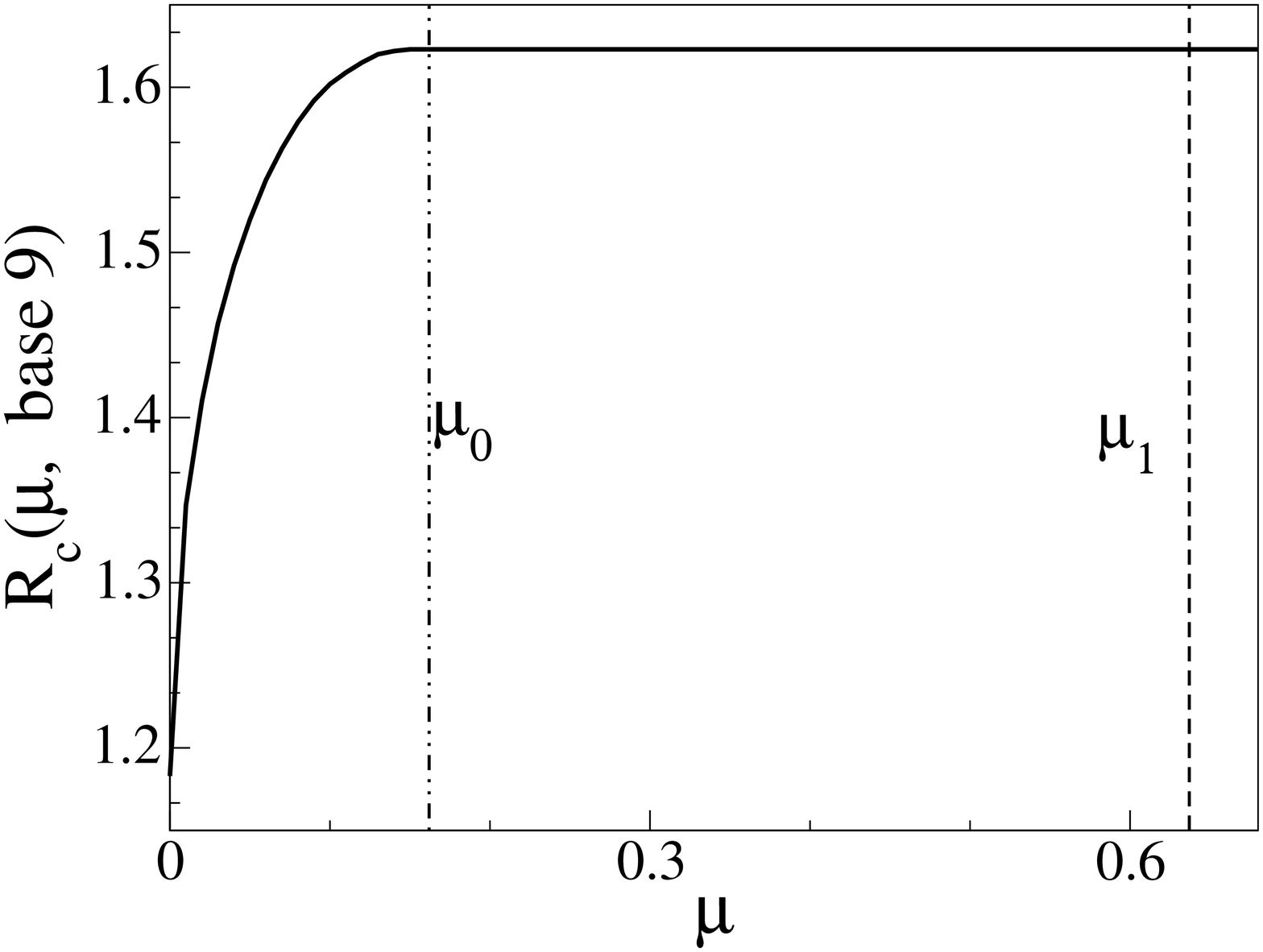,width=7cm} \vskip0.5cm
\caption{{\bf A}. rate function $\omega(\hat{u})$ governing the large
deviations of the number $\hat{u}$ of openings of a base per
unzipping. $\omega$ vanishes when $\hat{u}$ equals its average value,
$\langle \hat{u}\rangle$, and is strictly negative otherwise.  {\bf
B}. $R_c$ vs. logarithm of the number of samples, $\mu=\ln M/R$, for
the $9^{th}$ base of the $\lambda$-phage sequence.}
\label{lade}
\end{center}
\end{figure}

\begin{figure}
\begin{center}
\epsfig{file=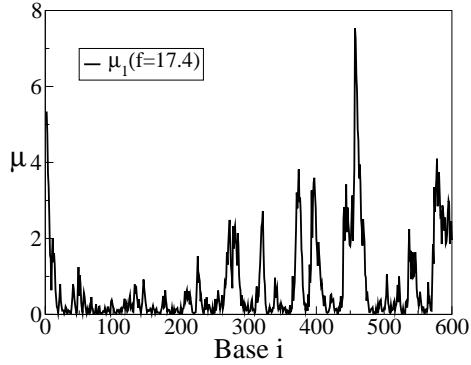,width=7cm}
\caption{Logarithm $\mu_1$ of the number of samples (divided by $R$)  to
obtain a good estimate of $R_c(i)$ vs base pair index $i$. $\mu_1$
strongly depends on the base index $i$ e.g. we need to sample over 
$M\sim e^{8 \times R}$ to accurately estimate $R_c$ for all bases. The
force is $f=17.4$ pN.}
\label{mu}
\end{center}
\end{figure}

We have calculated in Section (\ref{rcfiniteforcesec}) the average
fraction of mispredicted bases within the hypothesis of exact sampling
of the distribution of the number $u_i$ of openings. Let us turn to
the more realistic case of a finite number of samples, $M$. As $M$
decreases, the values of $u_i$ with exponentially small-in-$R$
probabilities are less and less likely to be sampled, leading to large
deviations corrections. Let us fix on a base pair dropping the base
index $i$ to shorten the notation. The values of $u$ which can be
found in a sample of size $M$ are the ones such that
\begin{equation} \label{condsamp}
\rho _R( u) \times M \gg 1 \ ,
\end{equation}
where $\rho _R$ is given in eqn (\ref{rhor}).  Assume that we keep fix
$R$ and scale the number of samples according to $M \sim e^{R\, \mu}$.
Upon introduction of the rate function for $\hat{u}=u/R$,
\begin{equation} 
\omega\left(\hat{u}\right)=\lim_{R\rightarrow \infty} \frac{1}{R} \ln
\rho_R \big ( R \, \hat{u}\big) =
(1-\hat{u})\,\ln\left(\frac{\hat{u}-1}{\langle
\hat{u}\rangle-1}\right)+\hat{u}\,\ln\left(\frac{\hat{u}}{\langle{\hat{u}}\rangle}\right)
,
\end{equation}
we rewrite condition (\ref{condsamp}) into
\begin{equation}\label{umu}
\omega\left(\hat{u}\right)\geq -\mu\ ,
\end{equation}
This condition is graphically solved in Fig \ref{lade}A. At fixed
$\mu$ a compact range of available values for $\hat{u}$ is obtained,
centered around the average number $\langle \hat{u}\rangle$ of
openings of a bp per unzipping. For instance, the smallest accessible
value, $\hat{u}(\mu)$, is obtained when solving condition (\ref{umu})
as an equality (Fig. \ref{lade}A).

For each sample $m=1,\ldots,M$ the measured error $\epsilon_R^m$ takes
value $v=0$ (if the base is correctly predicted) and $1$ otherwise,
with probabilities
\begin{equation}
P_v = \int _{\omega(\hat{u}) \ge - \mu} d\hat{u}\;
e^{R\;\omega(\hat{u})}\left[\left(1-e^{-R \hat{u}/{R_c}} \right)\;
\delta_{v,0}+ e^{-R \hat{u}/{R_c}} \; \delta_{v,1} \right] \ .
\end{equation}
We evaluate this probability through a saddle--point approximation,
\begin{equation}\label{sapo}
P_v = \left(1-e^{-R/R_c(\mu)}\right) \; \delta_{v,0} +
e^{-R/R_c(\mu)}\; \delta_{v,1}\ ,
\end{equation}
where
\begin{equation} \label{rcmu}
R_c(\mu)=\max _{\hat{u}\geq \hat{u}(\mu)} \left[ \frac{R_c}{ \hat{u}-R_c\,
\omega\left(\hat{u}\right)} \right] \ .
\end{equation}
Let us call $\mu_0=\omega(\hat u_0)$ where $\hat u_0$ is the root
of $\omega'(\hat u_0) = \frac 1{R_c}$, and
$\mu_1=\frac{1}{R_c(\mu_0)}=\mu_0+\frac{\hat{u}(\mu_0)}{R_c}>\mu_0$.
As $\omega$ depends on the bp $i$ so do
$\mu _0,\mu_1$. Then, 
\begin{itemize}
\item when $\mu<\mu_0$ the maximum on the r.h.s. of (\ref{rcmu}) is
  reached in $\hat u(\mu)$  fulfilling the equality (\ref{umu}), and
  is an increasing function of $\mu$ (Fig~\ref{lade}B).
\item when $\mu_0\leq\mu\leq \mu_1$ $R_c(\mu)=R_c(\mu_0)=\hat{R}_c$ 
does not depend on $\mu$ anymore (Fig \ref{lade}B). The average number of
erroneous samples reads 
\begin{equation} \label{merr}
M_{err} = M\; e^{-R/R_c(\mu)} = e^{R\, (\mu - 1/R_c(\mu_0))}
\end{equation}
and is exponentially small in $R$ by the very definition $\mu_1$,
Hence no erroneous sample is detected and no estimate of $R_c$ can be made.
\item when $\mu>\mu_1$ $M_{err}$ is exponentially large (\ref{merr}),
 and the decay constant of the error can be safely measured and
 estimated to be $R_c(\mu_0)$. 
\end{itemize}
Figure \ref{mu} shows $\mu_1$, the logarithm (divided by $R$) 
of the number of samples needed to accurately estimate $R_c$, as a function of
the base index $i$. We observe that $\mu _1$ varies a lot from base to
base.

\end{document}